%% file: Manuscript.tex
\documentclass[a4paper,11pt]{article}
\usepackage[margin=2.5cm]{geometry}
\usepackage{jheppub}
\usepackage[utf8]{inputenc}
\usepackage{microtype}
\usepackage{lmodern}
\usepackage{inconsolata}
\usepackage{relsize,exscale}
\usepackage{mathtools}
\usepackage{mathrsfs}
\usepackage{tensor}
\usepackage{cleveref}
\usepackage{xspace}
\usepackage{hhline}
\AddToHook{cmd/appendix/before}{\crefalias{section}{appendix}}

\usepackage[dvipsnames]{xcolor}
\usepackage{fontawesome}
\usepackage[breakable]{tcolorbox}

\input{Macros.tex}
\input{MacrosNames.tex}
\input{MacrosLstListings.tex}
\hypersetup{
 colorlinks,
 linkcolor={Blue},
 citecolor={Blue},
 urlcolor={Blue}
}

\numberwithin{equation}{section}
\allowdisplaybreaks

\makeatletter
\newcommand{\dalembertian}{\mathop{\mathpalette\dalembertian@\relax}}
\newcommand{\dalembertian@}[2]{%
 \begingroup
 \sbox\z@{$\m@th#1\square$}%
 \dimen0=\fontdimen8
 \ifx#1\displaystyle\textfont\else
 \ifx#1\textstyle\textfont\else
 \ifx#1\scriptstyle\scriptfont\else
 \scriptscriptfont\fi\fi\fi3
 \makebox[\wd\z@]{%
 \hbox to \ht\z@{%
 \vrule width \dimen0
 \kern-\dimen0
 \vbox to \ht\z@{
 \hrule height \dimen0 width \ht\z@
 \vss
 \hrule height 2\dimen0
 }%
 \kern-2.5\dimen0
 \vrule width 2.5\dimen0
 }%
 }%
 \endgroup
}
\makeatother

\title{Fast Poisson brackets and constraint algebras in canonical gravity}

\author{Will Barker}
\emailAdd{barker@fzu.cz}
\affiliation{Central European Institute for Cosmology and Fundamental Physics, Institute of Physics of the Czech Academy of Sciences, Na Slovance 1999/2, 182 00 Prague 8, Czechia}

\abstract{
In the study of alternative or extended theories of gravity, Dirac's Hamiltonian constraint algorithm is invaluable for enumerating the propagating modes and gauge symmetries. For gravity, this canonical approach is frequently applied as a means for finding pathologies such as strongly coupled modes; more generally it facilitates the reconstruction of gauge symmetries and the quantization of gauge theories. For gravity, however, the algorithm can become notoriously arduous to implement. We present a simple computer algebra package for efficiently computing Poisson brackets and reconstructing constraint algebras. The tools are stress-tested against pure general relativity and modified gravity, including the order reduction of general relativity at two loops.}

\begin{document}
\maketitle

\section{Introduction}\label{Introduction}

\begin{figure}[t]
\centering
\includegraphics[width=\textwidth]{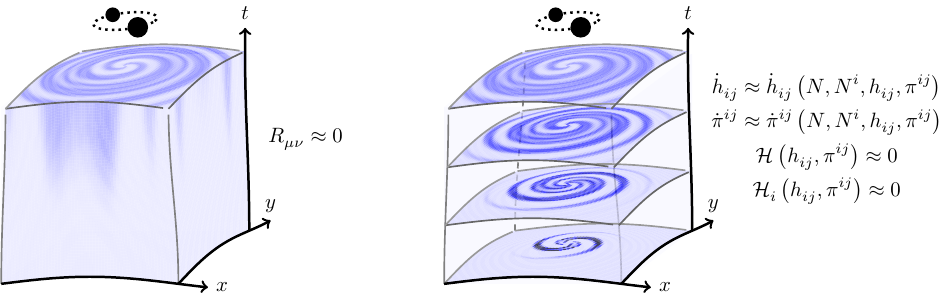}
\caption{The canonical formulation of a field theory such as general relativity allows the propagation of initial data to subsequent foliations by means of first-order equations for phase-space variables~($\MetricFoliation{_{ij}}$~and~$\ConjugateMomentumMetricFoliation{^{ij}}$), subject to additional constraints~($\SuperConstraint{}$~and~$\SuperConstraint{_i}$) and a gauge choice to fix the values of undetermined Lagrange multipliers~($\Lapse$~and~$\Shift{^i}$) on each foliation.}
\label{fig:CanonicalSchematic}
\end{figure}

\paragraph*{Hamiltonian formulation of gravity} In the study of classical field theories, the canonical (phase-space) formulation is reached by foliating the spacetime into equal-time hypersurfaces, and defining canonical variables (the field components and their conjugate momenta) which evolve between hypersurfaces according to first-order Hamilton equations. In the minimal transition to the canonical formulation, it frequently happens that not all of the velocities may be solved for in terms of canonical variables. In such cases, the definitions of the canonical momenta (as the variations of the action with respect to the velocities) lead to constraints. The minimal canonical formulation can be made consistent by enforcing these constraints with the aid of extra Lagrange multipliers, and further identifying all ancillary constraints which arise by enforcing the conservation (in time, via the Hamilton equations) of the original constraints. This algorithm leads to a counting of~$\NPhy{}$ pairs of Cauchy data, i.e., physical modes. The total number of constraints is partitioned into~$\NFirst{}$ of the first class (commuting with all the rest), and~$\NSecond{}$ of the second class (non-commuting with each other). Given the original number of phase-space variables~$\NCan{}$, the number of propagating modes is then
\begin{equation}
	\NPhy{} = \frac{1}{2}
	\Bigl( \NCan{} - 2 \NFirst{} - \NSecond{} \Bigr)
	\, .
\label{Nphy}
\end{equation}
In the case of general relativity (GR) shown in~\cref{fig:CanonicalSchematic},\footnote{Note that the Hamilton equations in~\cref{fig:CanonicalSchematic} omit the stress-energy sources; moreover, their weak hyperbolicity means that they are not immediately suitable for use in numerical relativity.} the induced metric~$\MetricFoliation{_{ij}}$ and its conjugate momentum~$\ConjugateMomentumMetricFoliation{^{ij}}$ propagate subject to first-class Hamiltonian and momentum constraints~$\SuperConstraint{}$ and~$\SuperConstraint{_i}$, whilst the lapse~$\Lapse$ and shift~$\Shift{^i}$ are undetermined multipliers whose values constitute a choice of gauge. For more general theories, the algorithm for determining and classifying all constraints is due to Dirac~\cite{Dirac:1950pj,Dirac:1958sq} and Bergmann~\cite{Bergmann:1955}, see also~\cite{Anderson:1951ta,Castellani:1982,Henneaux:1992ig,blagojevic2002gravitation}. This algorithm, and the formula in~\cref{Nphy}, are exceptionally useful in providing a definitive counting of modes when a new classical theory is proposed. Moreover, constraints of the first class following from momentum definitions may be identified as the generators of gauge symmetries, allowing the latter to be reconstructed. In turn, this first-class algebra facilitates the quantisation of the theory~\cite{Becchi:1975nq,Tyutin:1975qk}. Certain steps in canonical analysis would benefit from automation. These principally include the computation of Poisson brackets between arbitrary functionals of phase-space variables, and the re-expression of the results in terms of specified constraints.

\paragraph{In this paper} We present a set of computer algebra tools (\Hamilcar) which facilitate canonical analysis. The \Hamilcar package is built on top of the \xAct suite of packages for \Mathematica~\cite{Martin-Garcia:2007bqa,Martin-Garcia:2008ysv,Martin-Garcia:2008yei,Nutma:2013zea}, and automates the computation and simplification of Poisson brackets between arbitrary functionals of canonical variables in three spatial dimensions.\footnote{Note that \Hamilcar{} is maximally theory-agnostic. It accommodates any tensorial canonical physics on a (possibly curved) three-dimensional manifold. Thus, \Hamilcar{} completely deprecates the earlier software \emph{HiGGS}~\cite{Barker:2022kdk}, which was suitable only for computing Poisson brackets in Poincar\'e gauge theories of gravity.}

\paragraph{Structure of this paper} The remainder of this paper is set out as follows. In~\cref{Hamilcar} we describe the \Hamilcar package, and demonstrate its use by performing a full Dirac--Bergmann analysis of pure GR in~\cref{PureGR}, and pure~$\Curvature{}^2$ theory in~\cref{PureR2Theory} (see e.g.~\cite{Buchbinder:1987vp,Alvarez-Gaume:2015rwa,Hell:2023mph,Golovnev:2023zen,Karananas:2024hoh,Barker:2025gon}). We also reconstruct the constraint algebra of pure GR at two loops, as described by the Goroff--Sagnotti action, in~\cref{PureGRTwoLoops} (see e.g.~\cite{Goroff:1985sz,Goroff:1985th,vandeVen:1991gw,Glavan:2024cfs}). Brief conclusions follow in~\cref{Conclusions}. There is one technical appendix.

\paragraph{Notation and conventions} We closely follow the conventions of~\cite{Barker:2025gon}. We use natural units in which~$\hbar = c = 1$, Greek letters denote four-dimensional spacetime indices, while Roman letters denote three-dimensional spatial indices. The signature is~$(-,+,+,+)$, other conventions are introduced as needed.

\section{Explicit examples}\label{Hamilcar}

\subsection{Overview of the package}\label{HamilcarIntroduction}

\subsubsection{Installation}\label{subsubsec:Installation}

\paragraph*{Acquiring the sources} The \Hamilcar package can be installed once the \xAct suite of packages has been installed. The \xAct suite can be found at \href{http://www.xact.es/}{\texttt{xact.es}}. The \Hamilcar package is made available at the public \GitHub repository \href{https://github.com/wevbarker/Hamilcar}{\texttt{github.com/wevbarker/Hamilcar}}, along with installation guidelines for common operating systems, including \Windows and \Mac. Here, we only demonstrate a \Linux installation.\footnote{Note that the syntax highlighting for \Bash differs from \WolframLanguage highlighting used in other sections.} It is easiest to use \Bash to download \Hamilcar to the home directory as follows:
\begin{lstlisting}[language=Special]
[user@system ~]$ git clone https://github.com/wevbarker/Hamilcar
\end{lstlisting}

\paragraph*{Installing the sources} To make the installation, the sources should simply be copied alongside the other \xAct sources. If the installation of \xAct is global, one can use:
\begin{lstlisting}[language=Special]
[user@system ~]$ cd Hamilcar/xAct
[user@system xAct]$ sudo cp -r Hamilcar /usr/share/Wolfram/Applications/xAct/
\end{lstlisting}
Or, for a local installation of \xAct, one may use:
\begin{lstlisting}[language=Special]
[user@system xAct]$ cp -r Hamilcar ~/.Wolfram/Applications/xAct/
\end{lstlisting}

\paragraph*{Syntax highlighting} From this point, we will rely heavily on code listings in the \WolframLanguage, the syntax highlighting for which is now described. Symbols defined in the \WolframLanguage (\Mathematica) are \lstinline!BlackBold!, those defined in \xAct are \lstinline!NavyBlueBold!, those from \Hamilcar are \lstinline!BlueBold!, and those which are to be defined during the course of the user session are in \lstinline!SkyBlueBold!. Each fresh input cell in a \Mathematica notebook begins with the symbol `\lstinline!In[#]:=!', each output cell begins with the symbol `\lstinline!Out[#]=!'. Non-parsing comments within the \WolframLanguage appear in \lstinline!(*Gray*)!, and strings (which should not be confused with symbols) are shown in \lstinline!"GrayBold"!.

\paragraph*{Loading in the kernel} The software is loaded via the \lstinline!Get! command:
\lstinputoutput{code_listings_LoadHamilcar.tex}

\subsubsection{Usage}\label{subsubsec:Usage}

\paragraph*{Geometric environment} When~\lstinputcref{code_listings_LoadHamilcar} is initially run, the software defines a three-dimensional spatial hypersurface with the ingredients shown in~\cref{tab:predefined}. In the framework of \xAct, note that calls to \lstinline!DefManifold! and \lstinline!DefMetric! are made internally at this stage. The package establishes a spatial manifold \lstinline!M3!, creating the necessary geometric structure for canonical field theory calculations.

\begin{table}[h]
\centering
\begin{tabular}{l|l|l}
\hhline{=|=|=}
\WolframLanguage & Output format & Meaning \\
\hhline{=|=|=}
\lstinline!a!, \lstinline!b!, \lstinline!c!, \ldots, \lstinline!z! & $a$, $b$, $c$, \ldots, $z$ & Spatial coordinate indices \\
\lstinline!G[-a,-b]! & $\MetricFoliation{_{ab}}$ & Induced metric on the spatial hypersurface \\
\lstinline!CD[-a]@! & $\CD{_{a}}$ & Spatial covariant derivative \\
\lstinline!epsilonG[-a,-b,-c]! & $\tensor{\epsilon}{_{abc}}$ & Induced totally antisymmetric tensor \\
\hhline{=|=|=}
\end{tabular}
\caption{Pre-defined geometric objects in \Hamilcar. The spatial coordinate indices correspond to \emph{adapted} coordinates in the ADM prescription.}
\label{tab:predefined}
\end{table}

\paragraph*{Function \lstinline!DefCanonicalField!}
The command
\begin{lstlisting}
DefCanonicalField[<Fld>[]]
\end{lstlisting}
defines a scalar canonical field \lstinline!<Fld>! and its conjugate momentum \lstinline!ConjugateMomentum<Fld>!.
The command
\begin{lstlisting}
DefCanonicalField[<Fld>[<Ind1>,<Ind2>,...]]
\end{lstlisting}
defines a tensor canonical field \lstinline!<Fld>! and its conjugate momentum \lstinline!ConjugateMomentum<Fld>! with indices \lstinline!<Ind1>!, \lstinline!<Ind2>!, etc.
The command
\begin{lstlisting}
DefCanonicalField[<Fld>[<Ind1>,<Ind2>,...],<Symm>]
\end{lstlisting}
defines a tensor canonical field \lstinline!<Fld>! and its conjugate momentum \lstinline!ConjugateMomentum<Fld>! with indices \lstinline!<Ind1>!, \lstinline!<Ind2>!, etc.\ and symmetry \lstinline!<Symm>!. The syntax of \lstinline!DefCanonicalField! follows similar patterns to \lstinline!DefTensor! in \emph{xTensor}. Any number of comma-separated indices may be drawn from the contravariant \lstinline!a!, \lstinline!b!, \lstinline!c!, up to \lstinline!z!, the covariant \lstinline!-a!, \lstinline!-b!, \lstinline!-c!, up to \lstinline!-z!, or any admixture. The symmetry \lstinline!<Symm>! can be one of the following (or any admixture allowed by \lstinline!DefTensor!):
\begin{itemize}
\item \lstinline!Symmetric[{<SymmInd1>,<SymmInd2>,...}]! denotes symmetrized indices.
\item \lstinline!Antisymmetric[{<SymmInd1>,<SymmInd2>,...}]! denotes antisymmetrized indices.
\end{itemize}
Note that if the global variable \lstinline!$DynamicalMetric! is set to \lstinline!True!, then:
\begin{itemize}
\item The spatial metric \lstinline!G! is automatically registered, and \lstinline!ConjugateMomentumG! is defined.
\item All conjugate momenta are automatically defined as tensor densities of weight one (the effects of this are only seen by the \lstinline!G!-variations performed internally by \lstinline!PoissonBracket!).
\end{itemize}
The following options may be given:
\begin{itemize}
\item \lstinline!FieldSymbol! is the symbol that \lstinline!<Fld>! will use for display formatting.
\item \lstinline!MomentumSymbol! is the symbol that the conjugate momentum will use for display formatting.
\end{itemize}

\paragraph*{Function \lstinline!PoissonBracket!}
The command
\begin{lstlisting}
PoissonBracket[<Op1>,<Op2>]
\end{lstlisting}
computes the Poisson bracket between operators \lstinline!<Op1>! and \lstinline!<Op2>!. The operators \lstinline!<Op1>! and \lstinline!<Op2>! must be valid tensor expressions involving:
\begin{itemize}
\item Canonical fields and their conjugate momenta defined using \lstinline!DefCanonicalField!.
\item Tensors which have been defined on the manifold \lstinline!M3! using \lstinline!DefTensor!, and which are assumed always to be independent of the phase-space fields unless their expansion in terms of canonical fields and fundamental independent tensors has been registered using \lstinline!PrependTotalFrom!.
\item Derivatives via \lstinline!CD! of canonical and non-canonical quantities, the spatial metric \lstinline!G!, and the totally antisymmetric tensor \lstinline!epsilonG!.
\item Constant symbols defined using \lstinline!DefConstantSymbol! (or \lstinline!DefNiceConstantSymbol! from the \emph{xTras} package).
\end{itemize}
The function automatically generates smearing tensors unless \lstinline!$ManualSmearing! is set to \lstinline!True!, in which case \lstinline!<Op1>! and \lstinline!<Op2>! must be scalars (interpreted as functionals). When \lstinline!$DynamicalMetric! is set to \lstinline!True!, the \lstinline!G!-sector contributions are included. The function computes variational derivatives with respect to all registered fields and momenta. The following options may be given:
\begin{itemize}
\item \lstinline!Parallel! is a boolean which, when set to \lstinline!True!, causes the bracket computation to be parallelised across multiple CPU cores. Default is \lstinline!True!.
\end{itemize}

\paragraph*{Function \lstinline!TotalFrom!}
The command
\begin{lstlisting}
TotalFrom[<Expr>]
\end{lstlisting}
expands \lstinline!<Expr>! by applying all registered rules in the global list \lstinline!$FromRulesTotal!, which is populated by \lstinline!PrependTotalFrom!. The function converts specially registered composite tensors into expressions involving only canonical fields and conjugate momenta defined using \lstinline!DefCanonicalField! (including the pre-defined \lstinline!G! and its dependencies \lstinline!epsilonG! and \lstinline!CD!), fundamental tensors (defined using \lstinline!DefTensor! without \lstinline!PrependTotalFrom!), and constant symbols defined using \lstinline!DefConstantSymbol!. This command is of standalone utility, and is also used internally by \lstinline!PoissonBracket! and \lstinline!FindAlgebra!.

\paragraph*{Function \lstinline!TotalTo!}
The command
\begin{lstlisting}
TotalTo[<Expr>]
\end{lstlisting}
attempts to invert the expansion performed by \lstinline!TotalFrom! on \lstinline!<Expr>! by applying all registered rules in the global variable \lstinline!$ToRulesTotal!, which is populated by \lstinline!PrependTotalTo!. This function is occasionally expedient, but unlike \lstinline!TotalFrom! it is mostly provided for completeness. For re-expression of functional-valued quantities, the \lstinline!FindAlgebra! function is preferred.

\paragraph*{Function \lstinline!PrependTotalFrom!}
The command
\begin{lstlisting}
PrependTotalFrom[<Rle>]
\end{lstlisting}
registers the expansion rule \lstinline!<Rle>! for future use by \lstinline!TotalFrom!. The function adds \lstinline!<Rle>! to the front of the global list \lstinline!$FromRulesTotal!. It is preferred that \lstinline!<Rle>! be defined by \lstinline!MakeRule!, which is the safe method for defining tensor-valued replacement rules in \xAct in such a way that all possible index valences and traces are automatically taken into account. Typically, \lstinline!<Rle>! will expand a single tensor into a (possibly multi-term) tensor-valued expression. Note that \lstinline!<Rle>! does \emph{not} need to expand the single tensor in terms of canonical variables, fundamental tensors, and constant symbols. Rather, the full list of registered rules in \lstinline!$FromRulesTotal! should work together to achieve this effect. This allows the sequential building up of progressively more complicated composite tensors.

\paragraph*{Function \lstinline!PrependTotalTo!}
The command
\begin{lstlisting}
PrependTotalTo[<Rle>]
\end{lstlisting}
registers the expansion rule \lstinline!<Rle>! for future use by \lstinline!TotalTo!. The function adds \lstinline!<Rle>! to the front of the global list \lstinline!$ToRulesTotal!. The considerations for defining \lstinline!<Rle>! are similar to those give in \lstinline!PrependTotalFrom!.

\paragraph*{Function \lstinline!FindAlgebra!} The command
\begin{lstlisting}
FindAlgebra[<Expr>,{{<Fctr1>,<Fctr2>,...},...}]
\end{lstlisting}
seeks to express the scalar \lstinline!<Expr>! (interpreted as a functional) as a sum of any number of terms, where each term corresponds to one of the sub-lists and has factors corresponding to indexed tensors whose heads are \lstinline!<Fctr1>!, \lstinline!<Fctr2>!, etc., where those tensor heads were defined using \lstinline!DefCanonicalField! or \lstinline!DefTensor!. The re-expression is achieved automatically by means of any required number of integrations by parts. The command
\begin{lstlisting}
FindAlgebra[<Expr>,{{<Fctr1>,<Fctr2>,...,{CD,...,<Fctr3>,...}},...}]
\end{lstlisting}
additionally admits terms where one or more spatial covariant derivatives \lstinline!CD! acts on any of a select group of factors, here \lstinline!<Fctr3>!, etc. The following options may be given:
\begin{itemize}
	\item \lstinline!Constraints! is a list of special (and appropriately indexed) tensors which were passed as part of the ansatz, with respect to which the re-expression is expected to be homogeneously linear. The answer will be expressed with these tensors factored out.
	\item \lstinline!Verify! is a boolean which, when set to \lstinline!True!, causes the re-expression and \lstinline!<Expr>! to be varied internally with respect to any tensors which appear exactly to the first power in all terms, after an application of \lstinline!TotalFrom!. This usually includes smearing functions, but it may happen to include some canonical variables. The variations are compared to ensure that the re-expression is correct. Default is \lstinline!False!.
	\item \lstinline!DDIs! is a boolean which, when set to \lstinline!True!, causes all relevant dimensionally dependent identities (DDIs) such as the Cayley--Hamilton theorem to be taken into account when performing the re-expression. Default is \lstinline!False!.

	\item \lstinline!Method! allows either \lstinline!Solve! (default) or \lstinline!LinearSolve!.
\end{itemize}

\subsection{Case study: pure GR}\label{PureGR}

\paragraph*{Motivation} The first theory we consider is Einstein's GR in the absence of both matter and a cosmological constant, defined by the Einstein--Hilbert action
\begin{equation}
	\Action{}[\G{_{\mu\nu}}] = \int\! \mathrm{d}^4x \, \sqrt{-\G{}} \, \frac{\Curvature{}}{\kappa^2} \, ,
	\label{GRaction}
\end{equation}
where~$\kappa$ is the coupling constant,~$\G{_{\mu\nu}}$ is the spacetime metric, and~$\Curvature{}\equiv\G{^{\mu\nu}}\Curvature{_{\mu\nu}}$ is the Ricci scalar, following from the Riemannian curvature tensor by~$\Curvature{_{\mu\nu}}\equiv\Curvature{^\rho_{\mu\rho\nu}}$ (see~\cref{ADMRicciScalar} for further index conventions). General coordinate invariance requires the use of~$\G{}\equiv{\rm det}(\G{_{\mu\nu}})$. Note that~\cref{GRaction} is not taken to be physical until both matter and the cosmological constant are introduced; however its formal structure is well known, and it provides a useful initial test case.  

\subsubsection{Phase-space formulation}\label{subsec: ADM decomposition}

\paragraph*{Lapse, shift and induced metric} The Hamiltonian treatment is to be performed using the Arnowitt--Deser--Misner (ADM) prescription according to 
\begin{equation}
	\G{^{00}} = - \frac{1}{\Lapse{}^2} \, ,
\qquad
\G{_{0i}} = \Shift{_i} \, ,
\qquad
\G{_{ij}} = \MetricFoliation{_{ij}} \, .
\label{ADMmetric1}
\end{equation}
The quantities in~\cref{ADMmetric1} are the lapse and shift (respectively~$\Lapse{}$ and~$\Shift{_i}$), and the residual metric on the foliations~$\MetricFoliation{_{ij}}$. We introduce~$\CD{_i}$ as the induced covariant derivative, which satisfies the metricity condition for~$\MetricFoliation{_{ij}}$. The three-dimensional curvature~$\R{^i_{jkl}}$ is associated with~$\CD{_i}$, and should not be confused with the four-dimensional curvature~$\Curvature{^\mu_{\nu\sigma\rho}}$. The three-dimensional curvature scalar is defined by~$\R{} \equiv \MetricFoliation{^{ij}} \R{_{ij}}$, where~$\R{_{ij}} \equiv \R{^k_{ikj}}$.\footnote{Apart from the change from Greek to Roman indices, our conventions for the curvature defined in three and four dimensions are identical.} From~\cref{ADMmetric1} we may also deduce
\begin{equation}
	\G{_{00}} = - \Lapse{}^2 + \Shift{_i} \Shift{^i} \, ,
\qquad
\G{^{0i}} = \frac{\Shift{^i}}{\Lapse{}^2} \, ,
\qquad
\G{^{ij}} = \MetricFoliation{^{ij}} - \frac{\Shift{^i} \Shift{^j}}{\Lapse{}^2} \, .
\label{ADMmetric2}
\end{equation}
In~\cref{ADMmetric2},~$\MetricFoliation{^{ij}}$ becomes the inverse metric on equal-time slices, also defined by the identity~$\MetricFoliation{_{ij}}\MetricFoliation{^{jk}} \equiv \Kronecker{_i^k}$. From~\cref{ADMmetric1} it also follows that the measure can be written as~$\sqrt{-\G{}} \equiv \Lapse{} \sqrt{\MetricFoliation{}}$.\footnote{From this separation, and in order for $\G{^{\mu\nu}}$ to be defined, it is important that~$\Lapse{} \!\neq\! 0$.} The induced spatial metric \lstinline!G[-a,-b]! and its conjugate momentum \lstinline!ConjugateMomentumG[a,b]! are already pre-defined once~\lstinputcref{code_listings_LoadHamilcar} has been executed; the latter is denoted by~$\ConjugateMomentumG{^{ij}}$. It will be useful, however, to additionally define the trace~$\ConjugateMomentumG{} \equiv \MetricFoliation{_{ij}} \ConjugateMomentumG{^{ij}}$ of the conjugate momentum using the standard \lstinline!DefTensor! command from \xAct:
\lstinputoutput{code_listings_Hamilcar-TraceConjugateMomentumG.tex}
At the point of definition, \Hamilcar{} is not aware that this trace has any special dependence on the canonical variables. We therefore need to teach \Hamilcar{} how to expand this trace in terms of the canonical variables. The programmatic rule \lstinline!FromTraceConjugateMomentumG!, which performs this expansion, is defined in~\lstinputcref{code_listings_Hamilcar-TraceConjugateMomentumG} using the \lstinline!MakeRule! command from \xAct; this rule is then registered with \Hamilcar{} using the \lstinline!PrependTotalFrom! command:
\lstinputoutput{code_listings_Hamilcar-PrependTotalFromTraceConjugateMomentumG.tex}
Having executed~\lstinputcref{code_listings_Hamilcar-PrependTotalFromTraceConjugateMomentumG}, it will be possible to feed expressions containing the trace~$\ConjugateMomentumG{}$ to other \Hamilcar{} functions, which will expand it correctly in terms of the canonical variables. Next, the lapse function~$\Lapse{}$ is defined using:
\lstinputoutput{code_listings_Hamilcar-DefineLapse.tex}
The shift vector~$\Shift{^i}$ is defined using:
\lstinputoutput{code_listings_Hamilcar-DefineShift.tex}
Note that, once again, we use \lstinline!DefTensor! from \xAct to perform the definitions. Properly, these quantities are themselves canonical fields for which \lstinline!DefCanonicalField! from \Hamilcar{} should be used. Since, however, the lapse and shift are non-dynamical Lagrange multipliers in the case of GR (and other theories considered in this paper), we do not need to compute their Poisson brackets, and so the simpler \lstinline!DefTensor! command suffices. Finally, the gravitational coupling constant~$\kappa$ from~\cref{GRaction} is defined using:
\lstinputoutput{code_listings_Hamilcar-DefineGravitationalCoupling.tex}
Note that the standard command \lstinline!DefConstantSymbol! from \xAct is used.

\paragraph*{Extrinsic curvature} Having defined~$\MetricFoliation{_{ij}}$, we now need to set up some helpful notation for the velocity~$\MetricFoliationDot{_{ij}} \equiv \partial_t \MetricFoliation{_{ij}}$. The velocity is encoded by
\begin{equation}
	\ExtrinsicCurvatureActual{_{ij}} \equiv - \frac{1}{2\Lapse{}} \Bigl( \MetricFoliationDot{_{ij}} - \CD{_i} \Shift{_j} - \CD{_j} \Shift{_i} \Bigr) \, ,
\label{Kdef}
\end{equation}
where~\cref{Kdef} is termed the extrinsic curvature. Similarly, the acceleration --- i.e., the velocity of the extrinsic curvature~$\ExtrinsicCurvatureActualDot{_{ij}} \equiv \partial_t \ExtrinsicCurvatureActual{_{ij}}$ --- is well captured by (see a similar construction in~\cite{Buchbinder:1987vp}, adapted in~\cite{Barker:2025gon})
\begin{equation}
	\FFunctionActual{_{ij}} \equiv - \frac{1}{\Lapse{}} \ExtrinsicCurvatureActualDot{_{ij}}
	- \ExtrinsicCurvatureActual{_{ik}} \ExtrinsicCurvatureActual{^k_j}
	+ \frac{\Shift{^k}}{\Lapse{}} \CD{_k} \ExtrinsicCurvatureActual{_{ij}}
	+ \frac{2}{\Lapse{}} \ExtrinsicCurvatureActual{_{k(i}} \CD{_{j)}} \Shift{^k}
	- \frac{1}{\Lapse{}} \CD{_i} \CD{_j} \Lapse{} \, .
\label{Fdef}
\end{equation}
The quantity in~\cref{Fdef} will be useful for the purpose of dealing with higher-order theories in~\cref{PureR2Theory}, and we denote the trace~$\FFunctionActual{}\equiv \MetricFoliation{^{ij}} \FFunctionActual{_{ij}}$. The four-dimensional curvature scalar is found to be
\begin{equation}
	\Curvature{} \equiv 2\FFunctionActual{} + \ExtrinsicCurvatureActual{}^2 - \ExtrinsicCurvatureActual{^{ij}}\ExtrinsicCurvatureActual{_{ij}} + \R{} \, ,
\label{WellKnownFormula}
\end{equation}
where~\cref{WellKnownFormula} results from some standard manipulations (see~\cref{ADMRicciScalar}). We also define another trace~$\ExtrinsicCurvatureActual{}\equiv\ExtrinsicCurvatureActual{^i_i}$.

\paragraph*{Hiding the velocities} By substituting the expansion in~\cref{WellKnownFormula} into~\cref{GRaction} we obtain
\begin{equation}
	\Action{}[ \Lapse{}, \Shift{^i}, \MetricFoliation{_{ij}} ]=\frac{1}{\kappa^2}\int\! \mathrm{d}^{4} x \, \Lapse{} \sqrt{\MetricFoliation{}} \, \Bigl( 2\FFunctionActual{} + \ExtrinsicCurvatureActual{}^2 - \ExtrinsicCurvatureActual{^{ij}} \ExtrinsicCurvatureActual{_{ij}} + \R{} \Bigr) \, .
	\label{ADMactionGR}
\end{equation}
By examining~\cref{Fdef} we see that the action in~\cref{ADMactionGR} makes explicit reference to the velocity of the extrinsic curvature,~$\ExtrinsicCurvatureActualDot{_{ij}}$ and hence the acceleration of the spatial metric~$\MetricFoliation{_{ij}}$, which is not desirable for the canonical approach. This term is, however, amenable to integration by parts, using the manipulation
\begin{align}
\Action{}[ \Lapse{}, \Shift{^i}, \MetricFoliation{_{ij}} ]&\supset
-\frac{2}{\kappa^2}\int\! \mathrm{d}^{4} x \, \sqrt{\MetricFoliation{}} \, \MetricFoliation{^{ij}} \ExtrinsicCurvatureActualDot{_{ij}} \nonumber\\
							  &= \frac{1}{\kappa^2}\int\! \mathrm{d}^{4} x \sqrt{\MetricFoliation{}} \, \left(\ExtrinsicCurvatureActual{}\MetricFoliation{^{ij}} - 2\ExtrinsicCurvatureActual{^{ij}}\right)\left(2\CD{_{(i}} \Shift{_{j)}} - 2 \Lapse{} \ExtrinsicCurvatureActual{_{ij}}\right) ,
\label{IntegrationByParts}
\end{align}
where we use~\cref{Kdef} in the second line. When~\cref{IntegrationByParts} is substituted back into~\cref{ADMactionGR}, the resulting action contains only velocities, and simplifies to
\begin{equation}\label{ADMactionGRFinal}
	\Action{}[ \Lapse{}, \Shift{^i}, \MetricFoliation{_{ij}} ]=
	\frac{1}{\kappa^2}\int\! \mathrm{d}^{4} x \, \Lapse{} \sqrt{\MetricFoliation{}} \, \left( \R{} - \ExtrinsicCurvatureActual{}^2 + \ExtrinsicCurvatureActual{^{ij}} \ExtrinsicCurvatureActual{_{ij}} \right) \, .
\end{equation}
To progress towards the phase-space formulation, we introduce an auxiliary variable~$\ExtrinsicCurvature{_{ij}}$ which is forced on-shell to equal the extrinsic curvature~$\ExtrinsicCurvature{_{ij}}\approx\ExtrinsicCurvatureActual{_{ij}}$ by means of a Lagrange multiplier~$\ConjugateMomentumG{^{ij}}$ --- we know that this multiplier is precisely the momentum conjugate to~$\MetricFoliation{_{ij}}$, but this is an identification that we will allow to arise naturally. The action is then
\begin{align}
	\Action{}\bigl[ \Lapse{}, \Shift{^i}, \MetricFoliation{_{ij}}, \ExtrinsicCurvature{_{ij}}, \ConjugateMomentumG{^{ij}} \bigr] = \int\! \mathrm{d}^{4} x \, \biggl[ &
		\frac{\Lapse{} \sqrt{\MetricFoliation{}}}{\kappa^2}  \, \left( \R{} - \ExtrinsicCurvature{}^2 + \ExtrinsicCurvature{^{ij}} \ExtrinsicCurvature{_{ij}} \right)
		\nonumber \\ &
	+ \ConjugateMomentumG{^{ij}} \Bigl( \MetricFoliationDot{_{ij}} - 2 \CD{_{(i}} \Shift{_{j)}} + 2 \Lapse{} \ExtrinsicCurvature{_{ij}} \Bigr) \biggr] \, .
	\label{CanonicalActionGR}
\end{align}
A further application of integration by parts, this time on the foliation, allows~\cref{CanonicalActionGR} to be separated into a symplectic part, and two further terms
\begin{align}
	\Action{}\bigl[ \Lapse{}, \Shift{^i}, \MetricFoliation{_{ij}}, \ExtrinsicCurvature{_{ij}}, \ConjugateMomentumG{^{ij}} \bigr] = \int\! \mathrm{d}^{4} x \, \biggl[ &
		\ConjugateMomentumG{^{ij}} \MetricFoliationDot{_{ij}}
		+ 2 \Shift{^i} \CD{_j} \ConjugateMomentumG{^{j}_i}
		\nonumber \\ &
	+ 2 \Lapse{} \biggl( 2\ConjugateMomentumG{^{ij}} \ExtrinsicCurvature{_{ij}} +\frac{\sqrt{\MetricFoliation{}}}{\kappa^2}  \, \left( \R{} - \ExtrinsicCurvature{}^2 + \ExtrinsicCurvature{^{ij}} \ExtrinsicCurvature{_{ij}} \right) \biggr) \biggr] \, .
	\label{CanonicalActionGRFinal}
\end{align}
These terms reveal~$\Lapse{}$ and~$\Shift{_i}$ to be themselves acting as Lagrange multipliers.

\paragraph*{Phase-space action} It is moreover possible in~\cref{CanonicalActionGRFinal} for~$\ConjugateMomentumG{^{ij}}$ to give up its role as a Lagrange multiplier, since~$\ExtrinsicCurvature{_{ij}}$ appears only algebraically and may be integrated out as~$\ExtrinsicCurvature{_{ij}} \approx \kappa^2\left(\ConjugateMomentumG{}\MetricFoliation{_{ij}} - 2\ConjugateMomentumG{_{ij}}\right)/2\sqrt{\MetricFoliation{}}$. By substituting this solution back into~\cref{CanonicalActionGRFinal}, one obtains the usual phase-space action of GR
\begin{align}\label{CanonicalActionGRFinal2}
	\Action{}\bigl[ \Lapse{}, \Shift{^i}, \MetricFoliation{_{ij}}, \ConjugateMomentumG{^{ij}} \bigr] = \int\! \mathrm{d}^{4} x \, \biggl[ &
		\ConjugateMomentumG{^{ij}} \MetricFoliationDot{_{ij}}
		-\Shift{^i}\SuperConstraint{_i}
	-\Lapse{} \SuperConstraint{} \biggr] \, ,
\end{align}
where the two constraints --- the super-Hamiltonian and super-momentum --- are defined as
\begin{equation}
	\SuperConstraint{} \equiv \frac{\kappa^2}{ \sqrt{\MetricFoliation{}}} \left( 2\ConjugateMomentumG{^{ij}} \ConjugateMomentumG{_{ij}} - \ConjugateMomentumG{}^2- \frac{\MetricFoliation{} \R{}}{\kappa^4} \right) \, ,
	\qquad
	\SuperConstraint{_i} \equiv -2 \CD{_j} \ConjugateMomentumG{^j_i} \, .
	\label{GRconstraints}
\end{equation}
Referring to~\cref{GRconstraints}, we progress by defining~$\SuperConstraint{}$ as the variable \lstinline!SuperHamiltonian[]!, along with a rule \lstinline!FromSuperHamiltonian! which expands it in terms of the canonical variables:
\lstinputoutput{code_listings_Hamilcar-DefineSuperHamiltonian.tex}
The syntax is again based on \xAct's \lstinline!DefTensor! command, and we use \lstinline!MakeRule! to define the expansion rule. Next, we define the momentum constraint~$\SuperConstraint{_i}$ as the variable \lstinline!SuperMomentum[-i]!, along with a rule \lstinline!FromSuperMomentum! which expands it in terms of the canonical variables:
\lstinputoutput{code_listings_Hamilcar-DefineSuperMomentum.tex}
At the moment, both \lstinline!FromSuperHamiltonian! and \lstinline!FromSuperMomentum! are simply variables in the user session, so as with the trace of the conjugate momentum in~\lstinputcref{code_listings_Hamilcar-PrependTotalFromTraceConjugateMomentumG}, we need to register these rules with \Hamilcar{}. For the super-Hamiltonian we use:
\lstinputoutput{code_listings_Hamilcar-PrependTotalFromSuperHamiltonian.tex}
For the super-momentum we use:
\lstinputoutput{code_listings_Hamilcar-PrependTotalFromSuperMomentum.tex}
With these commands, the kernel is now in a state where we can compute Poisson brackets of the constraints, and so implement the Dirac--Bergmann algorithm.

\subsubsection{Dirac--Bergmann algorithm}\label{subsec: Dirac-Bergmann algorithm}

\paragraph*{Total Hamiltonian} The variation of~\cref{CanonicalActionGRFinal2} with respect to~$\Lapse{}$ and~$\Shift{^i}$ reveals the constrained nature of the super-Hamiltonian and super-momentum
\begin{equation}
	\SuperConstraint{} \approx 0 \, ,
	\qquad
	\SuperConstraint{_i} \approx 0 \, ,
	\label{GRprimaryconstraints}
\end{equation}
meanwhile the usual Legendre transformation is apparent in the simple format of~\cref{CanonicalActionGRFinal2}, leading directly to the total Hamiltonian
\begin{equation}
	\TotalHamiltonian = \int\! \mathrm{d}^{3} x \, \left( \Lapse{} \SuperConstraint{} + \Shift{^i} \SuperConstraint{_i} \right) \, .
	\label{TotalHamiltonianGR}
\end{equation}
Evidently,~\cref{TotalHamiltonianGR} is a functional obtained by integrating canonical variables over equal-time slices. The Dirac--Bergmann algorithm is concerned with the maintenance of the constraints in~\cref{GRprimaryconstraints} under time evolution generated by~$\TotalHamiltonian{}$. This time evolution is expressed in terms of Poisson brackets between the constraints and the total Hamiltonian. What follows is seen also in~\cref{PureR2Theory,PureGRTwoLoops}, and is the standard scenario in canonical gravity theories: since~$\TotalHamiltonian{}$ is itself expressed entirely in terms of the constraints, the Dirac--Bergmann algorithm reduces to computing Poisson brackets among the constraints themselves.

\paragraph*{Smearing functions} Since Poisson brackets are the central currency, we introduce smearing functions to make them easier to handle. As an example of how smeared quantities are to be denoted, we follow~\cite{Barker:2025gon} by writing
\begin{equation}\label{SmearingDef}
	\SuperConstraint{}[\SmearingF{}] \equiv \int\! \mathrm{d}^3x \, \SmearingF{}(x) \SuperConstraint{}(x) \, ,
\qquad
	\SuperConstraint{^i}[\SmearingF{_i}] \equiv \int\! \mathrm{d}^3x \, \SmearingF{_i}(x) \SuperConstraint{^i}(x) \, .
\end{equation}
Whenever smearing functions appear inside Poisson brackets, they are taken to not depend on the phase-space variables \emph{in their presented index configuration}. This means, for example, that when~$\SmearingF{^i}$ and~$\SmearingF{_i}$ appear inside two different brackets, we cannot simply conclude~$\SmearingF{^i}=\MetricFoliation{^{ij}}\SmearingF{_j}$. The `$[\dots]$' notation of~\cref{SmearingDef} is also used more loosely outside of Poisson brackets to denote smearing by functions which may or may not depend on the canonical variables; for clarity, constraints are factored out of such brackets as far as possible. For the super-Hamiltonian, we define a scalar smearing function:
\lstinputoutput{code_listings_Hamilcar-DefineScalarSmearingS.tex}
A second scalar smearing function allows us to compute brackets between two independently smeared super-Hamiltonian constraints:
\lstinputoutput{code_listings_Hamilcar-DefineScalarSmearingF.tex}
For the super-momentum, we firstly define a covariant vector smearing function:
\lstinputoutput{code_listings_Hamilcar-DefineVectorSmearingCovariantS.tex}
We then add a second independent covariant smearing function:
\lstinputoutput{code_listings_Hamilcar-DefineVectorSmearingCovariantF.tex}
Similarly, we define a contravariant smearing function:
\lstinputoutput{code_listings_Hamilcar-DefineVectorSmearingContravariantS.tex}
And a second independent contravariant smearing function:
\lstinputoutput{code_listings_Hamilcar-DefineVectorSmearingContravariantF.tex}
By default, \Hamilcar{} automatically smears constraints with internal, single-use variables when computing Poisson brackets. For more control over the smearing process, however, we can enable manual smearing:
\lstinputoutput{code_listings_Hamilcar-EnableManualSmearing.tex}
Fundamentally, all the Poisson brackets we compute in this section can be derived from the formal identity\footnote{We introduce tensor smearing functions in~\cref{subsec: Constraint analysis}, but the notation is obvious as used in~\cref{FundamentalPoissonBracket}.}
\begin{equation}\label{FundamentalPoissonBracket}
	\bigl\{ \MetricFoliation{_{ij}}[\SmearingF{^{ij}}] , \ConjugateMomentumG{^{kl}}[\SmearingS{_{kl}}] \bigr\}
	\equiv
	\big[\SmearingF{^{(ij)}}\SmearingS{_{(ij)}}\big]
	\, .
\end{equation}
Only the identity in~\cref{FundamentalPoissonBracket} is required, because~$\MetricFoliation{_{ij}}$ and~$\ConjugateMomentumG{^{kl}}$ are the only canonical variables appearing in the symplectic part of the action in~\cref{CanonicalActionGRFinal2}, and in the constraints in~\cref{GRconstraints} --- extra fundamental commutators will appear in~\cref{PureR2Theory}. As with the variables \lstinline!G[-a,-b]! and \lstinline!ConjugateMomentumG[a,b]!, the bracket in~\cref{FundamentalPoissonBracket} is already pre-defined once~\lstinputcref{code_listings_LoadHamilcar} has been executed, so no further action is required.

\paragraph*{Super-Hamiltonian auto-commutator} We can now compute the auto-commutator of the super-Hamiltonian constraint. We begin by setting up the smeared expression using two independent scalar smearing functions:
\lstinputoutput{code_listings_Hamilcar-SetupSuperHamiltonianAutocommutator.tex}
We then compute the Poisson bracket by feeding~\lstinputcref{code_listings_Hamilcar-SetupSuperHamiltonianAutocommutator} into \lstinline!PoissonBracket!:
\lstinputoutput[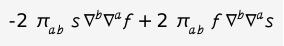]{code_listings_Hamilcar-ComputeSuperHamiltonianAutocommutator.tex}
The result of \lstinputcref{code_listings_Hamilcar-ComputeSuperHamiltonianAutocommutator} is, by default, expressed in terms of the canonical variables and their spatial derivatives, as seen in the output \lstoutputcref{code_listings_Hamilcar-SuperHamiltonianSuperHamiltonianPoissonBracket-output}. However, the goal of reconstructing the constraint algebra requires us to express this bracket in terms of the constraints themselves. The \lstinline!FindAlgebra! function from \Hamilcar{} performs this reconstruction automatically:
\lstinputoutput[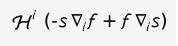]{code_listings_Hamilcar-FindAlgebraSuperHamiltonianAutocommutator.tex}
We see that the \lstinline!FindAlgebra! function requires us to specify in its second argument a schematic ansatz (with indices suppressed) for the final form of the bracket. In practice, such ans\"atze are usually easy to guess on dimensional grounds, although the specific index configurations can be tedious to enumerate in detail. In this case, we know that the final answer should be proportional to~$\SuperConstraint{^i}$, and each of the smearing functions~$\SmearingF{}$ and~$\SmearingS{}$, with one factor of~$\CD{_i}$ acting on either smearing function to provide the necessary spatial derivative. The additional structuring of the ansatz by means of nested lists signals that the~$\CD{_i}$ should be confined to act on the smearing functions only, and not on the constraint itself. The option \lstinline!Constraints! allows the user to specify the constraints that should be factored out of the final result, which is remarkably helpful for formatting. The output \lstoutputcref{code_listings_Hamilcar-SuperHamiltonianAlgebra-output} reflects
\begin{equation}\label{SuperHamiltonianAutocommutator}
	\bigl\{ \SuperConstraint{}[\SmearingF{}],
	\SuperConstraint{}[\SmearingS{}]\bigr\}
	=
	\SuperConstraint{^{i}}\bigg[
		\SmearingF{}\CD{_i}\SmearingS{}
		-\SmearingS{}\CD{_i}\SmearingF{}
	\bigg]
	\, ,
\end{equation}
and~\cref{SuperHamiltonianAutocommutator} is indeed the expected result.

\paragraph*{Super-Hamiltonian and super-momentum} Next, we compute the Poisson bracket between the super-Hamiltonian and the super-momentum, using a scalar smearing for the super-Hamiltonian and the contravariant vector smearing for the super-momentum:
\lstinputoutput{code_listings_Hamilcar-SetupSuperHamiltonianSuperMomentumBracket.tex}
The computation progresses by feeding~\lstinputcref{code_listings_Hamilcar-SetupSuperHamiltonianSuperMomentumBracket} into \lstinline!PoissonBracket! as before:
\lstinputoutput[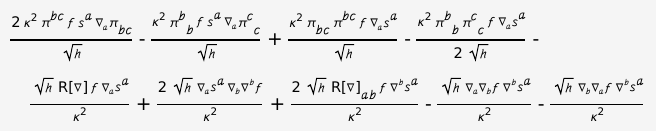]{code_listings_Hamilcar-ComputeSuperHamiltonianSuperMomentumBracket.tex}
We use \lstinline!FindAlgebra! to re-express the result in \lstoutputcref{code_listings_Hamilcar-SuperHamiltonianSuperMomentumPoissonBracket-output} cleanly in terms of the constraints:
\lstinputoutput[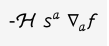]{code_listings_Hamilcar-FindAlgebraSuperHamiltonianSuperMomentumBracket.tex}
Once again,~\lstoutputcref{code_listings_Hamilcar-SuperHamiltonianMomentumAlgebra-output} reflects
\begin{equation}\label{SuperHamiltonianSuperMomentumBracket}
	\bigl\{ \SuperConstraint{}[\SmearingF{}],
	\SuperConstraint{_i}[\SmearingS{^i}]\bigr\}
	=\SuperConstraint{}\bigg[
		-\SmearingS{^i}\CD{_i}\SmearingF{}
	\bigg]
	\, ,
\end{equation}
and~\cref{SuperHamiltonianSuperMomentumBracket} is again the expected result.

\paragraph*{Super-momentum auto-commutator} The final bracket to compute is the auto-commutator of the super-momentum constraint. It is helpful to compute this using both contravariant \emph{and} covariant vector smearing functions, and to compare the results. We begin with the contravariant case, which requires the second contravariant vector smearing function:
\lstinputoutput{code_listings_Hamilcar-SetupSuperMomentumAutocommutatorContravariant.tex}
The computation proceeds by feeding~\lstinputcref{code_listings_Hamilcar-SetupSuperMomentumAutocommutatorContravariant} into \lstinline!PoissonBracket! as before:
\lstinputoutput[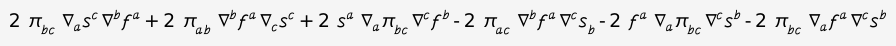]{code_listings_Hamilcar-ComputeSuperMomentumAutocommutatorContravariant.tex}
The result in~\lstoutputcref{code_listings_Hamilcar-SuperMomentumSuperMomentumPoissonBracket-output} is then re-expressed in terms of constraints using \lstinline!FindAlgebra!:
\lstinputoutput[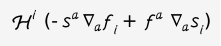]{code_listings_Hamilcar-FindAlgebraSuperMomentumAutocommutatorContravariant.tex}
The output~\lstoutputcref{code_listings_Hamilcar-SuperMomentumAlgebraContravariant-output} reveals how the super-momentum auto-commutator is proportional to the super-momentum itself:
\begin{equation}\label{SuperMomentumAutocommutatorCovariantIndices}
	\bigl\{ \SuperConstraint{_i}[\SmearingF{^i}],
	\SuperConstraint{_j}[\SmearingS{^j}]\bigr\}
	=\SuperConstraint{_{i}}\bigg[
		\SmearingF{^j}\CD{_j}\SmearingS{^i}
		-\SmearingS{^j}\CD{_j}\SmearingF{^i}
	\bigg]
	\, .
\end{equation}
Alternatively, we can set up essentially the same bracket using covariant vector smearing functions:
\lstinputoutput{code_listings_Hamilcar-SetupSuperMomentumAutocommutatorCovariant.tex}
We then feed~\lstinputcref{code_listings_Hamilcar-SetupSuperMomentumAutocommutatorCovariant} into \lstinline!PoissonBracket!:
\lstinputoutput[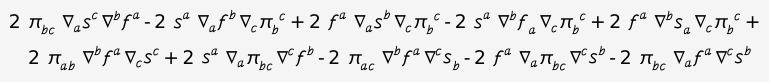]{code_listings_Hamilcar-ComputeSuperMomentumAutocommutatorCovariant.tex}
The result in~\lstoutputcref{code_listings_Hamilcar-SuperMomentumSuperMomentumCovariantPoissonBracket-output} is then re-expressed using \lstinline!FindAlgebra!:
\lstinputoutput[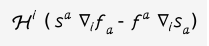]{code_listings_Hamilcar-FindAlgebraSuperMomentumAutocommutatorCovariant.tex}
This result is also correct, corresponding to
\begin{equation}\label{SuperMomentumAutocommutatorContravariantIndices}
	\bigl\{ \SuperConstraint{^i}[\SmearingF{_i}],
	\SuperConstraint{^j}[\SmearingS{_j}]\bigr\}
	=
	\SuperConstraint{^{i}}\bigg[
		\SmearingS{^j}\CD{_i}\SmearingF{_j}
		-\SmearingF{^j}\CD{_i}\SmearingS{_j}
	\bigg]
	\, .
\end{equation}
Both computations in~\lstoutputcref{code_listings_Hamilcar-SuperMomentumAlgebraContravariant-output} and~\lstoutputcref{code_listings_Hamilcar-SuperMomentumAlgebraCovariant-output} yield the same fundamental result, as seen also in~\cref{SuperMomentumAutocommutatorCovariantIndices,SuperMomentumAutocommutatorContravariantIndices}: the super-momentum auto-commutator is proportional to the super-momentum itself. The apparent difference is due to the point mentioned above, that the smearing functions are taken not to depend on the phase-space variables, including the spatial metric. At the level of \xAct{}, this is a feature already of how \lstinline!DefTensor! works: the defined index configuration is recorded as a property of the tensor. Thus, the choice of covariant versus contravariant smearing functions leads to an implicit extra factor of the spatial metric in the arguments of \lstinline!PoissonBracket!, and interactions with the momentum conjugate to the spatial metric then lead to the different final forms.

\paragraph*{Final summary} Taken together, the brackets in~\cref{SuperHamiltonianAutocommutator,SuperHamiltonianSuperMomentumBracket,SuperMomentumAutocommutatorCovariantIndices} constitute the Dirac algebra of the constraints in GR. Since each bracket is proportional to the original constraints, it follows that the constraints are all first class. It is not possible, therefore, for further constraints to arise in the Dirac--Bergmann algorithm. Taken together, the independent constraints of the first class contained in~$\SuperConstraint{}$ and~$\SuperConstraint{_i}$ are seen to be~$\NFirst{}\!=\!4$, meanwhile the independent constraints of the second class are~$\NSecond{} \!=\! 0$. The full number of phase-space variables distributed over the independent components of the spatial metric~$\MetricFoliation{_{ij}}$ and the conjugate momentum~$\ConjugateMomentumG{^{ij}}$ is~$\NCan{} \!=\! 12$, the final number of propagating modes is found by referring back to~\cref{Nphy} to be
\begin{equation}\label{PhysicalDegreesOfFreedom}
	\NPhy{} = \frac{1}{2}
	\Bigl( 12 - 2 \!\times\! 4 \Bigr)
	=
	2
	\, ,
\end{equation}
which are accounted for by the two polarizations of the massless graviton.

\subsection{Case study: pure~$R^2$ theory}\label{PureR2Theory}

\paragraph*{Motivation} In the space of quadratic gravity theories, the pure~$\Curvature{}^2$ theory is somewhat complementary to~\cref{GRaction}, in that it contains no Einstein--Hilbert term. As shown in~\cite{Barker:2025gon}, this feature prohibits a perturbative treatment on flat spacetime, but the canonical analysis can still be performed non-perturbatively. The action for the pure~$\Curvature{}^2$ theory is 
\begin{equation}
	\Action{}[\G{_{\mu\nu}}] = \int\! \mathrm{d}^4x \, \sqrt{-\G{}} \, \Curvature{}^2 \, .
\label{R2action}
\end{equation}
Unlike for the case of~\cref{GRaction}, there is no dimensionful coupling, and so we avoid introducing any coupling whatever in~\cref{R2action}. Note that the theory in~\cref{R2action} has --- apart from its pedagogical value as an exercise in field theory --- attracted attention in recent literature~\cite{Alvarez-Gaume:2015rwa,Hell:2023mph,Golovnev:2023zen,Karananas:2024hoh,Barker:2025gon}, including in the Palatini formulation~\cite{Glavan:2023cuy,Karananas:2024qrz}. The fully non-linear Hamiltonian analysis of the theory was presented for the first time in~\cite{Barker:2025gon}; in this section we use \Hamilcar{} to reproduce the results of~\cite{Barker:2025gon} in somewhat more detail.

\subsubsection{Phase-space formulation}\label{subsec: Canonical action}

\paragraph*{Hiding the accelerations} Once again, the result in~\cref{WellKnownFormula} allows~\cref{R2action} to be broken down into a more suitable notation, as
\begin{equation}
	\Action{}[ \Lapse{}, \Shift{^i}, \MetricFoliation{_{ij}} ] 
	= 
	\int\! \mathrm{d}^{4\!} x \, N \sqrt{h} \,
	\Bigl[ 2F + K^2 - K^{ij} K_{ij} + \mathcal{R} \Bigr]
	\Bigl[ 2F + K^2 - K^{kl} K_{kl} + \mathcal{R} \Bigr]
	\, .
\label{ADMaction}
\end{equation}
The theory in~\cref{ADMaction} contains both velocities and accelerations, and the trick in~\cref{IntegrationByParts} cannot be used to change this character. In order to make progress towards the phase-space formulation, we again introduce auxiliary fields: this time we need~$\ExtrinsicCurvature{_{ij}}\approx\ExtrinsicCurvatureActual{_{ij}}$ and~$\FFunction{_{ij}}\approx\FFunctionActual{_{ij}}$, with the fixed relation to the definitions in~\cref{Kdef,Fdef} ensured by Lagrange multipliers. This procedure was already used in moving from~\cref{ADMactionGRFinal} to~\cref{CanonicalActionGR}, but this time the introduction of the additional variable~$\FFunction{_{ij}}$ requires an extra Lagrange multiplier that we call~$\ConjugateMomentumExtrinsicCurvature{^{ij}}$. Thus, with reference to~\cref{Kdef,Fdef} we write
\begin{align}
\Action{}\bigl[ \Lapse{}, \Shift{^i}, \MetricFoliation{_{ij}}, \ExtrinsicCurvature{_{ij}}, \FFunction{_{ij}}, \ConjugateMomentumG{^{ij}}, \ConjugateMomentumExtrinsicCurvature{^{ij}} \bigr] 
	=
	\int\! \mathrm{d}^{4\!} x \, \biggl[ &
		\Lapse{} \sqrt{\MetricFoliation{}} \, \Bigl( 2\FFunction{} + \ExtrinsicCurvature{}^2 
		- \ExtrinsicCurvature{^{ij}} \ExtrinsicCurvature{_{ij}} + \R{} \Bigr)^{\!2}
		\nonumber \\ &
		+ \ConjugateMomentumG{^{ij}} \Bigl( \MetricFoliationDot{_{ij}} - 2 \CD{_{(i}} \Shift{_{j)}}
+ 2 \Lapse{} \ExtrinsicCurvature{_{ij}} \Bigr)
		\nonumber \\ &
+ \ConjugateMomentumExtrinsicCurvature{^{ij}} \Bigl( \ExtrinsicCurvatureDot{_{ij}}
	+ \Lapse{} \ExtrinsicCurvature{_{ik}} \ExtrinsicCurvature{^k_j}
	- \Shift{^k} \CD{_k} \ExtrinsicCurvature{_{ij}}
		\nonumber \\ &
	- 2 \ExtrinsicCurvature{_{k(i}} \CD{_{j)}} \Shift{^k}
+ \CD{_i} \CD{_j} \Lapse{} + \Lapse{} \FFunction{_{ij}} \Bigr)
\biggr]
\, .
\label{ExtendedAction}
\end{align}
This time, due to the presence of the new multiplier, the field~$\ExtrinsicCurvature{_{ij}}$ cannot be integrated out as when moving from~\cref{CanonicalActionGRFinal} to~\cref{CanonicalActionGRFinal2} without producing explicit squares of velocities. We thus retain~$\ExtrinsicCurvature{_{ij}}$ as a phase-space variable, and identify~$\ConjugateMomentumExtrinsicCurvature{^{ij}}$ as its conjugate momentum. In \Hamilcar{}, we define this canonical pair using \lstinline!DefCanonicalField!, which automatically defines the conjugate momentum based on the field's index structure and symbolic name:
\lstinputoutput{code_listings_Private-DefineExtrinsicCurvature.tex}
The conjugate momentum is \lstinline!ConjugateMomentumExtrinsicCurvature[a,b]!. We also define the traces of the extrinsic curvature and its conjugate momentum, along with rules for expanding and contracting these traces. First, for the extrinsic curvature:
\lstinputoutput{code_listings_Private-DefineTraceExtrinsicCurvature.tex}
Similarly, for its conjugate momentum:
\lstinputoutput{code_listings_Private-DefineTraceConjugateMomentumExtrinsicCurvature.tex}
These trace definitions will be useful for simplifying expressions involving contracted indices in the subsequent analysis.

\paragraph*{Phase-space action} Whilst~$\ExtrinsicCurvature{_{ij}}$ cannot be integrated out, partial success is found with the second auxiliary field~$\FFunction{_{ij}}$ whose contraction~$\FFunction{} \equiv \MetricFoliation{^{ij}} \FFunction{_{ij}}$ can be determined on-shell according to the field equation 
\begin{align}\label{FtraceSolution}
\mathcal{F} \approx
	- \frac{1}{2} \Bigl( \ExtrinsicCurvature{}^2 
	- \ExtrinsicCurvature{^{ij}} \ExtrinsicCurvature{_{ij}} + \R{} \Bigr)
	- \frac{1}{24} \frac{\ConjugateMomentumExtrinsicCurvature{}}{ \sqrt{\MetricFoliation{}} }
	\, .
\end{align}
When~\cref{FtraceSolution} is substituted back into~\cref{ExtendedAction}, we notice how the effect of the remaining trace-free portion~$\FFunction{_{ij}} - \tfrac{1}{3} \MetricFoliation{_{ij}} \FFunction{}$ is to enforce~$\PrimaryConstraint{^{ij}} \approx 0$, where we define the quantity 
\begin{equation}
\PrimaryConstraint{^{ij}}
 \equiv
	\ConjugateMomentumExtrinsicCurvature{^{ij}} - \frac{\ConjugateMomentumExtrinsicCurvature{}}{3}\MetricFoliation{^{ij}} \, .
\label{TracelessConstraint}
\end{equation}
Since it vanishes on-shell, we see that~$\PrimaryConstraint{^{ij}}$ is a new constraint. The corresponding \Hamilcar{} definition is:
\lstinputoutput{code_listings_Private-DefinePrimaryConstraint.tex}
It is useful to define rules that impose the primary constraint shell, setting the trace-free part of~$\ConjugateMomentumExtrinsicCurvature{^{ij}}$ to zero:
\lstinputoutput{code_listings_Private-DefineToPrimaryShell.tex}
The rule~\lstinputcref{code_listings_Private-DefineToPrimaryShell} will be used later when computing equations of motion, allowing us to simplify expressions by assuming the constraint to be satisfied. We also define an `explicit' rule that retains the primary constraint:
\lstinputoutput{code_listings_Private-DefineToPrimaryShellExplicit.tex}
The rule~\lstinputcref{code_listings_Private-DefineToPrimaryShellExplicit} is useful when we need to manipulate expressions on the constraint shell while keeping the constraint explicit in the result. Thus,~$\FFunction{_{ij}} - \tfrac{1}{3} \MetricFoliation{_{ij}} \FFunction{}$ is acting as a multiplier field, albeit one with an implicit dependence on the metric~$\MetricFoliation{_{ij}}$ which ensures its trace-free property at all points on the foliation. To avoid such a dependence, which renders the evaluation of time derivatives tedious, it is convenient to introduce an alternative full-rank Lagrange multiplier field~$\Multiplier{_{ij}}$ which directly multiplies~$\PrimaryConstraint{^{ij}}$ in the action.\footnote{Note that this step does not alter the physical content of the theory.} Collecting the constraints enforced by~$\Multiplier{_{ij}}$, and the lapse and shift, and so separating out the symplectic structure as we did in~\cref{CanonicalActionGRFinal}, the phase-space action may then be obtained. Contrary to the form given in~\cref{CanonicalActionGRFinal2}, however, we will follow~\cite{Barker:2025gon} by opting here to collect with respect to~$\Shift{_i}$ rather than~$\Shift{^i}$, so that the momentum constraint is taken to be contravariant --- as discussed already in~\cref{subsec: ADM decomposition} this choice is not necessarily ideal, and will result in complications later on. The phase-space action is thus found to be
\begin{align}
	\Action{} \bigl[ \Lapse{}, \Shift{_i}, \Multiplier{_{ij}}, \MetricFoliation{_{ij}}, \ConjugateMomentumG{^{ij}}, \ExtrinsicCurvature{_{ij}}, & \ConjugateMomentumExtrinsicCurvature{^{ij}} \bigr] 
	\nonumber\\
&=	
\int\! \mathrm{d}^4x \, \Bigl[
	\ConjugateMomentumG{^{ij}} \MetricFoliationDot{_{ij}}
		+
		\ConjugateMomentumExtrinsicCurvature{^{ij}} \ExtrinsicCurvatureDot{_{ij}}
		-
		\Lapse{} \bigl( \SuperConstraint{} + \Multiplier{_{ij}} \PrimaryConstraint{^{ij}} \bigr)
		-
		\Shift{_i} \SuperConstraint{^i}
		\Bigr]
		\, ,
\label{CanonicalAction}
\end{align}
where the new super-Hamiltonian and super-momentum (which are still constraints) are defined as
\begin{subequations}
\begin{align}
\SuperConstraint{}
	&\equiv
	\frac{1}{144}\frac{ \ConjugateMomentumExtrinsicCurvature{}^2 }{ \sqrt{\MetricFoliation{}} }
	-2\ExtrinsicCurvature{_{ij}} \ConjugateMomentumG{^{ij}}
	+\frac{1}{6} \Bigl( \ExtrinsicCurvature{}^2 - \ExtrinsicCurvature{^{ij}} \ExtrinsicCurvature{_{ij}} + \R{} \Bigr) \ConjugateMomentumExtrinsicCurvature{}
	-\ExtrinsicCurvature{_{ik}} \ExtrinsicCurvature{^k_j} \ConjugateMomentumExtrinsicCurvature{^{ij}}
	-\CD{_i}\CD{_j}\ConjugateMomentumExtrinsicCurvature{^{ij}}
	\, ,
	\label{PreHamiltonianConstraint}
\\
\SuperConstraint{^i}
	&\equiv
	-2 \CD{_j} \ConjugateMomentumG{^{ij}}
	+\ConjugateMomentumExtrinsicCurvature{^{kl}}\CD{^i}\ExtrinsicCurvature{_{kl}}
		-
		2 \CD{^k} \Bigl(
		\ExtrinsicCurvature{^{il}} \ConjugateMomentumExtrinsicCurvature{_{kl}}
		\Bigl)
	\, .
	\label{PreMomentumConstraint}
\end{align}
\end{subequations}
Evidently, the constraints in~\cref{PreHamiltonianConstraint,PreMomentumConstraint} differ from those in~\cref{GRconstraints}. They can be further refined by eliminating any dependence on~$\PrimaryConstraint{^{ij}}$, which is constrained, and this is equivalent to redefining the Lagrange multiplier~$\Multiplier{_{ij}}$. Thus, a more compact (and physically equivalent) form for~\cref{PreHamiltonianConstraint,PreMomentumConstraint} is
\begin{subequations}
\begin{align}
\SuperConstraint{} &\equiv
	\frac{1}{144}\frac{ \ConjugateMomentumExtrinsicCurvature{}^2 }{ \sqrt{\MetricFoliation{}} }
	-2\ExtrinsicCurvature{_{ij}} \ConjugateMomentumG{^{ij}}
	+\frac{1}{6} \Bigl( \ExtrinsicCurvature{}^2 - 3\ExtrinsicCurvature{^{ij}} \ExtrinsicCurvature{_{ij}} + \R{} \Bigr) \ConjugateMomentumExtrinsicCurvature{}
	-\frac{1}{3}\CD{_i}\CD{^i}\ConjugateMomentumExtrinsicCurvature{}
	\, ,
\label{HamiltonianConstraint}
\\
\SuperConstraint{^i} &\equiv
	-2 \CD{_j} \ConjugateMomentumG{^{ij}}
	+\frac{1}{3}\ConjugateMomentumExtrinsicCurvature{}\CD{^i}\ExtrinsicCurvature{}
	-\frac{2}{3} \CD{_j} \Bigl(
		\ExtrinsicCurvature{^{ij}} \ConjugateMomentumExtrinsicCurvature{}
		\Bigl)
	\, .
\label{MomentumConstraint}
\end{align}
\end{subequations}
The corresponding \Hamilcar{} definitions are as follows. The super-Hamiltonian~$\SuperConstraint{}$ is:
\lstinputoutput{code_listings_Private-DefineSuperHamiltonian.tex}
The super-momentum~$\SuperConstraint{^i}$ is defined as:
\lstinputoutput{code_listings_Private-DefineSuperMomentum.tex}
Taken together, the construction in~\cref{CanonicalAction,TracelessConstraint,HamiltonianConstraint,MomentumConstraint} is now in the correct format for the Dirac--Bergmann algorithm to be applied.

\subsubsection{Dirac--Bergmann algorithm}\label{subsec: Constraint analysis}

\paragraph*{Total Hamiltonian} By taking variations of~\cref{CanonicalAction} with respect to~$\Lapse{}$,~$\Shift{_i}$, and~$\Multiplier{_{ij}}$, three sets of constraints emerge, to be compared with the two sets in~\cref{GRprimaryconstraints} for GR, namely
\begin{equation}
	\SuperConstraint{} \approx 0 \, ,
\qquad
	\SuperConstraint{^i} \approx 0 \, ,
\qquad
	\PrimaryConstraint{^{ij}} \approx 0 \, .
\label{PrimaryConstraints}
\end{equation}
By analogy with~\cref{TotalHamiltonianGR}, the Legendre transformation is now apparent in the simple format of~\cref{CanonicalAction}, leading directly to the total Hamiltonian
\begin{equation}\label{TotalHamiltonianR2}
	\TotalHamiltonian{} = \int\! \mathrm{d}^3x \, \Bigl[ \Lapse{} \bigl( \SuperConstraint{} + \Multiplier{_{ij}} \PrimaryConstraint{^{ij}} \bigr)
		+
	\Shift{_i} \SuperConstraint{^i} \Bigr]\, ,
\end{equation}
and the Dirac--Bergmann algorithm is again concerned with the maintenance of the constraints in~\cref{PrimaryConstraints} under time evolution generated by~\cref{TotalHamiltonianR2}. Once again, the problem reduces to computing Poisson brackets among the various constraints. It will be necessary momentarily, for the purposes of computing the equations of motion, to define the total Hamiltonian density in~\cref{TotalHamiltonianR2}:
\lstinputoutput{code_listings_Private-DefineTotalHamiltonian.tex}
Note that in setting up~\lstinputcref{code_listings_Private-DefineTotalHamiltonian}, we used~\lstinputcref{code_listings_Hamilcar-DefineSuperHamiltonian},~\lstinputcref{code_listings_Hamilcar-DefineSuperMomentum} and~\lstinputcref{code_listings_Private-DefinePrimaryConstraint}.

\paragraph*{Smearing functions} It is also necessary to extend the smearing notation in~\cref{SmearingDef} to accommodate the new constraint in~\cref{TracelessConstraint}. Accordingly, we write
\begin{equation}\label{SmearingDef2}
	\PrimaryConstraint{^{ij}}[\SmearingF{_{ij}}] \equiv \int\! \mathrm{d}^3x \, \SmearingF{_{ij}}(x) \PrimaryConstraint{^{ij}}(x) \, .
\end{equation}
Based on~\cref{SmearingDef2}, we denote e.g.~$\SmearingF{}\equiv\SmearingF{_{ij}}\MetricFoliation{^{ij}}$ in expressions where the rank-two~$\SmearingF{_{ij}}$ has already been introduced inside a bracket; elsewhere~$\SmearingF{}$ denotes a smearing scalar that is completely independent of~$\MetricFoliation{_{ij}}$, but this abuse of notation does not give rise to conflicts. The corresponding \Hamilcar{} definitions for tensor smearing functions are:
\lstinputoutput{code_listings_Private-DefineTensorSmearing.tex}
We will need the additional `fundamental' Poisson bracket
\begin{equation}\label{FundamentalPB}
	\bigl\{ \ExtrinsicCurvature{_{ij}}[\SmearingF{^{ij}}] , \ConjugateMomentumExtrinsicCurvature{^{kl}}[\SmearingS{_{kl}}] \bigr\}
	\equiv
	\big[\SmearingF{^{(ij)}}\SmearingS{_{(ij)}}\big]
	\, ,
\end{equation}
where~\cref{FundamentalPB} and~\cref{FundamentalPoissonBracket} are both specified by the symplectic part of~\cref{CanonicalAction}. To verify that the fundamental Poisson bracket is as expected, following the definition in~\lstinputcref{code_listings_Private-DefineExtrinsicCurvature}, we first set up the bracket between the field and its momentum:
\lstinputoutput{code_listings_Private-SetupPoissonBracketExtrinsicCurvature.tex}
We then feed~\lstinputcref{code_listings_Private-SetupPoissonBracketExtrinsicCurvature} into \lstinline!PoissonBracket! and simplify:
\lstinputoutput[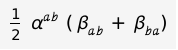]{code_listings_Private-ComputePoissonBracketExtrinsicCurvature.tex}
As expected, the result~\lstoutputcref{code_listings_Private-ComputePoissonBracketExtrinsicCurvature} confirms the canonical commutation relation between~$\ExtrinsicCurvature{_{ij}}$ and its conjugate momentum. Notice that for this basic check we did not take care to manually specify any smearing functions in~\lstinputcref{code_listings_Private-SetupPoissonBracketExtrinsicCurvature}, and so \lstinline!PoissonBracket! provides them for us.

\paragraph*{Hamilton equations} Since they happen to have been derived already in~\cite{Barker:2025gon}, it is worth using \Hamilcar{} to verify the equations of motion for the canonical variables in~\cref{CanonicalAction} explicitly --- note that this was not done in~\cref{PureGR} for GR. These equations follow immediately from the variations of~\cref{CanonicalAction} with respect to all canoncial variables, but they can also be obtained by computing their Poisson brackets with the total Hamiltonian in~\cref{TotalHamiltonianR2} according to
\begin{equation}\label{HamiltonEqs}
\begin{gathered}
	\MetricFoliationDot{_{ij}}\approx \frac{\delta}{\delta\SmearingF{^{ij}}} 
	\bigl\{ \MetricFoliation{_{kl}}\bigl[\SmearingF{^{kl}}\bigr] , \TotalHamiltonian{} \bigr\} \, ,
\qquad
	\ConjugateMomentumGDot{^{ij}}\approx \frac{\delta}{\delta\SmearingF{_{ij}}}
	\bigl\{ \ConjugateMomentumG{^{kl}}\bigl[\SmearingF{_{kl}}\bigr] , \TotalHamiltonian{} \bigr\} \, ,
	\\
	\ExtrinsicCurvatureDot{_{ij}}\approx \frac{\delta}{\delta\SmearingF{^{ij}}}
	\bigl\{ \ExtrinsicCurvature{_{kl}}\bigl[\SmearingF{^{kl}}\bigr] , \TotalHamiltonian{} \bigr\} \, ,
\qquad
\ConjugateMomentumExtrinsicCurvatureDot{^{ij}}\approx \frac{\delta}{\delta\SmearingF{_{ij}}}
	\bigl\{ \ConjugateMomentumExtrinsicCurvature{^{kl}}\bigl[\SmearingF{_{kl}}\bigr] , \TotalHamiltonian{} \bigr\} \, .
\end{gathered}
\end{equation}
The format in~\cref{HamiltonEqs} allows us to use \Hamilcar{} to compute the equations of motion straightforwardly. We set up the Poisson bracket as follows:
\lstinputoutput{code_listings_Private-SetupEquationOfMotionG.tex}
Now we feed~\lstinputcref{code_listings_Private-SetupEquationOfMotionG} into \lstinline!PoissonBracket!:
\lstinputoutput[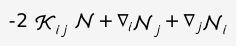]{code_listings_Private-ComputeEquationOfMotionG.tex}
The velocity of the induced metric is found from~\lstoutputcref{code_listings_Private-ComputeEquationOfMotionG-output} to be
\begin{equation}
	\MetricFoliationDot{_{ij}} \approx
	- 2 \Lapse{}\ExtrinsicCurvature{_{ij}}
	+ 2 \CD{_{(i}} \Shift{_{j)}}
	\, ,
\label{EOM1}
\end{equation}
which simply confirms~\cref{Kdef} under the on-shell condition~$\ExtrinsicCurvature{_{ij}}\approx\ExtrinsicCurvatureActual{_{ij}}$. Next, the velocity of the conjugate momentum to the induced metric is:
\lstinputoutput{code_listings_Private-SetupEquationOfMotionConjugateMomentumG.tex}
Now we feed~\lstinputcref{code_listings_Private-SetupEquationOfMotionConjugateMomentumG} into \lstinline!PoissonBracket!. We use~\lstinputcref{code_listings_Private-DefineToPrimaryShell} to simplify the result by imposing the primary constraint shell:
\lstinputoutput[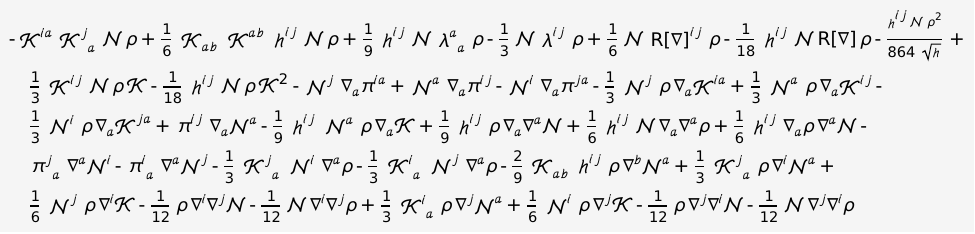]{code_listings_Private-ComputeEquationOfMotionConjugateMomentumG.tex}
Thus~\lstoutputcref{code_listings_Private-ComputeEquationOfMotionConjugateMomentumG-output} allows us to conclude 
\begin{align}
\ConjugateMomentumGDot{^{ij}}&\approx
	-
	\frac{\Lapse{}}{864} \frac{ \ConjugateMomentumExtrinsicCurvature{}^2 }{ \sqrt{\MetricFoliation{}} } \MetricFoliation{^{ij}}
	-
	\Lapse{} \ConjugateMomentumExtrinsicCurvature{}
	\Bigl( \MetricFoliation{^{ik}} \MetricFoliation{^{jl}} - \frac{ 1 }{6} \MetricFoliation{^{ij}} \MetricFoliation{^{kl}} \Bigr)
	\Bigl( \ExtrinsicCurvature{_{km}} \ExtrinsicCurvature{_{l}^m}
	- \frac{\ExtrinsicCurvature{}}{3} \ExtrinsicCurvature{_{kl}} \Bigr)
	+
	\frac{ \Lapse{} \ConjugateMomentumExtrinsicCurvature{} }{6}
	\Bigl(
		\R{^{ij}}
	-
	\frac{\R{} }{3} \MetricFoliation{^{ij}}
	\Bigr)
\nonumber \\ &\hspace{20pt}
	-
	\frac{ \Lapse{} }{6} \,
	\Bigl( \CD{^{(i}} \CD{^{j)}} - \MetricFoliation{^{ij}} \CD{_k} \CD{^k} - \MetricFoliation{^{ij}} \big(\CD{_k} \Lapse{}\big) \CD{^k} \Bigr)
	\ConjugateMomentumExtrinsicCurvature{}
	-
	\frac{ \ConjugateMomentumExtrinsicCurvature{} }{6} \Bigl( \CD{^i} \CD{^j} - \frac{2}{3} \MetricFoliation{^{ij}} \CD{^k} \CD{_k} \Bigr) \Lapse{}
\nonumber \\ &\hspace{20pt}
	+
\, \CD{_k} \Bigl( \Shift{^k} \ConjugateMomentumG{^{ij}} - 2 \Shift{^{(i}} \ConjugateMomentumG{^{j)k}} \Bigr)
	+
	\frac{\ConjugateMomentumExtrinsicCurvature{}}{3} \Bigl(
		\Shift{^k} \CD{_k} \ExtrinsicCurvature{^{ij}}
	+
	2\ExtrinsicCurvature{^{k(i}} \CD{^{j)}} \Shift{_k}
	-
	\frac{2}{3} \MetricFoliation{^{ij}} \ExtrinsicCurvature{^{kl}} \CD{_k} \Shift{_l}
	\Bigr)
\nonumber \\ &\hspace{20pt}
	-
	\frac{2}{3} \, \Shift{^{(i}} \CD{_k}
	\Bigl( \ExtrinsicCurvature{^{j)k}} \ConjugateMomentumExtrinsicCurvature{} \Bigr)
	+
	\frac{\ConjugateMomentumExtrinsicCurvature{}}{3}
	\Bigl( \Shift{^{(i}} \CD{^{j)}}
		-
		\frac{1}{3} \MetricFoliation{^{ij}} \Shift{^k} \CD{_k}
		\Bigr)
		\ExtrinsicCurvature{}
		-\frac{\Lapse{}\ConjugateMomentumExtrinsicCurvature{}}{3}\Big(\Multiplier{^{ij}}-\frac{\Multiplier{}}{3}\MetricFoliation{^{ij}}\Big)
	\, .
\label{EOM2}
\end{align}
For the auxiliary extrinsic curvature:
\lstinputoutput{code_listings_Private-SetupEquationOfMotionExtrinsicCurvature.tex}
We feed~\lstinputcref{code_listings_Private-SetupEquationOfMotionExtrinsicCurvature} into \lstinline!PoissonBracket!:
\lstinputoutput[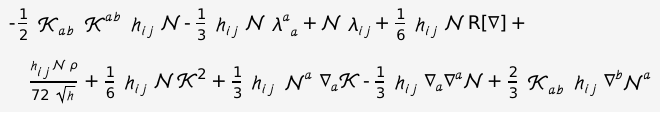]{code_listings_Private-ComputeEquationOfMotionExtrinsicCurvature.tex}
The velocity in~\lstoutputcref{code_listings_Private-ComputeEquationOfMotionExtrinsicCurvature-output} is thus found to be
\begin{align}
\ExtrinsicCurvatureDot{_{ij}} &\approx
\frac{\MetricFoliation{_{ij}} }{3}
	\biggl[
		\frac{\Lapse{}}{2} \biggl(
			\frac{ 1 }{ 12 } \frac{\ConjugateMomentumExtrinsicCurvature{}}{\sqrt{\MetricFoliation{}}}
	+
	\ExtrinsicCurvature{}^2
	\!-\! 3\ExtrinsicCurvature{^{kl}} \ExtrinsicCurvature{_{kl}} + \R{}
	\biggr)
	\!
	-
	\CD{_k} \CD{^k} \Lapse{}
	+
	\Shift{_k} \CD{^k} \ExtrinsicCurvature{}
	+
	2 \ExtrinsicCurvature{_{kl}} \CD{^k} \Shift{^l}
	\biggr]
	\!
\nonumber \\ &\hspace{20pt}
	+
	\Lapse{} \Big(\Multiplier{_{ij}}-\frac{\Multiplier{}}{3}\MetricFoliation{_{ij}}\Big)
	\, .
\label{EOM3}
\end{align}
Finally, for the conjugate momentum to the auxiliary extrinsic curvature:
\lstinputoutput{code_listings_Private-SetupEquationOfMotionConjugateMomentumExtrinsicCurvature.tex}
We feed~\lstinputcref{code_listings_Private-SetupEquationOfMotionConjugateMomentumExtrinsicCurvature} into \lstinline!PoissonBracket!:
\lstinputoutput[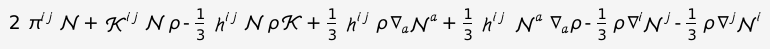]{code_listings_Private-ComputeEquationOfMotionConjugateMomentumExtrinsicCurvature.tex}
The velocity in~\lstoutputcref{code_listings_Private-ComputeEquationOfMotionConjugateMomentumExtrinsicCurvature-output} is thus found to be
\begin{equation}
\ConjugateMomentumExtrinsicCurvatureDot{^{ij}} \approx
2 \Lapse{} \ConjugateMomentumG{^{ij}}
	+
	\Bigl( \ExtrinsicCurvature{^{ij}} - \frac{\ExtrinsicCurvature{} }{3} \MetricFoliation{^{ij}} \Bigr)
	\ConjugateMomentumExtrinsicCurvature{}
	+
	\frac{\MetricFoliation{^{ij}} }{3} \CD{_k} \Bigl( \Shift{^k}\ConjugateMomentumExtrinsicCurvature{}  \Bigr)
	-
	\frac{2\ConjugateMomentumExtrinsicCurvature{}}{3}  \CD{_{(i}} \Shift{_{j)}}
	\, .
\label{EOM4}
\end{equation}
In principle, the velocities in~\crefrange{EOM1}{EOM4} can themselves be used to compute all future Poisson brackets. This is sometimes done, but it offers no computational advantage: in \Hamilcar{} it is the Poisson bracket operation itself that is taken as fundamental.

\paragraph*{Primary constraint algebra}

When computing the primary constraint algebra it will be useful to define the quantity
\begin{equation}
	\SecondaryConstraint{^{ij}} \equiv \ConjugateMomentumG{^{ij}} - \frac{\ConjugateMomentumG{}}{3}\MetricFoliation{^{ij}}
	+
	\frac{\ConjugateMomentumExtrinsicCurvature{}}{6}
	\Bigl(
	\ExtrinsicCurvature{^{ij}} - \frac{\ExtrinsicCurvature{}}{3}\MetricFoliation{^{ij}}
	\Bigr)
	\, ,
\label{SecondaryConstraint}
\end{equation}
which we will presently designate as a further constraint. The corresponding \Hamilcar{} definition is:
\lstinputoutput{code_listings_Private-DefineSecondaryConstraint.tex}
The auto-commutator of the super-Hamiltonian is the Poisson bracket:
\lstinputoutput{code_listings_Private-SetupSuperHamiltonianSuperHamiltonian.tex}
Now we feed~\lstinputcref{code_listings_Private-SetupSuperHamiltonianSuperHamiltonian} into \lstinline!PoissonBracket!:
\lstinputoutput[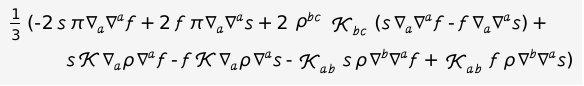]{code_listings_Private-ComputeSuperHamiltonianSuperHamiltonian.tex}
We then use \lstinline!FindAlgebra! to express~\lstoutputcref{code_listings_Private-ComputeSuperHamiltonianSuperHamiltonian} in terms of the constraints, making use of~\lstinputcref{code_listings_Private-DefineSecondaryConstraint} --- which defines~$\SecondaryConstraint{^{ij}}$ --- in our ansatz:
\lstinputoutput[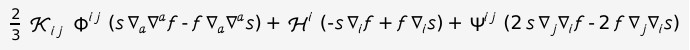]{code_listings_Private-FindAlgebraSuperHamiltonianSuperHamiltonian.tex}
The result~\lstoutputcref{code_listings_Private-FindAlgebraSuperHamiltonianSuperHamiltonian-output} is
\begin{align}
	\bigl\{ \SuperConstraint{}[\SmearingF{}] , \SuperConstraint{}[\SmearingS{}] \bigr\}
	&=
	\SuperConstraint{^i}\Big[
		\SmearingF{}\CD{_i}\SmearingS{}
		-\SmearingS{}\CD{_i}\SmearingF{}
	\Big]
	+\PrimaryConstraint{^{ij}}\bigg[
		\frac{2}{3}\ExtrinsicCurvature{_{ij}}\Big(
			\SmearingS{}\CD{_k}\CD{^k}\SmearingF{}
			-\SmearingF{}\CD{_k}\CD{^k}\SmearingS{}
		\Big)
	\bigg]
	\nonumber\\ &\ \ \
	+\SecondaryConstraint{^{ij}}\bigg[
		2\Big(
			\SmearingS{}\CD{_i}\CD{_j}\SmearingF{}
			-\SmearingF{}\CD{_i}\CD{_j}\SmearingS{}
		\Big)
	\bigg]
	\, .
\label{SuperHamiltonianSuperHamiltonian}
\end{align}
We progress similarly for the bracket between the super-Hamiltonian and super-momentum:
\lstinputoutput{code_listings_Private-SetupSuperHamiltonianSuperMomentum.tex}
Now we feed~\lstinputcref{code_listings_Private-SetupSuperHamiltonianSuperMomentum} into \lstinline!PoissonBracket!:
\lstinputoutput[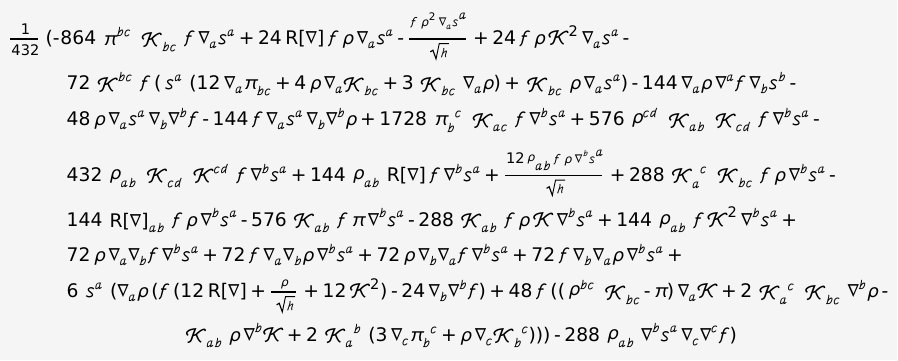]{code_listings_Private-ComputeSuperHamiltonianSuperMomentum.tex}
We then use \lstinline!FindAlgebra! to express~\lstoutputcref{code_listings_Private-ComputeSuperHamiltonianSuperMomentum} in terms of the constraints, again making use of~\lstinputcref{code_listings_Private-DefineSecondaryConstraint} in our ansatz:
\lstinputoutput[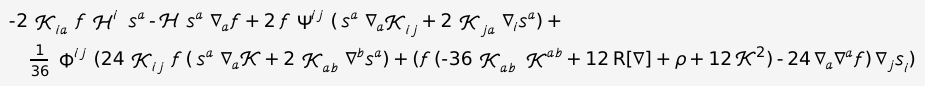]{code_listings_Private-FindAlgebraSuperHamiltonianSuperMomentum.tex}
The result~\lstoutputcref{code_listings_Private-FindAlgebraSuperHamiltonianSuperMomentum-output} is
\begin{align}
	\bigl\{ \SuperConstraint{}[\SmearingF{}] , \SuperConstraint{^i}[\SmearingS{_i}] \bigr\}
	&=
	\SuperConstraint{}\Big[
		-\SmearingS{^i}\CD{_i}\SmearingF{}
	\Big]
	+\SuperConstraint{^i}\Big[
		-2\SmearingF{}\ExtrinsicCurvature{_i^j}\SmearingS{_j}
	\Big]
	+\PrimaryConstraint{^{ij}}\Bigg[
		\frac{2}{3}\ExtrinsicCurvature{_{ij}}\SmearingF{}\Big(
			\SmearingS{^k}\CD{_k}\ExtrinsicCurvature{}
			+2\ExtrinsicCurvature{_{kl}}\CD{^k}\SmearingS{^l}
		\Big)
	\nonumber\\ &\hspace{20pt}
		-\bigg(
			\frac{2}{3}\CD{_k}\CD{^k}\SmearingF{}
			+\Big(
				\ExtrinsicCurvature{_{kl}}\ExtrinsicCurvature{^{kl}}
				-\frac{\R{}}{3}
				-\frac{\ConjugateMomentumExtrinsicCurvature{}}{36\sqrt{\MetricFoliation{}}}
				-\frac{\ExtrinsicCurvature{}^2}{3}
			\Big)\SmearingF{}
		\bigg)\CD{_i}\SmearingS{_j}
	\Bigg]
	\nonumber\\ &\ \ \
	+\SecondaryConstraint{^{ij}}\bigg[
		2\SmearingF{}\Big(
			\SmearingS{^k}\CD{_k}\ExtrinsicCurvature{_{ij}}
			+2\ExtrinsicCurvature{_i^k}\CD{_j}\SmearingS{_k}
		\Big)
	\bigg]
	\, .
\label{SuperHamiltonianSuperMomentum}
\end{align}
For the auto-commutator of the super-momentum, we set up the Poisson bracket:
\lstinputoutput{code_listings_Private-SetupSuperMomentumSuperMomentum.tex}
Now we feed~\lstinputcref{code_listings_Private-SetupSuperMomentumSuperMomentum} into \lstinline!PoissonBracket!:
\lstinputoutput[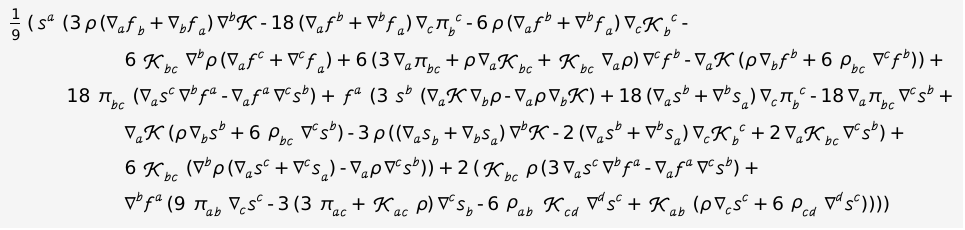]{code_listings_Private-ComputeSuperMomentumSuperMomentum.tex}
We then use \lstinline!FindAlgebra! to express~\lstoutputcref{code_listings_Private-ComputeSuperMomentumSuperMomentum} in terms of the constraints, again making use of~\lstinputcref{code_listings_Private-DefineSecondaryConstraint} in our ansatz:
\lstinputoutput[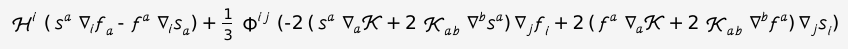]{code_listings_Private-FindAlgebraSuperMomentumSuperMomentum.tex}
The result~\lstoutputcref{code_listings_Private-FindAlgebraSuperMomentumSuperMomentum-output} is
\begin{align}
	\bigl\{ \SuperConstraint{^i}[\SmearingF{_i}] , \SuperConstraint{^j}[\SmearingS{_j}] \bigr\}
	&=
	\SuperConstraint{^i}\Big[
		\SmearingS{^j}\CD{_i}\SmearingF{_j}
		-\SmearingF{^j}\CD{_i}\SmearingS{_j}
	\Big]
	+\PrimaryConstraint{^{ij}}\Bigg[
		\frac{2}{3}\bigg(
			\CD{_k}\ExtrinsicCurvature{}\Big(
				\SmearingF{^k}\CD{_i}\SmearingS{_j}
				-\SmearingS{^k}\CD{_i}\SmearingF{_j}
			\Big)
	\nonumber\\ &\hspace{20pt}
			+2\ExtrinsicCurvature{_{kl}}\Big(
				\CD{^k}\SmearingF{^l}\CD{_i}\SmearingS{_j}
				-\CD{^k}\SmearingS{^l}\CD{_i}\SmearingF{_j}
			\Big)
		\bigg)
	\Bigg]
	\, .
\label{SuperMomentumSuperMomentum}
\end{align}
Next, we compute the commutator of the primary constraint with the super-Hamiltonian:
\lstinputoutput{code_listings_Private-SetupPrimaryConstraintSuperHamiltonian.tex}
Now we feed~\lstinputcref{code_listings_Private-SetupPrimaryConstraintSuperHamiltonian} into \lstinline!PoissonBracket!:
\lstinputoutput[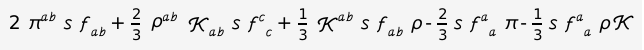]{code_listings_Private-ComputePrimaryConstraintSuperHamiltonian.tex}
We then use \lstinline!FindAlgebra! to express~\lstoutputcref{code_listings_Private-ComputePrimaryConstraintSuperHamiltonian} in terms of the constraints, making use of~\lstinputcref{code_listings_Private-DefineSecondaryConstraint} in our ansatz:
\lstinputoutput[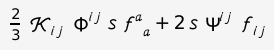]{code_listings_Private-FindAlgebraPrimaryConstraintSuperHamiltonian.tex}
The result~\lstoutputcref{code_listings_Private-FindAlgebraPrimaryConstraintSuperHamiltonian-output} is
\begin{equation}
\bigl\{ \PrimaryConstraint{^{ij}}[\SmearingF{_{ij}}] , \SuperConstraint{}[\SmearingS{}] \bigr\}
	=
	\PrimaryConstraint{^{ij}}\bigg[
		\frac{2}{3}\SmearingS{}\SmearingF{}
			\ExtrinsicCurvature{_{ij}}
	\bigg]
	+\SecondaryConstraint{^{ij}}\Big[
		2\SmearingS{}\SmearingF{_{ij}}
	\Big]
	\, .
\label{PrimaryConstraintSuperHamiltonian}
\end{equation}
Similarly, we compute the commutator of the primary constraint with the super-momentum:
\lstinputoutput{code_listings_Private-SetupPrimaryConstraintSuperMomentum.tex}
Now we feed~\lstinputcref{code_listings_Private-SetupPrimaryConstraintSuperMomentum} into \lstinline!PoissonBracket!:
\lstinputoutput[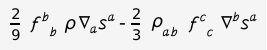]{code_listings_Private-ComputePrimaryConstraintSuperMomentum.tex}
We then use \lstinline!FindAlgebra! to express~\lstoutputcref{code_listings_Private-ComputePrimaryConstraintSuperMomentum} in terms of the constraints:
\lstinputoutput[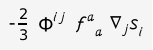]{code_listings_Private-FindAlgebraPrimaryConstraintSuperMomentum.tex}
The result~\lstoutputcref{code_listings_Private-FindAlgebraPrimaryConstraintSuperMomentum-output} is
\begin{equation}
\bigl\{ \PrimaryConstraint{^{ij}}[\SmearingF{_{ij}}] , \SuperConstraint{^k}[\SmearingS{_k}] \bigr\}
	=
	\PrimaryConstraint{^{ij}}\bigg[
		-\frac{2}{3}\SmearingF{}\CD{_i}\SmearingS{_j}
	\bigg]
	\, .
\label{PrimaryConstraintSuperMomentum}
\end{equation}
Finally, we can assert without detailed computation that the auto-commutator of the primary constraint vanishes, due to the definition in~\cref{TracelessConstraint}, i.e.,
\begin{equation}
\bigl\{ \PrimaryConstraint{^{ij}}[\SmearingF{_{ij}}] , \PrimaryConstraint{^{kl}}[\SmearingS{_{kl}}] \bigr\}
	=
	0
	\, .
\label{PrimaryConstraintPrimaryConstraint}
\end{equation}

\paragraph*{Secondary constraint algebra} As promised, the results in~\cref{SuperHamiltonianSuperHamiltonian,SuperHamiltonianSuperMomentum,SuperMomentumSuperMomentum,PrimaryConstraintSuperHamiltonian,PrimaryConstraintSuperMomentum,PrimaryConstraintPrimaryConstraint} reveal that~\cref{SecondaryConstraint} must also be constrained, i.e.,
\begin{equation}\label{SecondaryConstraintDefinition}
	\SecondaryConstraint{^{ij}} \approx 0 \, ,
\end{equation}
since~\cref{SecondaryConstraintDefinition} alone forces the full algebra to weakly disappear. Let us now compute the Poisson bracket of the secondary constraint with the primary constraint:
\lstinputoutput{code_listings_Private-SetupSecondaryConstraintPrimaryConstraint.tex}
Now we feed~\lstinputcref{code_listings_Private-SetupSecondaryConstraintPrimaryConstraint} into \lstinline!PoissonBracket! and simplify on the primary constraint shell using~\lstinputcref{code_listings_Private-DefineToPrimaryShellExplicit}. We don't bother to use \lstinline!FindAlgebra! for this task:
\lstinputoutput[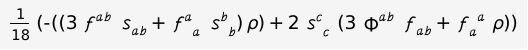]{code_listings_Private-ComputeSecondaryConstraintPrimaryConstraint.tex}
From~\lstoutputcref{code_listings_Private-ComputeSecondaryConstraintPrimaryConstraint} the bracket between the two new constraints is
\begin{equation}
	\bigl\{ \SecondaryConstraint{^{ij}}[\SmearingF{_{ij}}] , \PrimaryConstraint{^{kl}}[\SmearingS{_{kl}}] \bigr\}
	=
	\PrimaryConstraint{^{ij}}\bigg[
		\frac{\SmearingS{}}{3}\SmearingF{_{ij}}
	\bigg]
	+
	\ConjugateMomentumExtrinsicCurvature{}
	\bigg[
		-
	\frac{1}{6}
	\Bigl(
	\SmearingF{_{ij}}\SmearingS{^{ij}}
	-
	\frac{\SmearingF{}\SmearingS{}}{3}
	\Bigr)
	\bigg]
	\, .
\label{TracelessPoisson}
\end{equation}
An important feature of~\cref{TracelessPoisson}, which is discussed in~\cite{Barker:2025gon}, is that it disappears for phase-space configurations where~$\ConjugateMomentumExtrinsicCurvature{}= 0$. In the bulk of the phase space, the bracket does not disappear, and so the two new constraints are generally taken to be second class. The auto-commutator of the secondary constraint self-evidently vanishes due to the definition in~\cref{TracelessConstraint}, i.e.,
\begin{equation}
	\bigl\{ \SecondaryConstraint{^{ij}}[\SmearingF{_{ij}}] , \SecondaryConstraint{^{kl}}[\SmearingS{_{kl}}] \bigr\}
	=
	0
	\, .
\label{SecondarySecondary}
\end{equation}
From this point onwards for the~$R^2$ model, we take the functionality of \Hamilcar{} to be well-demonstrated, and we proceed directly by stating results.

\paragraph*{Determination of the multiplier} Because of the second class property following from~\cref{TracelessPoisson}, the vanishing of the velocity of~$\SecondaryConstraint{^{ij}}$ is not expected to lead to any more constraints. Instead, the conservation law is satisfied so long as the trace-free part of the Lagrange multiplier~$\Multiplier{_{ij}}$ obeys the following solution:
\begin{align}
	\Multiplier{_{ij}}-\frac{\Multiplier{}}{3}\MetricFoliation{_{ij}}\approx 
	\Big(\Kronecker{^k_{(i}}\Kronecker{^l_{j)}} - \frac{1}{3}\MetricFoliation{^{kl}}\MetricFoliation{_{ij}}\Big)
	\Bigg[&
		\CurvatureFoliation{_{kl}}
		-2\Big(\ExtrinsicCurvature{_k^m}\ExtrinsicCurvature{_{lm}}
		-\frac{\ExtrinsicCurvature{}}{3}\ExtrinsicCurvature{_{kl}}\Big)
	    \nonumber\\&
	-\frac{1}{\Lapse{}}\Big(
		\CD{_k}\CD{_l}\Lapse{}
		-2\ExtrinsicCurvature{_k^m}\CD{_l}\Shift{_m}
		-\Shift{^m}\CD{_m}\ExtrinsicCurvature{_{kl}}
		\Big)
	    \nonumber\\&
	-\frac{1}{\ConjugateMomentumExtrinsicCurvature{}}\Big(
		2\ConjugateMomentumG{}\ExtrinsicCurvature{_{kl}}
		+\CD{_k}\CD{_l}\ConjugateMomentumExtrinsicCurvature{}
		\Big)
	\Bigg].\label{MultiplierSolution}
\end{align}
Once again, we note that~\cref{MultiplierSolution} has a relevant feature, which is a direct consequence of~\cref{TracelessPoisson}: a \emph{negative} power of~$\ConjugateMomentumExtrinsicCurvature{}$ appears in the final term. As discussed in~\cite{Barker:2025gon}, this suggests that the model may have some curious properties in the vicinity of~$\ConjugateMomentumExtrinsicCurvature{}=0$. We did not encounter any solutions such as~\cref{MultiplierSolution} in the case of GR in~\cref{PureGR}. In that case, the multipliers were restricted to~$\Lapse{}$ and~$\Shift{^i}$, which remained entirely undetermined due to the diffeomorphism invariance of the theory. The theory~\cref{R2action} is also diffeomorphism-invariant, and so no further solutions are forthcoming.

\paragraph*{Dressing of the constraints} In order for~$\Lapse{}$ and~$\Shift{^i}$ to remain undetermined, it is sufficient, but not necessary, that the~$\SuperConstraint{^i}$ and~$\SuperConstraint{}$ be first class. This would, however, give rise to an inconsistency, since the dynamical evolution of~$\Multiplier{_{ij}}$ in~\cref{MultiplierSolution} is non-trivial. At an even more prosaic level, commutators of first class functions are necessarily first class, whilst~\cref{SuperHamiltonianSuperHamiltonian,SuperHamiltonianSuperMomentum,SuperMomentumSuperMomentum} are strongly linear in~$\PrimaryConstraint{^{ij}}$ and~$\SecondaryConstraint{^{ij}}$, which are currently thought to be second class. Accordingly, we define candidate first class constraints~$\DressedSuperConstraint{^i}$ and~$\DressedSuperConstraint{}$ by dressing the~$\SuperConstraint{^i}$ and~$\SuperConstraint{}$. The dressed constraints are defined as follows:
\begin{subequations}
\begin{align}
	\DressedSuperConstraint{}
	&\equiv
	\SuperConstraint{}
	+\PrimaryConstraint{^{ij}}\bigg(
		\R{_{ij}}
		-2\Big(
			\ExtrinsicCurvature{_i^k}
			-\frac{\ExtrinsicCurvature{}}{3}\Kronecker{^k_i}
		\Big)\ExtrinsicCurvature{_{jk}}
	\bigg)
	-\frac{\PrimaryConstraint{^{ij}}}{\ConjugateMomentumExtrinsicCurvature{}}\left(
		2\ConjugateMomentumG{}\ExtrinsicCurvature{_{ij}}
		+\CD{_i}\CD{_j}\ConjugateMomentumExtrinsicCurvature{}
		\right)
	-\CD{_i}\CD{_j}\PrimaryConstraint{^{ij}}
	,
	\label{DressedSuperHamiltonian}\\
	\DressedSuperConstraint{^i}
	&\equiv
	\SuperConstraint{^i}
	+\PrimaryConstraint{^{jk}}\CD{^i}\ExtrinsicCurvature{_{jk}}
	-2\CD{_k}\big(\PrimaryConstraint{^{jk}}\ExtrinsicCurvature{^i_j}\big).
	\label{DressedSuperMomentum}
\end{align}
\end{subequations}
Notice that~\cref{DressedSuperMomentum} simply restores the portion of~\cref{PreMomentumConstraint} that was lost in the shift to~\cref{MomentumConstraint}, revealing a false economy in eliminating~$\PrimaryConstraint{^{ij}}$ from~$\SuperConstraint{^i}$. A similar effect is seen in~\cref{DressedSuperHamiltonian,PreHamiltonianConstraint,HamiltonianConstraint}, though in this case the dressed~$\DressedSuperConstraint{}$ contains extra structures. In particular, the negative power of~$\ConjugateMomentumExtrinsicCurvature{}$ appears once again, as it did in~\cref{MultiplierSolution}. The dressed constraints in~\cref{DressedSuperHamiltonian,DressedSuperMomentum} have the following commutators with the secondary constraint:
\begin{subequations}
\begin{align}
	\bigl\{ \SecondaryConstraint{^{ij}}[\SmearingF{_{ij}}],
	\DressedSuperConstraint{}[\SmearingS{}]\bigr\}
	&=	\PrimaryConstraint{^{ij}}\bigg[
		2\SmearingS{}\Big(
			\SmearingF{^{kl}}
			-\frac{\SmearingF{}}{3}\MetricFoliation{^{kl}}
		\Big)\Big(
			\frac{1}{3}\ExtrinsicCurvature{_{kl}}\ExtrinsicCurvature{_{ij}}
			-\ExtrinsicCurvature{_{ik}}\ExtrinsicCurvature{_{jl}}
		\Big)
		-\frac{\ConjugateMomentumExtrinsicCurvature{}\SmearingS{}\SmearingF{_{ij}}}{48\sqrt{\MetricFoliation{}}}
	\nonumber\\ &\hspace{20pt}
		+\frac{1}{2\ConjugateMomentumExtrinsicCurvature{}}\CD{^k}\bigg(
			\ConjugateMomentumExtrinsicCurvature{}\SmearingS{}\Big(
				\CD{_k}\SmearingF{_{ij}}
				-2\CD{_i}\Big(
					\SmearingF{_{kj}}
					-\frac{\SmearingF{}}{3}\MetricFoliation{_{kj}}
				\Big)
			\Big)
		\bigg)
	\bigg]
	\nonumber\\ &\quad
	+\SecondaryConstraint{^{ij}}\bigg[
		\frac{2\SmearingS{}\SmearingF{}}{3}\ExtrinsicCurvature{_{ij}}
	\bigg]
	+\SuperConstraint{}\PrimaryConstraint{^{ij}}\bigg[
		\frac{\SmearingS{}}{\ConjugateMomentumExtrinsicCurvature{}}\SmearingF{_{ij}}
	\bigg]
	+\PrimaryConstraint{^{ij}}\SecondaryConstraint{^{kl}}\bigg[
		\frac{2\SmearingS{}}{\ConjugateMomentumExtrinsicCurvature{}}\Big(
		\ExtrinsicCurvature{_{ij}}\SmearingF{_{kl}}
		+\ExtrinsicCurvature{_{kl}}\SmearingF{_{ij}}
		\Big)
	\bigg]
	\nonumber\\ &\quad
	+\PrimaryConstraint{^{ij}}\PrimaryConstraint{^{kl}}\bigg[
		-\frac{\SmearingS{}}{\ConjugateMomentumExtrinsicCurvature{}^2}\SmearingF{_{ij}}\Big(
			2\ExtrinsicCurvature{_{kl}}\ConjugateMomentumG{}
			+\CD{_k}\CD{_l}\ConjugateMomentumExtrinsicCurvature{}
		\Big)
	\bigg]
	\nonumber\\ &\quad
	+\PrimaryConstraint{^{ij}}\bigg[
		\SmearingF{_{ij}}\CD{_k}\CD{_l}\Big(
			\frac{\SmearingS{}}{\ConjugateMomentumExtrinsicCurvature{}}\PrimaryConstraint{^{kl}}
		\Big)
	\bigg]
	\, ,
\label{SecondaryConstraintDressedSuperHamiltonian}\\
	\bigl\{ \SecondaryConstraint{^{ij}}[\SmearingF{_{ij}}],
	\DressedSuperConstraint{^k}[\SmearingS{_k}]\bigr\}
	&=	\SuperConstraint{^i}\bigg[
		\SmearingF{_i^j}\SmearingS{_j}
		-\frac{\SmearingF{}}{3}\SmearingS{_i}
	\bigg]
	+\PrimaryConstraint{^{ij}}\bigg[
		\Big(
			\SmearingF{_l^k}
			-\frac{\SmearingF{}}{3}\Kronecker{^k_l}
		\Big)\SmearingS{^l}\CD{_k}\ExtrinsicCurvature{_{ij}}
	\nonumber\\ &\hspace{20pt}
		+2\ExtrinsicCurvature{_i^l}\CD{_j}\left(\Big(
			\SmearingF{_l^k}
			-\frac{\SmearingF{}}{3}\Kronecker{^k_l}
		\Big)\SmearingS{_k}\right)
	\bigg]
	\nonumber\\ &\quad
	+\SecondaryConstraint{^{ij}}\bigg[
		-\SmearingS{^k}\CD{_k}\SmearingF{_{ij}}
		-2\SmearingF{_j^k}\CD{_i}\SmearingS{_k}
	\bigg]
	\, .
\label{SecondaryConstraintDressedSuperMomentum}
\end{align}
\end{subequations}
As required, the commutators with~$\SecondaryConstraint{^{ij}}$ in~\cref{SecondaryConstraintDressedSuperHamiltonian,SecondaryConstraintDressedSuperMomentum} vanish weakly. Moreover, the dressing corrections in~\cref{DressedSuperHamiltonian,DressedSuperMomentum} are all linear in~$\PrimaryConstraint{^{ij}}$, so that~$\DressedSuperConstraint{}$ and~$\DressedSuperConstraint{^i}$ preserve the property of~$\SuperConstraint{}$ and~$\SuperConstraint{^i}$ --- as seen in~\cref{PrimaryConstraintSuperHamiltonian,PrimaryConstraintSuperMomentum} --- of still weakly commuting with~$\PrimaryConstraint{^{ij}}$. Taken together, these observations support the hope that~$\DressedSuperConstraint{}$ and~$\DressedSuperConstraint{^i}$ may indeed be first class.

\paragraph*{The deformed Dirac algebra} This property is confirmed by the strong algebroid
\begin{subequations}
\begin{align}
	\bigl\{ \DressedSuperConstraint{}[\SmearingF{}],
	\DressedSuperConstraint{}[\SmearingS{}]\bigr\}
	&=	\DressedSuperConstraint{^{i}}\bigg[
		\SmearingF{}\CD{_i}\SmearingS{}
		-\SmearingS{}\CD{_i}\SmearingF{}
	\bigg]
	+\DressedSuperConstraint{^k}\PrimaryConstraint{^i_k}\bigg[
		\frac{6}{\ConjugateMomentumExtrinsicCurvature{}}\Big(
			\SmearingF{}\CD{_i}\SmearingS{}
			-\SmearingS{}\CD{_i}\SmearingF{}
		\Big)
	\bigg]
	\nonumber\\ &\quad
	+\PrimaryConstraint{^{ij}}\bigg[
		\frac{12}{\ConjugateMomentumExtrinsicCurvature{}}\CD{_k}\SecondaryConstraint{^k_j}\Big(
			\SmearingF{}\CD{_i}\SmearingS{}
			-\SmearingS{}\CD{_i}\SmearingF{}
		\Big)
	\bigg]
	\nonumber\\ &\quad
	+\PrimaryConstraint{^{ij}}\PrimaryConstraint{^{kl}}\bigg[
		\frac{6}{\ConjugateMomentumExtrinsicCurvature{}}\Big(
			\CD{_j}\ExtrinsicCurvature{_{kl}}
			+\frac{1}{\ConjugateMomentumExtrinsicCurvature{}}\ExtrinsicCurvature{_{kl}}\CD{_j}\ConjugateMomentumExtrinsicCurvature{}
		\Big)\Big(
			\SmearingS{}\CD{_i}\SmearingF{}
			-\SmearingF{}\CD{_i}\SmearingS{}
		\Big)
	\bigg]
	\nonumber\\ &\quad
	+\PrimaryConstraint{^{ij}}\bigg[
		\frac{2}{\ConjugateMomentumExtrinsicCurvature{}}\Big(
			6\CD{_l}\big(\ExtrinsicCurvature{_{jk}}\PrimaryConstraint{^{kl}}\big)
			+\CD{_j}\big(\ExtrinsicCurvature{_{kl}}\PrimaryConstraint{^{kl}}\big)
		\Big)\Big(
			\SmearingF{}\CD{_i}\SmearingS{}
			-\SmearingS{}\CD{_i}\SmearingF{}
		\Big)
	\bigg]
	\, ,
\label{DressedSuperHamiltonianDressedSuperHamiltonian}
\\
	\bigl\{ \DressedSuperConstraint{}[\SmearingF{}],
	\DressedSuperConstraint{^i}[\SmearingS{_i}]\bigr\}
	&=	\DressedSuperConstraint{}\bigg[
		-\SmearingS{^i}\CD{_i}\SmearingF{}
	\bigg]
	+\DressedSuperConstraint{^i}\bigg[
		-2\SmearingF{}\ExtrinsicCurvature{_i^j}\SmearingS{_j}
	\bigg]
	+\DressedSuperConstraint{^i}\PrimaryConstraint{^{jk}}\bigg[
		-\frac{2\SmearingF{}}{\ConjugateMomentumExtrinsicCurvature{}}\SmearingS{_i}\ExtrinsicCurvature{_{jk}}
	\bigg]
	\, ,
\label{DressedSuperHamiltonianDressedSuperMomentum}
\\
	\bigl\{ \DressedSuperConstraint{^i}[\SmearingF{_i}],
	\DressedSuperConstraint{^j}[\SmearingS{_j}]\bigr\}
	&=	\DressedSuperConstraint{^{i}}\bigg[
		\SmearingS{^j}\CD{_i}\SmearingF{_j}
		-\SmearingF{^j}\CD{_i}\SmearingS{_j}
	\bigg]
	\, ,
\label{DressedSuperMomentumDressedSuperMomentum}
\end{align}
\end{subequations}
which vanishes weakly. At the strong level, although the commutators in~\cref{DressedSuperHamiltonianDressedSuperHamiltonian,DressedSuperHamiltonianDressedSuperMomentum,DressedSuperMomentumDressedSuperMomentum} are between first class constraints, they do not form a \emph{closed} algebroid, as is the case in GR.\footnote{See also~\cite{Alexandrov:2021qry} for another example of closure being spoiled by dressing of the Hamiltonian constraint.} Since, however, they are strongly equal to expressions which are linear in first class constraints and quadratic in constraints of either class, they do themselves have the expected property of being first class, in contrast to~\cref{SuperHamiltonianSuperHamiltonian,SuperHamiltonianSuperMomentum,SuperMomentumSuperMomentum}. In particular, the latter two terms in~\cref{DressedSuperHamiltonianDressedSuperMomentum} are superfluous, and together with the format in~\cref{DressedSuperMomentumDressedSuperMomentum} are a result of our choice of convention in which the contravariant components~$\DressedSuperConstraint{^i}$ are used in the basis. It is the covariant components~$\DressedSuperConstraint{_i}$ which carry physical significance as the generators of transverse diffeomorphisms, and indeed~\cref{DressedSuperHamiltonianDressedSuperMomentum,DressedSuperMomentumDressedSuperMomentum} can be readily replaced by
\begin{equation}\label{DiracAlgebra}
	\bigl\{ \DressedSuperConstraint{}[\SmearingF{}],
	\DressedSuperConstraint{_i}[\SmearingS{^i}]\bigr\}
	=\DressedSuperConstraint{}\bigg[
		-\SmearingS{^i}\CD{_i}\SmearingF{}
	\bigg],
	\quad
	\bigl\{ \DressedSuperConstraint{_i}[\SmearingF{^i}],
	\DressedSuperConstraint{_j}[\SmearingS{^j}]\bigr\}
	=\DressedSuperConstraint{_{i}}\bigg[
		\SmearingF{^j}\CD{_j}\SmearingS{^i}
		-\SmearingS{^j}\CD{_j}\SmearingF{^i}
	\bigg],
\end{equation}
where~\cref{DiracAlgebra} represents the part of the standard Dirac hypersurface deformation algebroid given already in~\cref{SuperHamiltonianSuperMomentumBracket,SuperMomentumAutocommutatorCovariantIndices}. By contrast, the structural difference between~\cref{DressedSuperHamiltonianDressedSuperHamiltonian} and~\cref{SuperHamiltonianAutocommutator} is more significant in nature, and distinguishes the~$R^2$ model from pure GR.

\paragraph*{Final summary} To collect our results together, we find that the full counting of independent components among the first class~$\DressedSuperConstraint{}$ and~$\DressedSuperConstraint{^i}$ is~$\NFirst{}\!=\!4$. The full counting of second class components among~$\PrimaryConstraint{^{ij}}$ and~$\SecondaryConstraint{^{ij}}$ is~$\NSecond{} \!=\! 10$, where we recall that each of these tensorial constraints is trace-free and symmetric. Compared to the case of GR, we have \emph{twice} as many phase-space variables distributed between the spatial metric~$\MetricFoliation{_{ij}}$ and the conjugate momentum~$\ConjugateMomentumG{^{ij}}$, and also the auxiliary extrinsic curvature~$\ExtrinsicCurvature{_{ij}}$ and its conjugate momentum~$\ConjugateMomentumExtrinsicCurvature{^{ij}}$ --- a full counting of~$\NCan{} \!=\! 24$. The final number of propagating modes is found from~\cref{Nphy} to be
\begin{equation}\label{PhysicalDegreesOfFreedomR2}
	\NPhy{} = \frac{1}{2}
	\Bigl( 24 - 2 \!\times\! 4 - 10 \Bigr)
	=
	3
	\, .
\end{equation}
These are accounted for by the two tensorial polarizations of the graviton, plus an additional scalar mode arising from the higher-derivative nature of the theory.

\subsection{Case study: pure GR at two loops}\label{PureGRTwoLoops}

\begin{figure}[h]
\centering
\includegraphics[width=\textwidth]{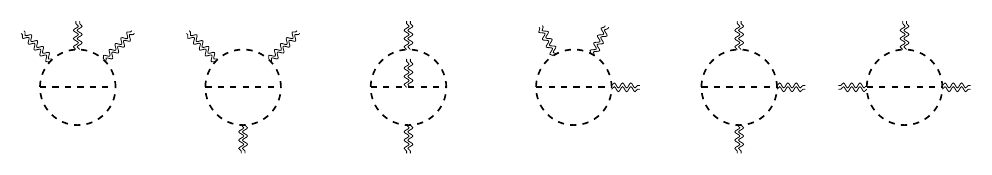}
\caption{Essential diagrams contributing to the two-loop divergence of pure GR in the background field approach, once the internal lines are carefully treated~\cite{Goroff:1985sz}.}
\label{fig:TwoLoopDiagrams}
\end{figure}

\paragraph*{Motivation} The final gravity theory we consider is an extension of~\cref{GRaction} with a cosmological constant and a cubic Riemann correction term, defined by the action
\begin{equation}
	\Action{}[\G{_{\mu\nu}}] = \int\! \mathrm{d}^4x \, \sqrt{-\G{}} \, \biggl[
		\frac{\Curvature{} - 2 \Lambda}{\kappa^2}
		+ \alpha \kappa^2 \Curvature{^{\mu\nu}_{\rho\sigma}}
			\Curvature{^{\rho\sigma}_{\alpha\beta}} \Curvature{^{\alpha\beta}_{\mu\nu}}
		\biggr]
		\, ,
	\label{GRTwoLoopsAction}
\end{equation}
where~$\Lambda$ is the cosmological constant and~$\alpha$ is a small coupling parameter. The specific cubic Riemann term considered in~\cref{GRTwoLoopsAction} arises as the leading quantum correction at two-loop order in perturbative quantum gravity~\cite{Goroff:1985sz,Goroff:1985th,vandeVen:1991gw}. Working from~\cref{GRaction}, the Gauss--Bonnet identity and field equation~$\Curvature{_{\mu\nu}}\approx 0$ together eliminate all possible one-loop divergences in the~$S$ matrix, and the correlators may be rendered finite by a field reparameterisation~\cite{tHooft:1974toh}. At two loops, however, there are insufficiently many such identities. It was shown in~\cite{Goroff:1985sz,Goroff:1985th} that the divergent part of the effective action at two loops may be computed via six contributions in the three-point correlator, as illustrated in~\cref{fig:TwoLoopDiagrams}. These divergences do not cancel, and cannot be removed by field redefinitions: upon further computation they can be seen to necessitate the cubic Riemann counterterm in~\cref{GRTwoLoopsAction}. This result was confirmed in~\cite{vandeVen:1991gw}. We did not include $\Lambda$ in~\cref{GRaction}, and without detailed calculation it might be expected that such an addition will lead to fewer cancellations and more curvature counterterms, even at the one-loop level~\cite{Christensen:1979iy,Goroff:1985th,Bastianelli:2023oca}. In fact, the only result is that $\Lambda$ itself is renormalised~\cite{Christensen:1979iy,Percacci:2017fkn}.

\paragraph{Need for order reduction} As a truncated effective field theory, the action in~\cref{GRTwoLoopsAction} is largely unsuitable for direct phenomenological study. This aspect has perhaps been addressed most thoroughly in recent work~\cite{Glavan:2024cfs}, though it comes four decades after the original formulation by Goroff and Sagnotti. Being higher-order in derivatives,~\cref{GRTwoLoopsAction} is expected to introduce additional degrees of freedom which (with or without $\Lambda$) do not smoothly connect to the particle spectrum of~\cref{GRaction}. A general-purpose order-reduction algorithm --- acting directly on the phase-space action --- was laid out by Glavan in~\cite{Glavan:2017srd}, based on earlier work in~\cite{Jaen:1986iz,Eliezer:1989cr}. When applied to~\cref{GRTwoLoopsAction} in~\cite{Glavan:2024cfs}, this method was found to require unusually lengthy manipulations. We take this \emph{tour de force} as a reasonable upper bound on the complexity of canonical operations in gravity, which therefore provides a natural opportunity to stress-test the \lstinline!PoissonBracket! and \lstinline!FindAlgebra! functionalities of \Hamilcar{}. In \Hamilcar{}, the cosmological constant is defined as:
\lstinputoutput{code_listings_ReproductionOfResults-DefineCosmologicalConstant.tex}
and the Wilson coefficient~$\alpha$ as:
\lstinputoutput{code_listings_ReproductionOfResults-DefineWilsonCoefficient.tex}
The gravitational coupling~$\kappa$ was already defined in~\cref{PureGR} via~\lstinputcref{code_listings_Hamilcar-DefineGravitationalCoupling}, so we do not redefine it here.

\subsubsection{Phase-space formulation}

\paragraph*{Shorthand notation} As shown in~\cite{Glavan:2024cfs}, the readability of the computations is improved by setting up definitions for three compact tensors, namely
\begin{equation}\label{KLQdefinitions}
\ShorthandK{_{ij}} \equiv
    -
    \frac{\kappa^2}{\sqrt{\MetricFoliation{}}} \biggl(
		\ConjugateMomentumG{_{ij}}
		- \frac{\ConjugateMomentumG{}}{2} \MetricFoliation{_{ij}}
		\biggr)
  \, ,
\quad
\ShorthandL{^i_{jk}}
	\equiv  2 \CD{_{[k}} \ShorthandK{_{j]}}^i
  \, ,
\quad
\ShorthandQ{^{ij}_{kl}}
	\equiv
	2\ShorthandK{^i_{[k}} \ShorthandK{^j_{l]}}
	+
	\R{^{ij}_{kl}}
	\, .
\end{equation}
Their traces are~$\ShorthandK{} \equiv \ShorthandK{^i_i}$,~$\ShorthandL{_i} \equiv \ShorthandL{^j_{ji}}$,~$\TraceShorthandQ{^i_j} \equiv \ShorthandQ{^{ik}_{jk}}$, and~$\TraceShorthandQ{} \equiv \TraceShorthandQ{^i_i}$.
The corresponding definitions in \Hamilcar{} are as follows. First, the trace-shifted momentum~$\ShorthandK{_{ij}}$ is defined as:
\lstinputoutput{code_listings_ReproductionOfResults-DefineScriptK.tex}
Next, the antisymmetric gradient~$\ShorthandL{^i_{jk}}$ is defined as:
\lstinputoutput{code_listings_ReproductionOfResults-DefineScriptL.tex}
Finally, the Riemann-like quantity~$\ShorthandQ{^{ij}_{kl}}$ is defined as:
\lstinputoutput{code_listings_ReproductionOfResults-DefineScriptQ.tex}
The corresponding traces are defined as follows. The scalar trace~$\ShorthandK{}$ is:
\lstinputoutput{code_listings_ReproductionOfResults-DefineTraceScriptK.tex}
The vector trace~$\ShorthandL{_i}$ is:
\lstinputoutput{code_listings_ReproductionOfResults-DefineScriptLContraction.tex}
The two-index trace~$\TraceShorthandQ{^i_j}$ is:
\lstinputoutput{code_listings_ReproductionOfResults-DefineScriptQSingleContraction.tex}
The full scalar trace~$\TraceShorthandQ{}$ is:
\lstinputoutput{code_listings_ReproductionOfResults-DefineTraceScriptQ.tex}

\paragraph*{Reduced phase-space action} The \emph{reduced} phase-space action was found in~\cite{Glavan:2024cfs}. The procedures used to obtain it can be implemented in \Hamilcar{}, but are not instructive for the purposes of this paper, so we simply state the final result as
\begin{align}
\MoveEqLeft[1.5]
\Action \bigl[ \Lapse, \Shift{_i}, \MetricFoliation{_{ij}}, \ConjugateMomentumG{^{ij}} \bigr]
    =\int\! d^{4\!} x \,
    \left(
    \ConjugateMomentumG{^{ij}}
    \dot{\MetricFoliation{}}_{ij}
	-
    \Lapse \ReducedSuperConstraint{}
    -
    \Shift{_i} \ReducedSuperConstraint{^i}
    \right)
\, ,
\label{ReducedCanonicalAction}
\end{align}
where the reduced analogues of the super-Hamiltonian and super-momentum are found to be modified to
\begin{subequations}
\begin{align}
\ReducedSuperConstraint{} ={}&
    \sqrt{\MetricFoliation{}} \biggl[
		- \frac{ ( \TraceShorthandQ{} - 2 \Lambda ) }{\kappa^2}
	+ 8\alpha \kappa^2 \TraceShorthandQ{^i_j} \Bigl(
        2 \TraceShorthandQ{^j_k} \TraceShorthandQ{^k_i}
        - 3 \ShorthandL{^j_{kl}} \ShorthandL{_i^{kl}}
        \Bigr)
\nonumber \\
&   \hspace{4cm}
	- 24\alpha \kappa^2 \Lambda \Bigl(
	    2 \TraceShorthandQ{^i_j} \TraceShorthandQ{^j_i}
	    - 2 \Lambda^2
        - \ShorthandL{_k^{ij}} \ShorthandL{^k_{ij}}
        \Bigr)
	\biggr]
    \, ,
\label{ReducedSuperHamiltonian}
\\
\ReducedSuperConstraint{^i} ={}&
    \sqrt{\MetricFoliation{}}
        \biggl[ - \frac{2 \ShorthandL{^i}}{\kappa^2} \biggr]
    \, .
\label{ReducedSuperMomentum}
\end{align}
\end{subequations}
It is an essential point that~\cref{ReducedCanonicalAction} has precisely the same structure as the phase-space action of GR given in~\cref{CanonicalActionGRFinal2}. The symplectic part is identical, and only the definitions of the super-Hamiltonian and super-momentum in~\cref{ReducedSuperHamiltonian,ReducedSuperMomentum} differ from their GR counterparts in~\cref{GRconstraints}. By contrast, the phase-space action of~$\Curvature{}^2$ theory in~\cref{CanonicalAction} has a richer symplectic structure and different constraints: the idea is that the higher derivative order in this latter theory is faithfully reflected in the canonical analysis, whereas the two-loop cubic Riemann theory in~\cref{GRTwoLoopsAction} must pass through a process of order reduction to produce~\cref{ReducedCanonicalAction}. The corresponding \Hamilcar{} definitions are as follows. The reduced super-Hamiltonian~$\ReducedSuperConstraint{}$ is defined as:
\lstinputoutput{code_listings_ReproductionOfResults-DefineReducedHamiltonianConstraint.tex}
The reduced super-momentum~$\ReducedSuperConstraint{^i}$ is defined as:
\lstinputoutput{code_listings_ReproductionOfResults-DefineReducedMomentumConstraint.tex}

\subsubsection{Dirac--Bergmann algorithm}

\paragraph*{Total Hamiltonian} The variation of~\cref{ReducedCanonicalAction} with respect to the two functions~$\Lapse{}$ and~$\Shift{_i}$ once again reveals the constrained nature of the reduced super-Hamiltonian and super-momentum
\begin{equation}
	\ReducedSuperConstraint{} \approx 0 \, ,
	\qquad
	\ReducedSuperConstraint{^i} \approx 0 \, ,
	\label{ReducedPrimaryConstraints}
\end{equation}
which is analogous to~\cref{GRprimaryconstraints}, meanwhile the usual Legendre transformation is apparent in the simple format of~\cref{ReducedCanonicalAction}, leading to the total Hamiltonian
\begin{equation}
	\TotalHamiltonian = \int\! \mathrm{d}^{3} x \, \left( \Lapse{} \ReducedSuperConstraint{} + \Shift{_i} \ReducedSuperConstraint{^i} \right) \, ,
	\label{TotalHamiltonianReduced}
\end{equation}
which is analogous to~\cref{TotalHamiltonianGR}.

\paragraph*{Expansion in the Wilson coefficient} To facilitate the constraint algebra analysis, we decompose the reduced Hamiltonian constraint into zeroth-order and first-order parts in the Wilson coefficient~$\alpha$, denoted as~$\ReducedSuperConstraint{} = \ZerothOrderReducedSuperConstraint{} + \FirstOrderReducedSuperConstraint{}$, where
\begin{subequations}
\begin{align}
\ZerothOrderReducedSuperConstraint{} \equiv{}&
	- \frac{\sqrt{\MetricFoliation{}}}{\kappa^2} \left(
		\TraceShorthandQ{} - 2 \Lambda
	\right)
	\, ,
\label{ZerothOrderReducedSuperHamiltonian}
\\
\FirstOrderReducedSuperConstraint{} \equiv{}&
	8\alpha \kappa^2 \sqrt{\MetricFoliation{}} \biggl[
		\TraceShorthandQ{^i_j} \Bigl(
			2 \TraceShorthandQ{^j_k} \TraceShorthandQ{^k_i}
			-
			3 \ShorthandL{^j_{kl}} \ShorthandL{_i^{kl}}
		\Bigr)
		-
		3 \Lambda \Bigl(
			2 \TraceShorthandQ{^i_j} \TraceShorthandQ{^j_i}
			-
			2 \Lambda^2
			-
			\ShorthandL{_{ijk}} \ShorthandL{^{ijk}}
		\Bigr)
	\biggr]
	\, .
\label{FirstOrderReducedSuperHamiltonian}
\end{align}
\end{subequations}
The corresponding \Hamilcar{} definition for the zeroth-order part~$\ZerothOrderReducedSuperConstraint{}$ is:
\lstinputoutput{code_listings_ReproductionOfResults-DefineReducedHamiltonianOrderUnity.tex}
The first-order part~$\FirstOrderReducedSuperConstraint{}$ is defined as:
\lstinputoutput{code_listings_ReproductionOfResults-DefineReducedHamiltonianOrderWilsonCoefficient.tex}
The only interesting part of the constraint algebra is the auto-commutator of the reduced super-Hamiltonian. To analyze it, we compute the Poisson bracket as a sum of three contributions, to first order in the Wilson coefficient~$\alpha$, namely
\begin{equation}
\left\{ \ReducedSuperConstraint{}\left[\SmearingF{}\right], \ReducedSuperConstraint{}\left[\SmearingS{}\right] \right\}
=
\left\{ \ZerothOrderReducedSuperConstraint{}\left[\SmearingF{}\right], \ZerothOrderReducedSuperConstraint{}\left[\SmearingS{}\right] \right\}
+
\left\{ \ZerothOrderReducedSuperConstraint{}\left[\SmearingF{}\right], \FirstOrderReducedSuperConstraint{}\left[\SmearingS{}\right] \right\}
+
\left\{ \FirstOrderReducedSuperConstraint{}\left[\SmearingF{}\right], \ZerothOrderReducedSuperConstraint{}\left[\SmearingS{}\right] \right\}
+
\mathcal{O}(\alpha^2)
\, .
\label{ReducedSuperHamiltonianAutocommutatorDecomposition}
\end{equation}
The first term in~\cref{ReducedSuperHamiltonianAutocommutatorDecomposition} is the auto-commutator of the zeroth-order part, while the second and third terms are the cross-brackets between zeroth and first-order parts. The term~$\left\{ \FirstOrderReducedSuperConstraint{}\left[\SmearingF{}\right], \FirstOrderReducedSuperConstraint{}\left[\SmearingS{}\right] \right\}$ is of order~$\alpha^2$, and is therefore safely dropped in this analysis. The computation progresses by first setting up each bracket with the appropriate smearing functions, then invoking the \lstinline!PoissonBracket! function from \Hamilcar{}. For the leading contribution in~\cref{ReducedSuperHamiltonianAutocommutatorDecomposition}, we set up the bracket:
\lstinputoutput{code_listings_ReproductionOfResults-SetupOrderUnityBracket.tex}
We then compute the Poisson bracket from~\lstinputcref{code_listings_ReproductionOfResults-SetupOrderUnityBracket} as follows:
\lstinputoutput[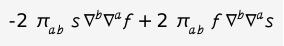]{code_listings_ReproductionOfResults-ComputeOrderUnityBracket.tex}
For the first cross-term in~\cref{ReducedSuperHamiltonianAutocommutatorDecomposition}, we set up the bracket:
\lstinputoutput{code_listings_ReproductionOfResults-SetupCrossBracket01.tex}
We then compute the bracket from~\lstinputcref{code_listings_ReproductionOfResults-SetupCrossBracket01} as follows:
\lstinputoutput[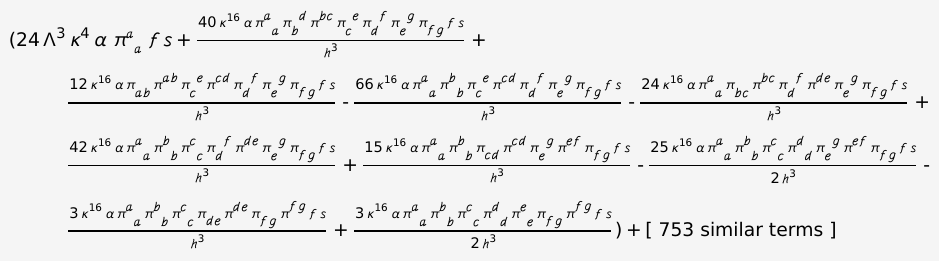]{code_listings_ReproductionOfResults-ComputeCrossBracket01.tex}
For the second cross-term in~\cref{ReducedSuperHamiltonianAutocommutatorDecomposition}, we set up the bracket:
\lstinputoutput{code_listings_ReproductionOfResults-SetupCrossBracket10.tex}
We then compute the bracket from~\lstinputcref{code_listings_ReproductionOfResults-SetupCrossBracket10} as follows:
\lstinputoutput[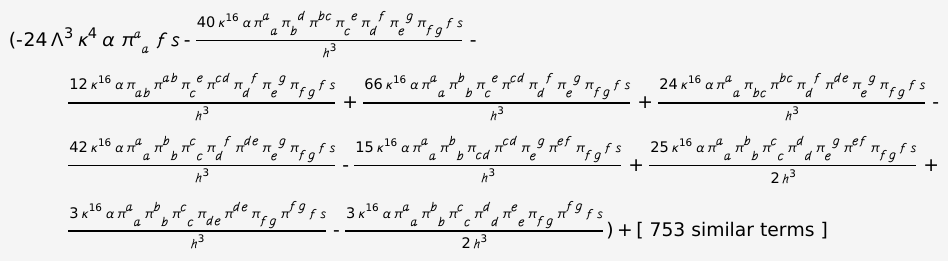]{code_listings_ReproductionOfResults-ComputeCrossBracket10.tex}
The outputs from~\lstinputcref{code_listings_ReproductionOfResults-ComputeCrossBracket01} and~\lstinputcref{code_listings_ReproductionOfResults-ComputeCrossBracket10} (shown truncated in~\lstoutputcref{code_listings_ReproductionOfResults-ComputeCrossBracket01} and~\lstoutputcref{code_listings_ReproductionOfResults-ComputeCrossBracket10}) are too lengthy to display in full.
Before combining the brackets, we define a utility function that will be used to simplify expressions to a canonical form, via a standard sequence of \xAct routines:
\lstinputoutput{code_listings_ReproductionOfResults-DefineStandardSimplify.tex}
With the aid of~\lstinputcref{code_listings_ReproductionOfResults-DefineStandardSimplify}, we combine all three brackets to obtain the total auto-commutator:
\lstinputoutput{code_listings_ReproductionOfResults-CombineBrackets.tex}
The result contains terms at various orders in both the cosmological constant~$\Lambda$ and the Wilson coefficient~$\alpha$. To analyze different parts of the constraint algebra systematically, we define another utility function that extracts specific orders from the total bracket using series expansions:
\lstinputoutput{code_listings_ReproductionOfResults-DefineExtractBracketAnatomy.tex}
The function in~\lstinputcref{code_listings_ReproductionOfResults-DefineExtractBracketAnatomy} allows us to isolate, for example, the zeroth-order contribution in~$\Lambda$ (which itself contains both zeroth and first-order terms in~$\alpha$), or the first-order contribution in~$\Lambda$, enabling systematic verification against the results of~\cite{Glavan:2024cfs}.

\paragraph*{Zeroth-order in~$\Lambda$}
The most complicated part of the total bracket is the contribution at zeroth-order in the cosmological constant~$\Lambda$, which contains both zeroth and first-order terms in the Wilson coefficient~$\alpha$. We extract this contribution using the utility function defined above:
\lstinputoutput{code_listings_ReproductionOfResults-ExtractZerothOrderBracket.tex}
The raw extracted bracket from~\lstinputcref{code_listings_ReproductionOfResults-ExtractZerothOrderBracket} is extremely lengthy and difficult to interpret directly; we do not display its output here. To make progress, we apply the \lstinline!FindAlgebra! function from \Hamilcar{} iteratively, each time specifying a different ansatz structure that we want the bracket to match.\footnote{It is not necessary that we do this, and we show all the possible configurations to indicate the power and flexibility of \lstinline!FindAlgebra!. One could skip ahead to the most compact form given in~\lstoutputcref{code_listings_ReproductionOfResults-FindAlgebraZerothOrderD}.} The \lstinline!FindAlgebra! function attempts to re-express the bracket through a combination of integration by parts (modifying boundary terms) and applications of Cayley--Hamilton-like identities in three dimensions. These algebraic manipulations are performed automatically; the user need only specify the desired structural form. We proceed through four progressively refined representations. First, we express the bracket in terms of first-order gradients of the shorthand tensorial momentum~$\ShorthandK{^i_j}$:
\lstinputoutput[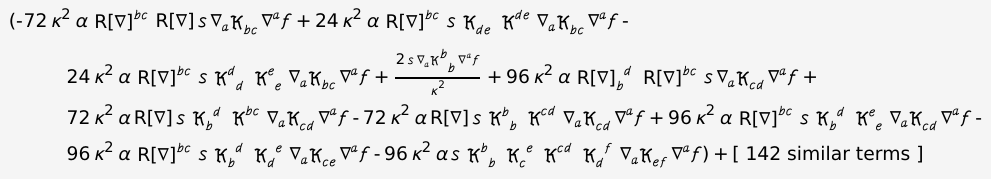]{code_listings_ReproductionOfResults-FindAlgebraZerothOrderA.tex}
While the output from~\lstinputcref{code_listings_ReproductionOfResults-FindAlgebraZerothOrderA} (shown truncated in~\lstoutputcref{code_listings_ReproductionOfResults-FindAlgebraZerothOrderA}) confirms that the bracket can be written using only first derivatives (rather than higher-order gradients), the result remains too complex for direct interpretation. Next, we express the bracket in terms of the shorthand momentum gradient tensor~$\ShorthandL{^i_{jk}}$ and the spatial Ricci curvature~$\R{_{ij}}$:
\lstinputoutput[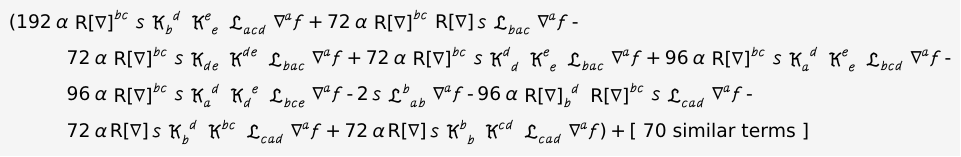]{code_listings_ReproductionOfResults-FindAlgebraZerothOrderB.tex}
The output from~\lstinputcref{code_listings_ReproductionOfResults-FindAlgebraZerothOrderB} (shown truncated in~\lstoutputcref{code_listings_ReproductionOfResults-FindAlgebraZerothOrderB}) is more compact, as it hides explicit gradients acting on non-smearing quantities, but further simplification is still possible. In the third refinement, we express the bracket entirely in terms of the shorthand functions~$\ShorthandL{^i_{jk}}$ and~$\ShorthandQ{^{ij}_{kl}}$:
\lstinputoutput[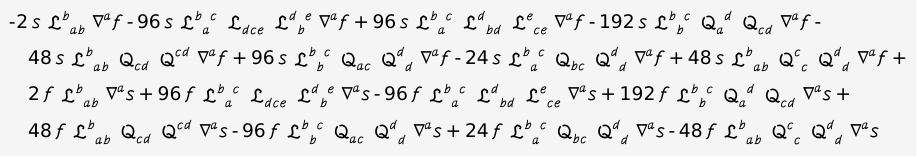]{code_listings_ReproductionOfResults-FindAlgebraZerothOrderC.tex}
The result~\lstoutputcref{code_listings_ReproductionOfResults-FindAlgebraZerothOrderC} provides a more geometric representation, though it still does not directly reveal the constraint algebra structure. Finally, we express the bracket in terms of the reduced momentum constraint~$\ReducedSuperConstraint{_i}$ and the zeroth-order reduced Hamiltonian constraint~$\ZerothOrderReducedSuperConstraint{}$. Since~$\ZerothOrderReducedSuperConstraint{}$ itself contains a term linear in~$\Lambda$, and we are working at zeroth-order in~$\Lambda$, we must wrap this computation in a \lstinline!Block! command that temporarily sets~$\Lambda = 0$. This prevents \lstinline!FindAlgebra! from attempting to expand terms that cannot appear at this order. Additionally, we use the \lstinline!Constraints! option to instruct \lstinline!FindAlgebra! to factor the result in terms of these constraint functions:
\lstinputoutput[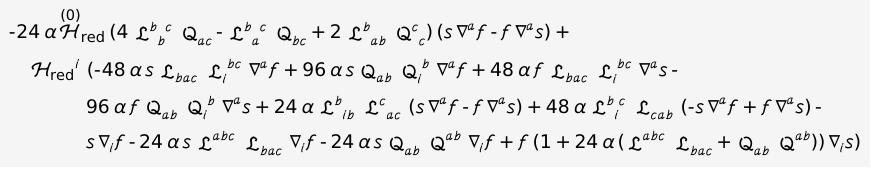]{code_listings_ReproductionOfResults-FindAlgebraZerothOrderD.tex}
The result~\lstoutputcref{code_listings_ReproductionOfResults-FindAlgebraZerothOrderD} reveals the structure of the constraint algebra at zeroth-order in~$\Lambda$ and facilitates direct comparison with~\cite{Glavan:2024cfs}.
\paragraph*{First-order in~$\Lambda$}
We next analyze the contribution at first-order in the cosmological constant~$\Lambda$. We extract this contribution using the same utility function from~\lstinputcref{code_listings_ReproductionOfResults-DefineExtractBracketAnatomy}:
\lstinputoutput{code_listings_ReproductionOfResults-ExtractFirstOrderBracket.tex}
The raw extracted bracket from~\lstinputcref{code_listings_ReproductionOfResults-ExtractFirstOrderBracket} is again too lengthy to display. Unlike the zeroth-order case, the first-order contribution cannot be cleanly expressed in terms of the reduced constraints alone, because the reduced Hamiltonian constraint itself contains both zeroth-order and first-order parts in~$\Lambda$. Instead, we express the bracket directly in terms of the shorthand variables~$\ShorthandL{^i_{jk}}$ and~$\ShorthandQ{^{ij}_{kl}}$, along with the spatial curvature:
\lstinputoutput[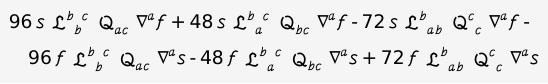]{code_listings_ReproductionOfResults-FindAlgebraFirstOrder.tex}
The result~\lstoutputcref{code_listings_ReproductionOfResults-FindAlgebraFirstOrder} provides the first-order contribution in a form suitable for comparison with~\cite{Glavan:2024cfs}.
\paragraph*{Second-order in~$\Lambda$}
Finally, we analyze the contribution at second-order in the cosmological constant~$\Lambda$. We extract this contribution with~\lstinputcref{code_listings_ReproductionOfResults-DefineExtractBracketAnatomy}:
\lstinputoutput[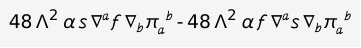]{code_listings_ReproductionOfResults-ExtractSecondOrderBracket.tex}
The extracted bracket~\lstoutputcref{code_listings_ReproductionOfResults-ExtractSecondOrderBracket} is remarkably simple compared to the lower-order contributions. With a single application of \lstinline!FindAlgebra!, we express it in terms of the reduced momentum constraint and the zeroth-order reduced Hamiltonian constraint, using the \lstinline!Constraints! option to factor the result:
\lstinputoutput[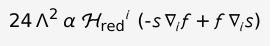]{code_listings_ReproductionOfResults-FindAlgebraSecondOrder.tex}
The result~\lstoutputcref{code_listings_ReproductionOfResults-FindAlgebraSecondOrder} consists of a single term proportional to the reduced momentum constraint, demonstrating the simplified structure at second-order in~$\Lambda$.

\paragraph*{Reduced super-Hamiltonian auto-commutator}
Having computed the constraint algebra at each order in~$\Lambda$, we now assemble the complete auto-commutator of the reduced super-Hamiltonian constraint to first order in~$\alpha$ by combining~\lstoutputcref{code_listings_ReproductionOfResults-FindAlgebraZerothOrderD},~\lstoutputcref{code_listings_ReproductionOfResults-FindAlgebraFirstOrder}, and~\lstoutputcref{code_listings_ReproductionOfResults-FindAlgebraSecondOrder}. According to~\cite{Glavan:2024cfs}, this should take the form:
\begin{align}
\bigl\{ \ReducedSuperConstraint{} [ \SmearingF{} ] , \ReducedSuperConstraint{} [\SmearingS{}] \bigr\}
	={}&
	\ReducedSuperConstraint{_i} \Bigr[ 
	\SmearingF{} \CD{^i} \SmearingS{} - \SmearingS{} \CD{^i} \SmearingF{} 
	+
	24 \alpha \kappa^4
	\ShorthandL{_k^{li}} \ShorthandL{_l^{kj}}
	\bigl( \SmearingF{}\CD{_j} \SmearingS{} - \SmearingS{} \CD{_j} \SmearingF{} \bigr)
	\nonumber \\
	&-
	24 \alpha \kappa^4
	\bigl(
	4 \ShorthandQ{^i_j} \ShorthandQ{^j_k}
	-
	2 ( \TraceShorthandQ{} + \Lambda ) \ShorthandQ{^{i}_k}
	+
	\ShorthandL{^i_{jk}} \ShorthandL{^j}
	\bigr)
	\bigl( \SmearingF{}\CD{^k} \SmearingS{} - \SmearingS{} \CD{^k} \SmearingF{} \bigr)
	\nonumber \\
	&+
	24 \alpha \kappa^4
	\bigl(
	2 \ShorthandQ{^j_{[k}} \ShorthandQ{^k_{j]}}
	+
	2 \Lambda \TraceShorthandQ{}
	+
	\ShorthandL{^j_{kl}} \ShorthandL{_j^{kl}}
	-
	\ShorthandL{^j} \ShorthandL{_j}
	\bigr)
	\bigl( \SmearingF{}\CD{^i} \SmearingS{} - \SmearingS{} \CD{^i} \SmearingF{} \bigr)
	\Bigr]
	\nonumber \\
	&+
	24 \alpha \kappa^4
	\ReducedSuperConstraint{}
	\Bigl[
	\bigl( \ShorthandQ{^i_j} \ShorthandL{^j_{ik}}
	- \Lambda \ShorthandL{_k} \bigr)
	(\SmearingF{} \CD{^{k}} \SmearingS{} - \SmearingS{} \CD{^{k}} \SmearingF{} )
	\Bigr]
	+
	\mathcal{O}(\alpha^2)
	\, . \label{TargetBracket}
\end{align}
It is easy to confirm using~\cref{ReducedSuperHamiltonian,ReducedSuperMomentum,ZerothOrderReducedSuperHamiltonian,FirstOrderReducedSuperHamiltonian} that our computed results match~\cref{TargetBracket} perfectly, up to additional DDIs that may themselves be resolved by \lstinline!FindAlgebra!. The deformed algebra results from the difference between the constraints in~\cref{ReducedSuperHamiltonian,ReducedSuperMomentum} and those of pure GR in~\cref{GRconstraints}. It was shown in~\cite{Glavan:2024cfs} that this difference is an essential property of the order-reduction of~\cref{GRTwoLoopsAction}, resulting in a \emph{minimally modified} theory of gravity~\cite{Lin:2017oow,Mukohyama:2019unx,Aoki:2018brq}. This concludes our discussion of the \Hamilcar{} suite of tools. 

\section{Conclusions}\label{Conclusions}

In this paper we have presented \Hamilcar{}, a \WolframLanguage package for computing and simplifying Poisson brackets in canonical field theory. The package was shown to be effective against highly non-trivial operations in the canonical formulation of gravity. This includes the automated reconstruction of closed-form constraint algebroids through integration-by-parts manipulations, and dimensionally-dependent identities such as the Cayley--Hamilton theorem. Many of the techniques used by \Hamilcar{} have been available for some years, though an effective implementation was lacking. It is expected that \Hamilcar will be useful for the order-reduction (via the algorithm in~\cite{Jaen:1986iz,Eliezer:1989cr,Glavan:2017srd}) of those modified gravity theories, and theories beyond the standard model of particle physics, which present as truncated effective theories. The expected outputs in such cases are quantum-corrected Hamiltonia which are suitable for phenomenological study. Current limitations of the package include the lack of support for fermionic fields, and for anonymous scalar-valued functions of scalars. The latter in particular are already part of the \xAct{} ecosystem, and could be integrated into \Hamilcar{} in future work.

\section*{Acknowledgements}\label{Acknowledgements}

This work was made possible by useful discussions with Boris Bolliet, Justin Feng, Dra\v{z}en Glavan, Will Handley, Carlo Marzo, Roberto Percacci, Syksy R\"as\"anen, Alessandro Santoni, Ignacy Sawicki, Richard Woodard and Tom Zlosnik.

WB is grateful for the support of Marie Sk\l odowska-Curie Actions and the Institute of Physics of the Czech Academy of Sciences.

WB was supported by the research environment and infrastructure of the Handley Lab at the University of Cambridge.

This work was performed using the Cambridge Service for Data Driven Discovery (CSD3), part of which is operated by the University of Cambridge Research Computing on behalf of the STFC DiRAC HPC Facility (\href{www.dirac.ac.uk}{www.dirac.ac.uk}). The DiRAC component of CSD3 was funded by BEIS capital funding via STFC capital grants ST/P002307/1 and ST/R002452/1 and STFC operations grant ST/R00689X/1. DiRAC is part of the National e-Infrastructure.

Co-funded by the European Union (Physics for Future -- Grant Agreement No. 101081515). Views and opinions expressed are however those of the author(s) only and do not necessarily reflect those of the European Union or European Research Executive Agency. Neither the European Union nor the granting authority can be held responsible for them.

\appendix

\section{Gauss--Codazzi and the Ricci scalar}\label{ADMRicciScalar}

In this appendix, we provide a more conventional introduction to the ADM formalism~\cite{Arnowitt:1962hi} than that given in~\cref{subsec: ADM decomposition}, allowing for a self-contained derivation of the formulae of~\cref{Fdef,WellKnownFormula} which lie at the heart of our approach --- we do not rely on computer manipulation~\cite{Peeters:2006kp,Peeters:2007wn,Glavan:2024cfs}. To begin with, our conventions for the curvature are as follows. Given some vector~$\Vector{^\mu}$, the Ricci tensor is~$\Curvature{_{\sigma\nu}}\equiv\Curvature{^\lambda_{\sigma\lambda\nu}}$ and the Riemann tensor~$\Curvature{^\rho_{\sigma\mu\nu}}$ is given by
\begin{equation}\label{Riemann}
	\CovD{_{[\mu}}\CovD{_{\nu]}}\Vector{^\rho}\equiv\frac{1}{2}\Curvature{^\rho_{\sigma\mu\nu}}\Vector{^\sigma},
	\quad
	\Curvature{^\rho_{\sigma\mu\nu}}\equiv\PD{_\mu}\Christoffel{^\rho_{\sigma\nu}}-\PD{_\nu}\Christoffel{^\rho_{\sigma\mu}}+\Christoffel{^\alpha_{\sigma\nu}}\Christoffel{^\rho_{\mu\alpha}}-\Christoffel{^\alpha_{\sigma\mu}}\Christoffel{^\rho_{\nu\alpha}},
\end{equation}
where the covariant derivative~$\CovD{_\mu}$ and Christoffel symbol~$\Christoffel{^\rho_{\sigma\mu}}$ are
\begin{equation}\label{CovDChristoffel}
	\CovD{_\mu}\Vector{^\nu}\equiv\PD{_\mu}\Vector{^\nu}+\Christoffel{^\nu_{\mu\sigma}}\Vector{^\sigma},
	\quad
	\Christoffel{^\rho_{\sigma\mu}}\equiv\frac{1}{2}\Metric{^{\rho\alpha}}\left(\PD{_\sigma}\Metric{_{\mu\alpha}}+\PD{_\mu}\Metric{_{\sigma\alpha}}-\PD{_\alpha}\Metric{_{\mu\sigma}}\right).
\end{equation}
It is a defining characteristic of the phase-space formulation that a definite sense of time is agreed upon in the calculations; this we parameterise with the coordinate~$t\equiv\Coordinate{^0}$. The remaining coordinates~$\Coordinate{^i}$, with holonomic Roman indices running from one to three, are taken to be `\emph{adapted}' in the sense that they span each spatial hypersurface for each respective value of~$t$. Having set up the coordinate system, the conventional starting point for the ADM formalism is to introduce the unit-timelike vector field~$\UnitTimelike{^\mu}\UnitTimelike{_\mu}\equiv -1$ with the property of being normal to spatial hypersurfaces~$\tensor{\left(\PD{_i}\right)}{^\mu}\UnitTimelike{_\mu}\equiv 0$, i.e., it is locally normal to the tangent space of the adapted~$\Coordinate{^i}$ at every point on each surface. The relationship between~$t$ and~$\UnitTimelike{^\mu}$ is variable, and parameterised by the lapse and shift via the formula
\begin{equation}\label{LapseShift}
	\PD{_t}\Coordinate{^\mu}\equiv\Lapse{}\UnitTimelike{^\mu}+\Shift{^i}\Kronecker{^\mu_i}.
\end{equation}
The extrinsic curvature in~\cref{Kdef} may then be defined in terms of the covariant derivative of the unit-timelike vector field according to
\begin{equation}\label{ImplicitExtrinsicCurvature}
	\ExtrinsicCurvatureActual{_{ij}}
	\equiv-\CovD{_j}\UnitTimelike{_i}
	\equiv-\frac{1}{2\Lapse{}}\left(
		\MetricFoliationDot{_{ij}}
		-2\CD{_{(i}}\Shift{_{j)}}
	\right).
\end{equation}
The conventional precursor to the canonical expansion of the Ricci scalar is the Gauss--Codazzi equation~\cite{Gauss:1827,Codazzi:1868} for the projected components of the Riemann tensor, which reads, together with its full contraction
\begin{equation}\label{GaussCodazzi}
	\Curvature{^i_{jkl}}\equiv\CurvatureFoliation{^i_{jkl}}-\ExtrinsicCurvatureActual{_{jk}}\ExtrinsicCurvatureActual{^i_{l}}+\ExtrinsicCurvatureActual{_{jl}}\ExtrinsicCurvatureActual{^i_{k}},
	\quad 
	\Curvature{^{ij}_{ij}}\equiv\CurvatureFoliation{}+\ExtrinsicCurvatureActual{}^2-\ExtrinsicCurvatureActual{_{ij}}\ExtrinsicCurvatureActual{^{ij}}.
\end{equation}
Since the latter expression in~\cref{GaussCodazzi} can also be given by~$\Curvature{^{ij}_{ij}}\equiv\Curvature{}+2\Curvature{^{\mu}_{\nu\mu\lambda}}\UnitTimelike{^\nu}\UnitTimelike{^\lambda}$, we can use~\cref{Riemann} to show that
\begin{equation}\label{ProjectedTrace}
	\Curvature{^{\mu}_{\nu\mu\lambda}}\UnitTimelike{^\nu}\UnitTimelike{^\lambda}\equiv
	2\UnitTimelike{^\lambda}\CovD{_{[\mu}}\CovD{_{\lambda]}}\UnitTimelike{^\mu}\equiv
	2\CovD{_{[\mu}}\left(\UnitTimelike{^\lambda}\CovD{_{\lambda]}}\UnitTimelike{^\mu}\right)
	-2\CovD{_{[\mu}}\UnitTimelike{^\lambda}\CovD{_{\lambda]}}\UnitTimelike{^\mu}.
\end{equation}
It follows from~\cref{ImplicitExtrinsicCurvature} that~$\CovD{_\mu}\UnitTimelike{^\mu}\equiv-\ExtrinsicCurvatureActual{}$ and~$\CovD{_{\mu}}\UnitTimelike{^\lambda}\CovD{_{\lambda}}\UnitTimelike{^\mu}\equiv\ExtrinsicCurvatureActual{_{ij}}\ExtrinsicCurvatureActual{^{ij}}$, so that~\cref{GaussCodazzi,ProjectedTrace} may be used to deduce
\begin{equation}\label{Intermediate}
\begin{aligned}
		\Curvature{}& \equiv\CurvatureFoliation{}-\ExtrinsicCurvatureActual{}^2+\ExtrinsicCurvatureActual{_{ij}}\ExtrinsicCurvatureActual{^{ij}}+4\CovD{_{[\mu}}\left(\UnitTimelike{^\mu}\CovD{_{\nu]}}\UnitTimelike{^\nu}\right)
		\\
		&\equiv
	\CurvatureFoliation{}+\ExtrinsicCurvatureActual{}^2-\ExtrinsicCurvatureActual{_{ij}}\ExtrinsicCurvatureActual{^{ij}}+2\UnitTimelike{^\mu}\CovD{_\mu}\ExtrinsicCurvatureActual{}-2\UnitTimelike{^\mu}\CovD{_\nu}\CovD{_\mu}\UnitTimelike{^\nu}.
\end{aligned}
\end{equation}
By similar reasoning, the final term in~\cref{Intermediate} is~$\UnitTimelike{^\mu}\CovD{_\nu}\CovD{_\mu}\UnitTimelike{^\nu}\equiv\CovD{_\nu}\big(\UnitTimelike{^\mu}\CovD{_\mu}\UnitTimelike{^\nu}\big)-\ExtrinsicCurvatureActual{_{ij}}\ExtrinsicCurvatureActual{^{ij}}$, whilst a consequence of~\cref{LapseShift} is that~$\UnitTimelike{^\nu}\CovD{_\nu}\UnitTimelike{_\mu}\equiv\big(\Kronecker{^\nu_\mu}+\UnitTimelike{^\nu}\UnitTimelike{_\mu}\big)\CovD{_\nu}\ln\Lapse{}$. After a tedious calculation in which the latter identity is used several times it may be shown that
\begin{equation}\label{LengthyCalculation}
	\UnitTimelike{^\mu}\CovD{_\nu}\CovD{_\mu}\UnitTimelike{^\nu}\equiv\frac{1}{\Lapse{}}\CD{_i}\CD{^i}\Lapse{}-\ExtrinsicCurvatureActual{_{ij}}\ExtrinsicCurvatureActual{^{ij}}.
\end{equation}
Meanwhile, we can use~\cref{LapseShift,Intermediate} to write~$\UnitTimelike{^\mu}\CovD{_\mu}\ExtrinsicCurvatureActual{}\equiv\left(\ExtrinsicCurvatureActualDot{}-\Shift{^i}\CD{_i}\ExtrinsicCurvatureActual{}\right)/\Lapse{}$. The explicit time derivative of the whole trace requires some further manipulation: since~$\ExtrinsicCurvatureActualDot{}\equiv\MetricFoliation{^{ij}}\ExtrinsicCurvatureActualDot{_{ij}}-\ExtrinsicCurvatureActual{^{ij}}\MetricFoliationDot{_{ij}}$ we can use~\cref{ImplicitExtrinsicCurvature} to write
\begin{equation}\label{ExplicitTimeDerivative}
	\UnitTimelike{^\mu}\CovD{_\mu}\ExtrinsicCurvatureActual{}\equiv
	\frac{1}{N}\left(\MetricFoliation{^{ij}}\ExtrinsicCurvatureActualDot{_{ij}}
	+2\Lapse{}\ExtrinsicCurvatureActual{_{ij}}\ExtrinsicCurvatureActual{^{ij}}
	-2\ExtrinsicCurvatureActual{^{ij}}\CD{_i}\Shift{_j}
	-\Shift{^i}\CD{_i}\ExtrinsicCurvatureActual{}\right).
\end{equation}
Finally,~\cref{Fdef,WellKnownFormula} are obtained by plugging~\cref{LengthyCalculation,ExplicitTimeDerivative} into~\cref{Intermediate}.

\bibliographystyle{apsrev4-2}
\bibliography{Manuscript.bib}

\end{document}

%% file: Macros.tex
\newrobustcmd{\NPhy}{%
	\tensor{N}{_{\text{Phy}}}%
}
\newrobustcmd{\NCan}{%
	\tensor{N}{_{\text{Can}}}%
}
\newrobustcmd{\NFirst}{%
	\tensor{N}{_{\text{1st}}}%
}
\newrobustcmd{\NSecond}{%
	\tensor{N}{_{\text{2nd}}}%
}

\newrobustcmd{\ScaleFactor}{%
	a
}
\newrobustcmd{\HubbleNumber}{%
	H
}
\newrobustcmd{\HubbleNumberDot}{%
	\dot{H}
}

\newrobustcmd{\Primary}[1]{\tensor{\Phi}{_{#1}}}
\newrobustcmd{\FirstOrderPrimary}[1]{\tensor{\vphantom{l^{l^2}}\smash{\stackrel{\raisebox{-3pt}{\scalebox{0.55}{$(1)$}}}{\scalebox{0.99}{$\Phi$}}}}{_{#1}}}
\newrobustcmd{\SecondOrderPrimary}[1]{\tensor{\vphantom{l^{l^2}}\smash{\stackrel{\raisebox{-3pt}{\scalebox{0.55}{$(2)$}}}{\scalebox{0.99}{$\Phi$}}}}{_{#1}}}
\newrobustcmd{\Secondary}[1]{\tensor{\Psi}{_{#1}}}
\newrobustcmd{\FirstOrderSecondary}[1]{\tensor{\vphantom{l^{l^2}}\smash{\stackrel{\raisebox{-3pt}{\scalebox{0.55}{$(1)$}}}{\scalebox{0.99}{$\Psi$}}}}{_{#1}}}
\newrobustcmd{\SecondOrderSecondary}[1]{\tensor{\vphantom{l^{l^2}}\smash{\stackrel{\raisebox{-3pt}{\scalebox{0.55}{$(2)$}}}{\scalebox{0.99}{$\Psi$}}}}{_{#1}}}
\newrobustcmd{\ActualSecondary}[1]{\tensor{\Theta}{_{#1}}}
\newrobustcmd{\FirstOrderActualSecondary}[1]{\tensor{\vphantom{l^{l^2}}\smash{\stackrel{\raisebox{-3pt}{\scalebox{0.55}{$(1)$}}}{\scalebox{0.99}{$\Theta$}}}}{_{#1}}}
\newrobustcmd{\SecondOrderActualSecondary}[1]{\tensor{\vphantom{l^{l^2}}\smash{\stackrel{\raisebox{-3pt}{\scalebox{0.55}{$(2)$}}}{\scalebox{0.99}{$\Theta$}}}}{_{#1}}}
\newrobustcmd{\QVar}[1]{\tensor{q}{_{#1}}}
\newrobustcmd{\BackgroundQVar}[1]{\tensor*{\overline{q}}{_{#1}}}
\newrobustcmd{\DeltaQVar}[1]{\tensor*{\delta q}{_{#1}}}
\newrobustcmd{\QVarDot}[1]{\tensor{\dot{q}}{_{#1}}}
\newrobustcmd{\DeltaQVarDot}[1]{\tensor*{\delta \dot{q}}{_{#1}}}
\newrobustcmd{\QVarDash}[1]{\tensor{q'}{_{#1}}}
\newrobustcmd{\DeltaQVarDash}[1]{\tensor*{\delta q'}{_{#1}}}
\newrobustcmd{\PVar}[1]{\tensor{p}{_{#1}}}
\newrobustcmd{\BackgroundPVar}[1]{\tensor*{\overline{p}}{_{#1}}}
\newrobustcmd{\DeltaPVar}[1]{\tensor*{\delta p}{_{#1}}}
\newrobustcmd{\PVarDot}[1]{\tensor{\dot{p}}{_{#1}}}
\newrobustcmd{\DeltaPVarDot}[1]{\tensor*{\delta \dot{p}}{_{#1}}}
\newrobustcmd{\PVarDash}[1]{\tensor{p'}{_{#1}}}
\newrobustcmd{\DeltaPVarDash}[1]{\tensor*{\delta p'}{_{#1}}}
\newrobustcmd{\Mul}[1]{\tensor{\ell}{_{#1}}}
\newrobustcmd{\BackgroundMul}[1]{\tensor*{\overline{\ell}}{_{#1}}}
\newrobustcmd{\DeltaMul}[1]{\tensor*{\delta\ell}{_{#1}}}
\newrobustcmd{\MulDot}[1]{\tensor{\dot{\ell}}{_{#1}}}
\newrobustcmd{\DeltaMulDot}[1]{\tensor*{\delta \dot{\ell}}{_{#1}}}
\newrobustcmd{\MulDash}[1]{\tensor{\ell'}{_{#1}}}
\newrobustcmd{\DeltaMulDash}[1]{\tensor*{\delta \ell'}{_{#1}}}

\newrobustcmd{\Action}{%
	\mathscr{S}%
}

\newrobustcmd{\FirstOrderAction}[1]{%
	\tensor{\vphantom{l^{l^2}}\smash{\stackrel{\raisebox{-3pt}{\scalebox{0.55}{$(1)$}}}{\scalebox{0.99}{$\mathscr{S}$}}}}{#1}%
}
\newrobustcmd{\SecondOrderAction}[1]{%
	\tensor{\vphantom{l^{l^2}}\smash{\stackrel{\raisebox{-3pt}{\scalebox{0.55}{$(2)$}}}{\scalebox{0.99}{$\mathscr{S}$}}}}{#1}%
}

\newrobustcmd{\CovD}[1]{%
	\tensor{D}{#1}
}
\newrobustcmd{\BackgroundCovD}[1]{%
	\tensor*{\overline{D}}{#1}
}
\newrobustcmd{\CD}[1]{%
	\tensor{\nabla}{#1}%
}
\newrobustcmd{\BackgroundCD}[1]{%
	\tensor{\overline{\nabla}}{#1}%
}
\newrobustcmd{\Bach}[1]{%
	\tensor*{B}{#1}%
}
\newrobustcmd{\Weyl}[1]{%
	\tensor{C}{#1}%
}
\newrobustcmd{\Schouten}[1]{%
	\tensor{S}{#1}%
}
\newrobustcmd{\Curvature}[1]{%
	\tensor{R}{#1}
}
\newrobustcmd{\DotCurvature}[1]{%
	\tensor*{\dot{R}}{#1}
}
\newrobustcmd{\DDotCurvature}[1]{%
	\tensor*{\ddot{R}}{#1}
}
\newrobustcmd{\BackgroundCurvature}[1]{%
	\tensor{\overline{R}}{#1}
}
\newrobustcmd{\R}[1]{%
	\tensor{\mathcal{R}}{#1}%
}
\newrobustcmd{\BackgroundR}[1]{%
	\tensor*{\overline{\mathcal{R}}}{#1}%
}
\newrobustcmd{\CurvatureFoliation}[1]{%
	\tensor{\mathcal{R}}{#1}
}
\newrobustcmd{\BackgroundCurvatureFoliation}[1]{%
	\tensor*{\overline{\mathcal{R}}}{#1}
}
\newrobustcmd{\PrimaryConstraint}[1]{%
	\tensor*{\Phi}{#1}%
}
\newrobustcmd{\FirstOrderPrimaryConstraint}[1]{%
	\tensor{\vphantom{l^{l^2}}\smash{\stackrel{\raisebox{-3pt}{\scalebox{0.55}{$(1)$}}}{\scalebox{0.99}{$\Phi$}}}}{#1}%
}
\newrobustcmd{\SecondOrderPrimaryConstraint}[1]{%
	\tensor{\vphantom{l^{l^2}}\smash{\stackrel{\raisebox{-3pt}{\scalebox{0.55}{$(2)$}}}{\scalebox{0.99}{$\Phi$}}}}{#1}%
}
\newrobustcmd{\Multiplier}[1]{%
	\tensor*{\lambda}{#1}%
}
\newrobustcmd{\BackgroundMultiplier}[1]{%
	\tensor*{\overline{\lambda}}{#1}%
}
\newrobustcmd{\DeltaMultiplier}[1]{%
	\tensor*{\delta\lambda}{#1}%
}
\newrobustcmd{\SecondaryConstraint}[1]{%
	\tensor*{\Psi}{#1}%
}
\newrobustcmd{\FirstOrderSecondaryConstraint}[1]{%
	\tensor{\vphantom{l^{l^2}}\smash{\stackrel{\raisebox{-3pt}{\scalebox{0.55}{$(1)$}}}{\scalebox{0.99}{$\Psi$}}}}{#1}%
}
\newrobustcmd{\SecondOrderSecondaryConstraint}[1]{%
	\tensor{\vphantom{l^{l^2}}\smash{\stackrel{\raisebox{-3pt}{\scalebox{0.55}{$(2)$}}}{\scalebox{0.99}{$\Psi$}}}}{#1}%
}
\newrobustcmd{\FirstOrderActualSecondaryConstraint}[1]{%
	\tensor{\vphantom{l^{l^2}}\smash{\stackrel{\raisebox{-3pt}{\scalebox{0.55}{$(1)$}}}{\scalebox{0.99}{$\Theta$}}}}{#1}%
}
\newrobustcmd{\SecondOrderActualSecondaryConstraint}[1]{%
	\tensor{\vphantom{l^{l^2}}\smash{\stackrel{\raisebox{-3pt}{\scalebox{0.55}{$(2)$}}}{\scalebox{0.99}{$\Theta$}}}}{#1}%
}
\newrobustcmd{\SmearingS}[1]{%
	\tensor*{s}{#1}%
}
\newrobustcmd{\SmearingF}[1]{%
	\tensor*{f}{#1}%
}
\newrobustcmd{\GenericPrimaryConstraint}[1]{%
	\tensor*{\Phi}{}\left[{#1}\right]%
}
\newrobustcmd{\FirstOrderGenericPrimaryConstraint}[1]{%
	\tensor*{\vphantom{l^{l^2}}\smash{\stackrel{\raisebox{-3pt}{\scalebox{0.55}{$(1)$}}}{\scalebox{0.99}{$\Phi$}}}}{}\left[{#1}\right]%
}
\newrobustcmd{\SecondOrderGenericPrimaryConstraint}[1]{%
	\tensor*{\vphantom{l^{l^2}}\smash{\stackrel{\raisebox{-3pt}{\scalebox{0.55}{$(2)$}}}{\scalebox{0.99}{$\Phi$}}}}{}\left[{#1}\right]%
}
\newrobustcmd{\TotalHamiltonian}{%
	\tensor*{\mathscr{H}}{}%
}
\newrobustcmd{\TotalHamiltonianGen}[1]{%
	\tensor*{\mathscr{H}}{}\left[{#1}\right]%
}
\newrobustcmd{\FirstOrderTotalHamiltonian}{%
	\tensor*{\vphantom{l^{l^2}}\smash{\stackrel{\raisebox{-3pt}{\scalebox{0.55}{$(1)$}}}{\scalebox{0.99}{$\mathscr{H}$}}}}{}%
}
\newrobustcmd{\SecondOrderTotalHamiltonian}[1]{%
	\tensor*{\vphantom{l^{l^2}}\smash{\stackrel{\raisebox{-3pt}{\scalebox{0.55}{$(2)$}}}{\scalebox{0.99}{$\mathscr{H}$}}}}{}\left[{#1}\right]%
}
\newrobustcmd{\ThirdOrderTotalHamiltonian}[1]{%
	\tensor*{\vphantom{l^{l^2}}\smash{\stackrel{\raisebox{-3pt}{\scalebox{0.55}{$(3)$}}}{\scalebox{0.99}{$\mathscr{H}$}}}}{}\left[{#1}\right]%
}
\newrobustcmd{\SuperConstraint}[1]{%
	\tensor{\mathcal{H}}{#1}%
}
\newrobustcmd{\ReducedSuperConstraint}[1]{%
	\tensor{\mathbb{H}}{#1}%
}
\newrobustcmd{\ZerothOrderReducedSuperConstraint}[1]{%
	\tensor{\vphantom{l^{l^2}}\smash{\stackrel{\raisebox{-3pt}{\scalebox{0.55}{$(0)$}}}{\scalebox{0.99}{$\mathbb{H}$}}}}{#1}%
}
\newrobustcmd{\FirstOrderReducedSuperConstraint}[1]{%
	\tensor{\vphantom{l^{l^2}}\smash{\stackrel{\raisebox{-3pt}{\scalebox{0.55}{$(1)$}}}{\scalebox{0.99}{$\mathbb{H}$}}}}{#1}%
}
\newrobustcmd{\SuperConstraintDot}[1]{%
	\tensor{\dot{\mathcal{H}}}{#1}%
}
\newrobustcmd{\DressedSuperConstraint}[1]{%
	\tensor{\mathfrak{H}}{#1}%
}
\newrobustcmd{\FirstOrderSuperConstraint}[1]{%
	\tensor{\vphantom{l^{l^2}}\smash{\stackrel{\raisebox{-3pt}{\scalebox{0.55}{$(1)$}}}{\scalebox{0.99}{$\mathcal{H}$}}}}{#1}%
}
\newrobustcmd{\SecondOrderSuperConstraint}[1]{%
	\tensor{\vphantom{l^{l^2}}\smash{\stackrel{\raisebox{-3pt}{\scalebox{0.55}{$(2)$}}}{\scalebox{0.99}{$\mathcal{H}$}}}}{#1}%
}
\newrobustcmd{\Lapse}{%
	N	
}
\newrobustcmd{\BackgroundLapse}{%
	\overline{N}	
}
\newrobustcmd{\DeltaLapse}{%
	\delta N	
}
\newrobustcmd{\Shift}[1]{%
	\tensor{N}{#1}
}
\newrobustcmd{\BackgroundShift}[1]{%
	\tensor*{\overline{N}}{#1}
}
\newrobustcmd{\DeltaShift}[1]{%
	\tensor*{\delta N}{#1}
}
\newrobustcmd{\Kronecker}[1]{%
	\tensor*{\delta}{#1}
}
\newrobustcmd{\G}[1]{%
	\tensor*{g}{#1}%
}
\newrobustcmd{\Metric}[1]{%
	\tensor{g}{#1}
}
\newrobustcmd{\BackgroundG}[1]{%
	\tensor*{\overline{g}}{#1}%
}
\newrobustcmd{\DeltaMetric}[1]{%
	\tensor*{\delta g}{#1}%
}
\newrobustcmd{\DeltaStressEnergy}[1]{%
	\tensor*{\delta T}{#1}%
}
\newrobustcmd{\DeltaG}[1]{%
	\tensor*{\delta h}{#1}%
}
\newrobustcmd{\DeltaMetricFoliation}[1]{%
	\tensor*{\delta h}{#1}%
}
\newrobustcmd{\DotDeltaMetricFoliation}[1]{%
	\tensor*{\delta \dot{h}}{#1}%
}
\newrobustcmd{\ConjugateMomentumG}[1]{%
	\tensor*{\pi}{#1}%
}
\newrobustcmd{\ConjugateMomentumGDot}[1]{%
	\tensor*{\dot{\pi}}{#1}%
}
\newrobustcmd{\BackgroundConjugateMomentumG}[1]{%
	\tensor*{\overline{\pi}}{#1}%
}
\newrobustcmd{\ConjugateMomentumDeltaG}[1]{%
	\tensor*{\delta\pi}{#1}%
}
\newrobustcmd{\ConjugateMomentumDDeltaG}[1]{%
	\tensor*{\delta\delta\pi}{#1}%
}
\newrobustcmd{\FFunctionActual}[1]{%
	\tensor*{F}{#1}%
}
\newrobustcmd{\FFunction}[1]{%
	\tensor*{\mathcal{F}}{#1}%
}
\newrobustcmd{\ExtrinsicCurvatureActual}[1]{%
	\tensor*{K}{#1}%
}
\newrobustcmd{\GaugeProjector}[1]{%
	\tensor{\mathscr{P}}{#1}%
}
\newrobustcmd{\SchoutenProjector}[1]{%
	\tensor{\mathfrak{P}}{#1}%
}
\newrobustcmd{\Killing}[1]{%
	\tensor*{\xi}{#1}%
}
\newrobustcmd{\ExtrinsicCurvatureActualDot}[1]{%
	\tensor*{\dot{K}}{#1}%
}
\newrobustcmd{\ExtrinsicCurvature}[1]{%
	\tensor*{\mathcal{K}}{#1}%
}
\newrobustcmd{\BackgroundExtrinsicCurvature}[1]{%
	\tensor*{\overline{\mathcal{K}}}{#1}%
}
\newrobustcmd{\DeltaExtrinsicCurvature}[1]{%
	\tensor*{\delta\mathcal{K}}{#1}%
}
\newrobustcmd{\DotDeltaExtrinsicCurvature}[1]{%
	\tensor*{\delta\dot{\mathcal{K}}}{#1}%
}
\newrobustcmd{\ConjugateMomentumExtrinsicCurvature}[1]{%
	\tensor*{\rho}{#1}%
}
\newrobustcmd{\ConjugateMomentumDDeltaExtrinsicCurvature}[1]{%
	\tensor*{\delta\delta\rho}{#1}%
}
\newrobustcmd{\ConjugateMomentumExtrinsicCurvatureDot}[1]{%
	\tensor*{\dot{\rho}}{#1}%
}
\newrobustcmd{\BackgroundConjugateMomentumExtrinsicCurvature}[1]{%
	\tensor*{\overline{\rho}}{#1}%
}
\newrobustcmd{\ConjugateMomentumDeltaExtrinsicCurvature}[1]{%
	\tensor*{\delta\rho}{#1}%
}
\newrobustcmd{\Cadabra}{\textit{Cadabra}\xspace}

\newrobustcmd{\UnitTimelike}[1]{%
	\tensor{n}{#1}
}
\newrobustcmd{\Coordinate}[1]{%
	\tensor{x}{#1}
}
\newrobustcmd{\Vector}[1]{%
	\tensor{A}{#1}
}
\newrobustcmd{\PD}[1]{%
	\tensor*{\partial}{#1}
}
\newrobustcmd{\Christoffel}[1]{%
	\tensor*{\Gamma}{#1}
}
\newrobustcmd{\GenQ}{%
	\tensor*{q}{}
}
\newrobustcmd{\GenP}{%
	\tensor*{p}{}
}
\newrobustcmd{\DeltaGenQ}{%
	\tensor*{\delta q}{}
}
\newrobustcmd{\DeltaGenP}{%
	\tensor*{\delta p}{}
}
\newrobustcmd{\DotDeltaGenQ}{%
	\tensor*{\delta\dot{q}}{}
}
\newrobustcmd{\DotDeltaGenP}{%
	\tensor*{\delta\dot{p}}{}
}
\newrobustcmd{\DDeltaGenQ}{%
	\tensor*{\delta\delta q}{}
}
\newrobustcmd{\DDeltaGenP}{%
	\tensor*{\delta\delta p}{}
}
\newrobustcmd{\DotDDeltaGenQ}{%
	\tensor*{\delta\delta \dot{q}}{}
}
\newrobustcmd{\DotDDeltaGenP}{%
	\tensor*{\delta\delta \dot{p}}{}
}
\newrobustcmd{\ConjugateMomentumMetricFoliation}[1]{%
	\tensor*{\pi}{#1}
}
\newrobustcmd{\ConjugateMomentumMetricFoliationDot}[1]{%
	\tensor*{\dot{\pi}}{#1}
}
\newrobustcmd{\MetricFoliation}[1]{%
	\tensor{h}{#1}
}
\newrobustcmd{\BackgroundMetricFoliation}[1]{%
	\tensor*{\overline{h}}{#1}
}
\newrobustcmd{\BackgroundMetricFoliationDot}[1]{%
	\tensor*{\dot{\overline{h}}}{#1}
}
\newrobustcmd{\MetricFoliationDot}[1]{%
	\tensor{\dot{h}}{#1}
}
\newrobustcmd{\KillingVector}[1]{%
	\tensor{\xi}{#1}
}
\newrobustcmd{\ExtrinsicCurvatureDot}[1]{%
	\tensor*{\dot{\mathcal{K}}}{#1}
}
\newrobustcmd{\Brinkmann}{%
	{\mathcal{G}}
}
\newrobustcmd{\SchwarzschildRadius}{%
	{r_{\text{g}}}
}
\newrobustcmd{\PsiFunc}{%
	{\psi}
}
\newrobustcmd{\GammaFunc}{%
	{\gamma}
}
\newrobustcmd{\UFunc}{%
	{\mathcal{U}}
}
\newrobustcmd{\DotUFunc}{%
	{\mathcal{\dot{U}}}
}
\newrobustcmd{\AngularMomentumPerUnitMass}{%
	{a}
}
\newrobustcmd{\KasnerA}{%
	{\alpha}
}
\newrobustcmd{\KasnerB}{%
	{\beta}
}
\newrobustcmd{\KasnerC}{%
	{\gamma}
}
\newrobustcmd{\ShorthandK}[1]{%
	\tensor*{\mathscr{K}}{#1}%
}
\newrobustcmd{\ShorthandL}[1]{%
	\tensor{\mathscr{L}}{#1}%
}
\newrobustcmd{\ShorthandQ}[1]{%
	\tensor{\mathscr{Q}}{#1}%
}
\newrobustcmd{\TraceShorthandQ}[1]{%
	\tensor*{\mathscr{Q}}{#1}%
}
\newrobustcmd{\VectorZeroPlus}[1]{%
	\tensor{A{\scalebox{0.75}{$\substack{\#1 \\ 0^+}$}}}{#1}%
}
\newrobustcmd{\VectorOneMinus}[1]{%
	\tensor{A{\scalebox{0.75}{$\substack{\#1 \\ 1^-}$}}}{#1}%
}
\newrobustcmd{\ConjugateMomentumVectorZeroPlus}[1]{%
	\tensor{\pi(A){\scalebox{0.75}{$\substack{\#1 \\ 0^+}$}}}{#1}%
}
\newrobustcmd{\ConjugateMomentumVectorOneMinus}[1]{%
	\tensor{\pi(A){\scalebox{0.75}{$\substack{\#1 \\ 1^-}$}}}{#1}%
}
\newrobustcmd{\LagrangeMultiplierVectorZeroPlus}[1]{%
	\tensor{\lambda(A){\scalebox{0.75}{$\substack{\#1 \\ 0^+}$}}}{#1}%
}
\newrobustcmd{\LagrangeMultiplierVectorOneMinus}[1]{%
	\tensor{\lambda(A){\scalebox{0.75}{$\substack{\#1 \\ 1^-}$}}}{#1}%
}
\newrobustcmd{\PrimaryConstraintVectorOneMinus}[1]{%
	\tensor{\phi(A){\scalebox{0.75}{$\substack{\#1 \\ 1^-}$}}}{#1}%
}
\newrobustcmd{\PrimaryConstraintVectorOneMinusDot}[1]{%
	\tensor{\dot{\phi}(A){\scalebox{0.75}{$\substack{\#1 \\ 1^-}$}}}{#1}%
}
\newrobustcmd{\SecondaryConstraintVectorZeroPlus}[1]{%
	\tensor{\chi(A){\scalebox{0.75}{$\substack{\#1 \\ 0^+}$}}}{#1}%
}
\newrobustcmd{\SecondaryConstraintVectorOneMinus}[1]{%
	\tensor{\chi(A){\scalebox{0.75}{$\substack{\#1 \\ 1^-}$}}}{#1}%
}
\newrobustcmd{\ScalarField}[1]{%
	\tensor{\phi}{#1}%
}
\newrobustcmd{\FirstVector}[1]{%
	\tensor{A}{#1}%
}
\newrobustcmd{\SecondVector}[1]{%
	\tensor{B}{#1}%
}
\newrobustcmd{\ThirdVector}[1]{%
	\tensor{C}{#1}%
}

%% file: MacrosNames.tex
\usepackage{xspace}

\newrobustcmd{\Hamilcar}{\textit{Hamilcar}\xspace}
\newrobustcmd{\Hasdrubal}{\textit{Hasdrubal}\xspace}
\newrobustcmd{\PSALTer}{\textit{PSALTer}\xspace}
\newrobustcmd{\xAct}{\textit{xAct}\xspace}
\newrobustcmd{\xTensor}{\textit{xTensor}\xspace}
\newrobustcmd{\xCoba}{\textit{xCoba}\xspace}
\newrobustcmd{\xPerm}{\textit{xPerm}\xspace}
\newrobustcmd{\xCore}{\textit{xCore}\xspace}
\newrobustcmd{\xTras}{\textit{xTras}\xspace}
\newrobustcmd{\xPlain}{\textit{xPlain}\xspace}
\newrobustcmd{\SymManipulator}{\textit{SymManipulator}\xspace}
\newrobustcmd{\Mathematica}{\textit{Mathematica}\xspace}
\newrobustcmd{\Wolfram}{\textit{Wolfram}\xspace}
\newrobustcmd{\WolframLanguage}{\textit{Wolfram Language}\xspace}
\newrobustcmd{\GitHub}{\textit{GitHub}\xspace}
\newrobustcmd{\GitLab}{\textit{GitLab}\xspace}
\newrobustcmd{\Bash}{\textit{Bash}\xspace}
\newrobustcmd{\Linux}{\textit{GNU/Linux}\xspace}
\newrobustcmd{\Windows}{\textit{Windows}\xspace}
\newrobustcmd{\Mac}{\textit{macOS}\xspace}
\newrobustcmd{\OpenAI}{\textit{OpenAI}\xspace}
\newrobustcmd{\AgentsSDK}{\textit{Agents SDK}\xspace}
\newrobustcmd{\WolframResearch}{\textit{Wolfram Research}\xspace}
\newrobustcmd{\Python}{\textit{Python}\xspace}

%% file: MacrosLstListings.tex
\usepackage{listings}
\usepackage{upquote}
\usepackage{graphicx}
\usepackage{adjustbox}
\usepackage{xstring}
\usepackage{catchfile}
\newrobustcmd{\dollar}{\mbox{\color{gray}\textbf\textdollar}}

\newcounter{lstoutput}
\newcommand{\lstoutputpath}{}

\makeatletter
\newrobustcmd{\lstinputoutput}[2][]{%
	\refstepcounter{lstoutput}%
	\edef\@lstcurrentnum{\thelstoutput}%
	\StrSubstitute{#2}{/}{-}[\@tempinput]%
	\StrSubstitute{\@tempinput}{.tex}{}[\@lstinputlabelname]%
	\expandafter\label\expandafter{\@lstinputlabelname}%
	\lstinputlisting[literate={In[\#]:=}{{\color{gray}\textbf{In[\@lstcurrentnum]:=}\hspace{0.5em}}}8]{\lstoutputpath#2}%
	\ifx&#1&%
	\else%
		\StrSubstitute{#1}{/}{-}[\@tempoutput]%
		\StrSubstitute{\@tempoutput}{.pdf}{}[\@lstoutputlabelname]%
		\ifx\@lstoutputlabelname\@lstinputlabelname\else%
			\expandafter\label\expandafter{\@lstoutputlabelname}%
		\fi%
		\vspace{-5pt}%
		\noindent%
		{\fboxsep=0.5pt\colorbox{backing}{%
			\begin{minipage}{\dimexpr\textwidth-2\fboxsep-2\fboxrule}
				\sbox0{{\ttfamily\color{gray}\textbf{Out[\thelstoutput]=}}\hspace{0.5em}}%
				\usebox{0}%
				\sbox2{\includegraphics{\lstoutputpath#1}}%
				\ifdim\wd2>\dimexpr\textwidth-2\fboxsep-2\fboxrule-\wd0\relax%
					\raisebox{-\height+\ht\strutbox}{%
						\resizebox{\dimexpr\textwidth-2\fboxsep-2\fboxrule-\wd0\relax}{!}{\usebox{2}}%
					}%
				\else%
					\raisebox{-\height+\ht\strutbox}{\usebox{2}}%
				\fi%
			\end{minipage}%
		}}%
		\par\vspace{\medskipamount}\noindent\ignorespaces%
	\fi%
}\ignorespaces%
\makeatother

\newrobustcmd{\lstinputcref}[1]{%
	{\ttfamily\color{gray}\textbf{In[\ref{#1}]}}%
}

\newrobustcmd{\lstoutputcref}[1]{%
	{\ttfamily\color{gray}\textbf{Out[\ref{#1}]}}%
}

\newrobustcmd{\hasdrubalsession}[1]{%
	\StrSubstitute{#1}{.jsonl}{_full.tex}[\@hasdrubalfile]%
	\input{code_listings_Hasdrubal-\@hasdrubalfile}%
}

\newrobustcmd{\hasdrubalsummary}[1]{%
	\par\vspace{\bigskipamount}\noindent%
	\StrSubstitute{#1}{.jsonl}{_summary.tex}[\@hasdrubalfile]%
	\begin{tcolorbox}[colback=backing, colframe=NavyBlue, boxrule=0pt, arc=0pt, leftrule=2pt, left=6pt, right=0pt, top=4pt, bottom=4pt, breakable, before skip=0pt, after skip=\medskipamount]%
	{\ttfamily\color{NavyBlue}\textbf{Hasdrubal:}}\hspace{0.5em}%
	\footnotesize\itshape\sloppy%
	\input{code_listings_Hasdrubal-\@hasdrubalfile}%
	\end{tcolorbox}%
	\noindent\ignorespaces%
}

\lstdefinestyle{plain}{
	language={},
	keywordstyle=\ttfamily,
	keywordstyle=[2]\ttfamily,
	keywordstyle=[3]\ttfamily,
	keywordstyle=[4]\ttfamily,
	keywordstyle=[5]\ttfamily,
	stringstyle=\ttfamily,
	commentstyle=\ttfamily
}

\lstdefinestyle{hasdrubalinline}{
	postbreak={},
	breakatwhitespace=true
}

 \lstdefinestyle{ascii-tree}{ 
    literate=
    {├}{{%
	\hphantom{\raisebox{0.5ex}{\rule{0.5ex}{1pt}}}%
	\smash{\raisebox{-1ex}{\rule{1pt}{1.1\baselineskip}}}%
	\raisebox{0.5ex}{\rule{0.5ex}{1pt}}%
	}}1 
    {│}{{%
	\hphantom{\raisebox{0.5ex}{\rule{0.5ex}{1pt}}}%
	\smash{\raisebox{-1ex}{\rule{1pt}{1.1\baselineskip}}}%
	\hphantom{\raisebox{0.5ex}{\rule{0.5ex}{1pt}}}%
	}}1 
    {─}{{%
	\smash{\raisebox{0.5ex}{\rule{1ex}{1pt}}}%
	}}1 
    {└}{{%
	\hphantom{\raisebox{0.5ex}{\rule{0.5ex}{1pt}}}%
	\smash{\raisebox{0.5ex}{\rule{1pt}{0.5\baselineskip}}}%
	\raisebox{0.5ex}{\rule{0.5ex}{1pt}}%
	}}1 
    { }{ }1 
 }

\input{MacrosLstKeywords.tex}

\lstdefinelanguage{Special}{%
morekeywords=[1]{%
ParticleSpectrographMassiveGravity.m,ValidateSymmetryField.m,ValidateSymmetryMode.m,ValidateTraceless.m,ValidateInverseField.m,ValidateInverseMode.m,ValidateSpatial.m,DefAllComponentValues.m,DefSummary.m,ValidateSO3Irreps.m,DefSymbol.m,MakeAutomaticallyTraceless.m,MakeUniquePartialDual.m,MakeUniqueQuadratic.m,RemoveContraction.m,MakeAutomaticallyNotAntisymmetric.m,RegisterFieldRank0.m,RegisterFieldRank1.m,RegisterFieldRank2.m,RegisterFieldRank2Antisymmetric.m,RegisterFieldRank2Symmetric.m,RegisterFieldRank3.m,RegisterFieldRank3Antisymmetric13.m,RegisterFieldRank3Symmetric12.m,RegisterFieldRank3Symmetric13.m,RegisterFieldRank3Symmetric23.m,RegisterFieldRank3TotallySymmetric.m,DecompositionTable.m,ExpansionTable.m,FieldMosaic.m,AllIndexConfigurations.m,AllocateTensorValues.m,RegisterFieldRank3Antisymmetric12.m,RegisterFieldRank3Antisymmetric23.m,RegisterFieldRank3TotallyAntisymmetric.m,AppendToField.m,DefFiducialField.m,DefSO3Irrep.m,PreComputeComponents.m,CombineRules.m,SummariseField.m,IsNegativeParitySpinTwo.m,CatalogueInvariant.m,CacheContexts.m,DefPlaceholderSpins.m,GenerateAnsatz.m,NormaliseRescalings.m,PoleToSquareMass.m,IsolatePoles.m,IrrationalQ.m,StripPoly.m,PartitionDeterminant.m,GaugeArtifactQ.m,MassiveGhost.m,MassiveAnalysisOfSector.m,SimplifyMasses.m,SquareMassQ.m,UnresolvedPoleQ.m,IndependentComponents.m,DefFreeSourceVariables.m,AllIndependentComponents.m,ConstraintComponentToLightcone.m,MakeConstraintComponentList.m,MakeFreeSourceVariables.m,RescaleNullVector.m,ExtractPart.m,ExtractReparameterisationMatrix.m,ExtractDenominator.m,DenominatorOfElement.m,ExtractSecularEquation.m,MasslessAnalysisOfTotal.m,AssistedResidue.m,NumericalLowestOrder.m,LowestOrder.m,ObtainTermReplacements.m,ProcessPart.m,ResidueFormula.m,NullResidue.m,PerformReduction.m,PrepareReduction.m,BasicNullResidue.m,TestDependency.m,ReparameteriseSources.m,ConstrainInLightcone.m,ExpressInLightcone.m,FullyCanonicalise.m,FullyExpandSources.m,MakeSaturatedMatrix.m,MatrixFromSymbolic.m,MatrixToSymbolic.m,Repartition.m,SourcesToComponents.m,ExaminePoleOrder.m,ConstructLightcone.m,ConvertLightcone.m,CarefullyOrthogonalise.m,ParameterisedNullVectorQ.m,GrabExpression.m,SimplifyIfSmall.m,GradualExpand.m,BatchExpanded.m,ConsolidateUnmakeSymbolic.m,GradualExpandSubTask.m,InitialExpand.m,ConsolidateFinalElement.m,IntermediateRules.m,MakeSymbolic.m,ManualPseudoInverse.m,DistributeConjugate.m,UnmakeSymbolic.m,ConjectureInverse.m,CreateList.m,IsNullVectorOfSpace.m,MinimalExampleCaseNullSpace.m,CommonNullVector.m,CleanNullVector.m,EnsureLinearInCouplings.m,RemoveReferencesToMomentum.m,SymbolicNullSpace.m,ConjectureNullSpace.m,NonTrivialDot.m,ToCovariantForm.m,GetHermitianPart.m,ConstructOperator.m,FourierLagrangian.m,TimedReduce.m,CombineAssociations.m,ConstructLinearAction.m,ConstructWaveOperator.m,ValidateMaxLaurentDepth.m,ValidateTheoryName.m,ConstructUnitarityConditions.m,ConstructSourceConstraints.m,GetDiagram.m,NRoot.m,ProtractList.m,StripFactors.m,ParticleRow.m,ParticleRows.m,CLICallStack.m,Status.m,GUICallStack.m,ShowIfSmall.m,TheoryRows.m,UnresolvedPoleRows.m,UnresolvedPoleRow.m,AbbreviationRow.m,SourceConstraintRows.m,UnitarityRow.m,Abbreviate.m,ReplaceRadicals.m,CarefulReplace.m,GetAbbreviationRules.m,ShortEnoughQ.m,StripRadicals.m,WignerGrid.m,AllParticleRows.m,ValidateLagrangian.m,ConstructMassiveAnalysis.m,ConstructMasslessAnalysis.m,ConstructSaturatedPropagator.m,ConstructSpectrograph.m,UpdateTheoryAssociation.m,Diagnostic.m,MonitorParallel.m,ParallelGrid.m,RaggedBlock.m,ReMagnify.m,NewFramed.m,NewParallelSubmit.m,Vectorize.m,CallStackEnd.m,NameOfFunction.m,StackStrip.m,ToNewCanonical.m,CallStackBegin.m,Colours.m,StackSetDelayed.m,DefGeometry.m,ReloadPackage.m,DefField.m,ParticleSpectrum.m,PSALTer.m},morekeywords=[2]{%
init.wl},morekeywords=[3]{%
},morekeywords=[4]{%
FieldKinematics.pdf,GitHubLogo.pdf,GitLabLogo.pdf,ParticleSpectrograph.pdf,FeynmanDiagramHexic.pdf,FeynmanDiagramQuadratic.pdf,FeynmanDiagramQuartic.pdf,FeynmanDiagramSpin0ParityEvenResolved.pdf,FeynmanDiagramSpin0ParityEvenUnresolvedDouble.pdf,FeynmanDiagramSpin0ParityEvenUnresolvedQuadruple.pdf,FeynmanDiagramSpin0ParityEvenUnresolvedQuintuple.pdf,FeynmanDiagramSpin0ParityEvenUnresolvedTriple.pdf,FeynmanDiagramSpin0ParityNoneResolved.pdf,FeynmanDiagramSpin0ParityNoneUnresolvedDouble.pdf,FeynmanDiagramSpin0ParityNoneUnresolvedQuadruple.pdf,FeynmanDiagramSpin0ParityNoneUnresolvedQuintuple.pdf,FeynmanDiagramSpin0ParityNoneUnresolvedTriple.pdf,FeynmanDiagramSpin0ParityOddResolved.pdf,FeynmanDiagramSpin0ParityOddUnresolvedDouble.pdf,FeynmanDiagramSpin0ParityOddUnresolvedQuadruple.pdf,FeynmanDiagramSpin0ParityOddUnresolvedQuintuple.pdf,FeynmanDiagramSpin0ParityOddUnresolvedTriple.pdf,FeynmanDiagramSpin1ParityEvenResolved.pdf,FeynmanDiagramSpin1ParityEvenUnresolvedDouble.pdf,FeynmanDiagramSpin1ParityEvenUnresolvedQuadruple.pdf,FeynmanDiagramSpin1ParityEvenUnresolvedQuintuple.pdf,FeynmanDiagramSpin1ParityEvenUnresolvedTriple.pdf,FeynmanDiagramSpin1ParityNoneResolved.pdf,FeynmanDiagramSpin1ParityNoneUnresolvedDouble.pdf,FeynmanDiagramSpin1ParityNoneUnresolvedQuadruple.pdf,FeynmanDiagramSpin1ParityNoneUnresolvedQuintuple.pdf,FeynmanDiagramSpin1ParityNoneUnresolvedTriple.pdf,FeynmanDiagramSpin1ParityOddResolved.pdf,FeynmanDiagramSpin1ParityOddUnresolvedDouble.pdf,FeynmanDiagramSpin1ParityOddUnresolvedQuadruple.pdf,FeynmanDiagramSpin1ParityOddUnresolvedQuintuple.pdf,FeynmanDiagramSpin1ParityOddUnresolvedTriple.pdf,FeynmanDiagramSpin2ParityEvenResolved.pdf,FeynmanDiagramSpin2ParityEvenUnresolvedDouble.pdf,FeynmanDiagramSpin2ParityEvenUnresolvedQuadruple.pdf,FeynmanDiagramSpin2ParityEvenUnresolvedQuintuple.pdf,FeynmanDiagramSpin2ParityEvenUnresolvedTriple.pdf,FeynmanDiagramSpin2ParityNoneResolved.pdf,FeynmanDiagramSpin2ParityNoneUnresolvedDouble.pdf,FeynmanDiagramSpin2ParityNoneUnresolvedQuadruple.pdf,FeynmanDiagramSpin2ParityNoneUnresolvedQuintuple.pdf,FeynmanDiagramSpin2ParityNoneUnresolvedTriple.pdf,FeynmanDiagramSpin2ParityOddResolved.pdf,FeynmanDiagramSpin2ParityOddUnresolvedDouble.pdf,FeynmanDiagramSpin2ParityOddUnresolvedQuadruple.pdf,FeynmanDiagramSpin2ParityOddUnresolvedQuintuple.pdf,FeynmanDiagramSpin2ParityOddUnresolvedTriple.pdf,FeynmanDiagramSpin3ParityEvenResolved.pdf,FeynmanDiagramSpin3ParityEvenUnresolvedDouble.pdf,FeynmanDiagramSpin3ParityEvenUnresolvedQuadruple.pdf,FeynmanDiagramSpin3ParityEvenUnresolvedQuintuple.pdf,FeynmanDiagramSpin3ParityEvenUnresolvedTriple.pdf,FeynmanDiagramSpin3ParityNoneResolved.pdf,FeynmanDiagramSpin3ParityNoneUnresolvedDouble.pdf,FeynmanDiagramSpin3ParityNoneUnresolvedQuadruple.pdf,FeynmanDiagramSpin3ParityNoneUnresolvedQuintuple.pdf,FeynmanDiagramSpin3ParityNoneUnresolvedTriple.pdf,FeynmanDiagramSpin3ParityOddResolved.pdf,FeynmanDiagramSpin3ParityOddUnresolvedDouble.pdf,FeynmanDiagramSpin3ParityOddUnresolvedQuadruple.pdf,FeynmanDiagramSpin3ParityOddUnresolvedQuintuple.pdf,FeynmanDiagramSpin3ParityOddUnresolvedTriple.pdf},morekeywords=[5]{%
FieldKinematics.png,GitHubLogo.png,GitLabLogo.png,ParticleSpectrograph.png},morekeywords=[6]{%
FeynmanDiagramHexic.tex,FeynmanDiagramQuadratic.tex,FeynmanDiagramQuartic.tex,Preamble.tex,FeynmanDiagramSpin0ParityEvenResolved.tex,FeynmanDiagramSpin0ParityEvenUnresolvedDouble.tex,FeynmanDiagramSpin0ParityEvenUnresolvedQuadruple.tex,FeynmanDiagramSpin0ParityEvenUnresolvedQuintuple.tex,FeynmanDiagramSpin0ParityEvenUnresolvedTriple.tex,FeynmanDiagramSpin0ParityNoneResolved.tex,FeynmanDiagramSpin0ParityNoneUnresolvedDouble.tex,FeynmanDiagramSpin0ParityNoneUnresolvedQuadruple.tex,FeynmanDiagramSpin0ParityNoneUnresolvedQuintuple.tex,FeynmanDiagramSpin0ParityNoneUnresolvedTriple.tex,FeynmanDiagramSpin0ParityOddResolved.tex,FeynmanDiagramSpin0ParityOddUnresolvedDouble.tex,FeynmanDiagramSpin0ParityOddUnresolvedQuadruple.tex,FeynmanDiagramSpin0ParityOddUnresolvedQuintuple.tex,FeynmanDiagramSpin0ParityOddUnresolvedTriple.tex,FeynmanDiagramSpin1ParityEvenResolved.tex,FeynmanDiagramSpin1ParityEvenUnresolvedDouble.tex,FeynmanDiagramSpin1ParityEvenUnresolvedQuadruple.tex,FeynmanDiagramSpin1ParityEvenUnresolvedQuintuple.tex,FeynmanDiagramSpin1ParityEvenUnresolvedTriple.tex,FeynmanDiagramSpin1ParityNoneResolved.tex,FeynmanDiagramSpin1ParityNoneUnresolvedDouble.tex,FeynmanDiagramSpin1ParityNoneUnresolvedQuadruple.tex,FeynmanDiagramSpin1ParityNoneUnresolvedQuintuple.tex,FeynmanDiagramSpin1ParityNoneUnresolvedTriple.tex,FeynmanDiagramSpin1ParityOddResolved.tex,FeynmanDiagramSpin1ParityOddUnresolvedDouble.tex,FeynmanDiagramSpin1ParityOddUnresolvedQuadruple.tex,FeynmanDiagramSpin1ParityOddUnresolvedQuintuple.tex,FeynmanDiagramSpin1ParityOddUnresolvedTriple.tex,FeynmanDiagramSpin2ParityEvenResolved.tex,FeynmanDiagramSpin2ParityEvenUnresolvedDouble.tex,FeynmanDiagramSpin2ParityEvenUnresolvedQuadruple.tex,FeynmanDiagramSpin2ParityEvenUnresolvedQuintuple.tex,FeynmanDiagramSpin2ParityEvenUnresolvedTriple.tex,FeynmanDiagramSpin2ParityNoneResolved.tex,FeynmanDiagramSpin2ParityNoneUnresolvedDouble.tex,FeynmanDiagramSpin2ParityNoneUnresolvedQuadruple.tex,FeynmanDiagramSpin2ParityNoneUnresolvedQuintuple.tex,FeynmanDiagramSpin2ParityNoneUnresolvedTriple.tex,FeynmanDiagramSpin2ParityOddResolved.tex,FeynmanDiagramSpin2ParityOddUnresolvedDouble.tex,FeynmanDiagramSpin2ParityOddUnresolvedQuadruple.tex,FeynmanDiagramSpin2ParityOddUnresolvedQuintuple.tex,FeynmanDiagramSpin2ParityOddUnresolvedTriple.tex,FeynmanDiagramSpin3ParityEvenResolved.tex,FeynmanDiagramSpin3ParityEvenUnresolvedDouble.tex,FeynmanDiagramSpin3ParityEvenUnresolvedQuadruple.tex,FeynmanDiagramSpin3ParityEvenUnresolvedQuintuple.tex,FeynmanDiagramSpin3ParityEvenUnresolvedTriple.tex,FeynmanDiagramSpin3ParityNoneResolved.tex,FeynmanDiagramSpin3ParityNoneUnresolvedDouble.tex,FeynmanDiagramSpin3ParityNoneUnresolvedQuadruple.tex,FeynmanDiagramSpin3ParityNoneUnresolvedQuintuple.tex,FeynmanDiagramSpin3ParityNoneUnresolvedTriple.tex,FeynmanDiagramSpin3ParityOddResolved.tex,FeynmanDiagramSpin3ParityOddUnresolvedDouble.tex,FeynmanDiagramSpin3ParityOddUnresolvedQuadruple.tex,FeynmanDiagramSpin3ParityOddUnresolvedQuintuple.tex,FeynmanDiagramSpin3ParityOddUnresolvedTriple.tex},morekeywords=[7]{%
convert_logos.sh,compile_diagrams.sh,generate_sources.sh},morekeywords=[8]{%
ASCIILogo.txt},morekeywords=[9]{%
GitHubLogo.svgz,GitLabLogo.svgz},morekeywords=[10]{%
LICENSE.md,README.md},morekeywords=[11]{cd,git,clone,r,cp,tree,mathematica,sudo,pacman,S},
keywordstyle=[1]\textbf,keywordstyle=[2]\texttt,keywordstyle=[3]\color{Gray}\textbf,keywordstyle=[4]\color{Gray}\textbf,keywordstyle=[5]\color{Gray}\textbf,keywordstyle=[6]\color{Gray}\textbf,keywordstyle=[7]\color{Gray}\textbf,keywordstyle=[8]\color{Black}\textbf,keywordstyle=[9]\color{Gray}\textbf,keywordstyle=[10]\color{Gray}\textbf,keywordstyle=[11]\color{black}\textbf,alsoletter={./},basicstyle=\color{gray}\ttfamily,moredelim=[s][\color{gray}\textbf]{[user@system }{]}}

\definecolor{backing}{rgb}{0.96,0.96,0.96}

%% file: MacrosLstKeywords.tex
\lstdefinelanguage[PSALTer]{Mathematica}[]{Mathematica}{
	alsoletter={<>},
	morekeywords=[2]{brown,PacletInstall,DrazinInverse,mx,wl,Association,KernelID,OptionsPattern,OptionValue,Private,Head,DistributeDefinitions,Portrait,BlackBold,Gray,GrayBold},
	morekeywords=[3]{SymmetryOf,xAct,xAct`,xTensor,xCore,xPerm,xTras,SymManipulator,AllowUpperDerivatives,Antisymmetric,ChangeCovD,Christoffel,ConstantSymbolQ,ContractMetric,ContractMetrics,DefConstantSymbol,DefNiceConstantSymbol,DefScalarFunction,DefTensor,delta,DependenciesOfTensor,IndicesOfVBundle,Labels,LI,LieDToCovD,MakeRule,NoScalar,OverDerivatives,ParamD,PD,PrintAs,Projected,ScalarFunctionQ,ScreenDollarIndices,SeparateMetric,SlotsOfTensor,Symmetric,SymmetryGroupOfTensor,ToCanonical,CommuteCovDs,xTensorQ,Zero,Tensors,ConstantSymbols,DefManifold,IndexRange,DefMetric,FlatMetric,SymCovDQ,DefCovD,MetricOn,Scalar,CollectTensors,DDIs,AutomaticRules,VarD,NavyBlueBold},
	morekeywords=[4]{BlueBold,DefCanonicalField,PoissonBracket,Parallel,FindAlgebra,Constraints,Verify,TotalFrom,TotalTo,PrependTotalFrom,PrependTotalTo,\$ToRulesTotal,\$FromRulesTotal,\$DynamicalMetric,\$ManualSmearing,a,b,c,CD,ChristoffelCD,ConjugateMomentumG,d,DetG,e,EinsteinCD,En,Eps,epsilonG,f,g,G,h,i,j,k,KretschmannCD,l,m,M3,n,o,p,P,q,r,RicciCD,RicciScalarCD,RiemannCD,s,SchoutenCD,SymRiemannCD,t,TangentM3,TetraG,TFRicciCD,u,v,w,WeylCD,x,y,z,FieldSymbol,MomentumSymbol},
	morekeywords=[5]{<Fld>,ConjugateMomentum<Fld>,<Expr>,<Ind1>,<Ind2>,<Symm>,<SymmInd1>,<SymmInd2>,<Op1>,<Op2>,<Rle>,<Fctr1>,<Fctr2>,<Fctr3>,ScriptK,FromScriptK,ScriptL,FromScriptL,ScriptQ,FromScriptQ,TraceScriptK,FromTraceScriptK,ScriptLContraction,FromScriptLContraction,ScriptQSingleContraction,FromScriptQSingleContraction,TraceScriptQ,FromTraceScriptQ,ReducedHamiltonianConstraint,FromReducedHamiltonianConstraint,ReducedHamiltonianOrderUnity,FromReducedHamiltonianOrderUnity,ReducedHamiltonianOrderWilsonCoefficient,FromReducedHamiltonianOrderWilsonCoefficient,WilsonCoefficient,ReducedMomentumConstraint,FromReducedMomentumConstraint,OrderUnityBracketValue,CrossBracket01Value,CrossBracket10Value,BracketKValue,ExtractBracketAnatomy,Result,CosmologicalConstantOrder,WilsonCoefficientOrder,StandardSimplify,CosmologicalConstant,EinsteinConstant,TraceConjugateMomentumG,FromTraceConjugateMomentumG,Lapse,Shift,GravitationalCoupling,SuperHamiltonian,FromSuperHamiltonian,SuperMomentum,FromSuperMomentum,ScalarSmearingS,ScalarSmearingF,VectorSmearingCovariantS,VectorSmearingCovariantF,VectorSmearingContravariantS,VectorSmearingContravariantF,Expr,Fctr1,Fctr2,Fctr3,Fld,Rle,ConjugateMomentumFld,Ind1,Ind2,Symm,SymmInd1,SymmInd2,Op1,Op2,BracketExpr,RulesExpr,PrimaryConstraint,ConjugateMomentumExtrinsicCurvature,TraceConjugateMomentumExtrinsicCurvature,FromPrimaryConstraint,ExtrinsicCurvature,TotalHamiltonian,FromTotalHamiltonian,TensorSmearingCovariantF,TensorSmearingContravariantF,ToPrimaryShell,TraceExtrinsicCurvature,ToTraceExtrinsicCurvature,FromTraceExtrinsicCurvature,FromTraceConjugateMomentumExtrinsicCurvature,ToTraceConjugateMomentumExtrinsicCurvature,SecondaryConstraint,FromSecondaryConstraint,VectorSmearingCovariantS,VectorSmearingContravariantS,VectorSmearingCovariantF,VectorSmearingContravariantF,ToPrimaryShellExplicit,TensorSmearingCovariantS,TensorSmearingContravariantS,SkyBlueBold,CanonicalField1ea68330,CanonicalField8eca3bb0,CanonicalFieldbc2ebc94,CanonicalFieldc8d16b6b,CanonicalFieldccb6bb11,CanonicalFieldf25eb593,CanonicalFieldfa1caa2d,ConjugateMomentumCanonicalField1ea68330,ConjugateMomentumCanonicalField8eca3bb0,ConjugateMomentumCanonicalFieldbc2ebc94,ConjugateMomentumCanonicalFieldc8d16b6b,ConjugateMomentumCanonicalFieldccb6bb11,ConjugateMomentumCanonicalFieldf25eb593,ConjugateMomentumCanonicalFieldfa1caa2d,CouplingConstant6d9a09c9,CouplingConstant750dbd1c,CouplingConstant83bce689,CouplingConstanta5eb6381,CouplingConstantad6be3ca,CouplingConstante5f53ecd,CouplingConstantef18d0ee,LagrangeMultiplier1ea68330,LagrangeMultiplier8eca3bb0,LagrangeMultiplierc8d16b6b,LagrangeMultiplierbc2ebc94,LagrangeMultiplierccb6bb11,LagrangeMultiplierf25eb593,LagrangeMultiplierfa1caa2d}}